\documentclass[fleqn,usenatbib]{mnras}

\usepackage{newtxtext,newtxmath}
\usepackage[T1]{fontenc}
\usepackage{graphicx}
\usepackage{dcolumn}

\usepackage[separate-uncertainty=true]{siunitx}
\DeclareSIUnit\parsec{pc}
\DeclareSIUnit\year{yr}
\DeclareSIUnit\years{yrs}
\DeclareSIUnit\erg{erg}
\DeclareSIUnit\gauss{G}

\usepackage{booktabs}
\usepackage{orcidlink}
\usepackage{xspace}
\usepackage{commath}
\usepackage{amsmath}
\usepackage{longtable}
\usepackage[capitalise]{cleveref}
\crefname{figure}{Fig.}{Figs.}
\crefname{section}{Sec.}{Secs.}
\crefname{appendix}{Apx.}{Apx.}
\crefname{equation}{Eq.}{Eqs.}
\crefname{align}{Eq.}{Eqs.}

\usepackage{hyperref}
\definecolor{ntnu}{RGB}{0, 80, 158}
\hypersetup{
   colorlinks = true,
   citecolor  = ntnu,
   urlcolor   = ntnu,
   linkcolor  = ntnu,
}

\graphicspath {{images/}}

\newcommand{\lbol}{L_\text{bol}}
\newcommand{\disc}{_\text{disc}}
\newcommand{\lhx}{L_{2-10\,\si{\kilo\electronvolt}}}
\newcommand{\ir}{_\text{IR}}
\newcommand{\rfs}{R_\text{fs}}
\newcommand{\rsh}{R_\text{sh}}
\newcommand{\emax}{E_\text{max}}
\newcommand{\vw}{v_\text{w}}

\title[UHECRs from UFOs in AGN]{Ultra-high-energy cosmic rays from ultra-fast outflows of active galactic nuclei}

\author[D. Ehlert et al.]{
Domenik Ehlert \orcidlink{0000-0002-4322-6400},$^{1}$\thanks{domenik.ehlert@ntnu.no}
Foteini Oikonomou \orcidlink{0000-0002-0525-3758},$^{1}$
Enrico Peretti \orcidlink{0000-0003-0543-0467}$^{2}$
\\
$^{1}$Institutt for fysikk, Norwegian University of Science and Technology, Høgskoleringen 5, NO-7491 Trondheim, Norway\\
$^{2}$Université Paris Cité, CNRS, Astroparticule et Cosmologie, 10 Rue Alice Domon et Léonie Duquet, F-75013 Paris, France
}
    
\date{Accepted XXX. Received YYY; in original form ZZZ}

\pubyear{2025}

\begin{document}
\label{firstpage}
\pagerange{\pageref{firstpage}--\pageref{lastpage}}
\maketitle

\begin{abstract}
We investigate ultra-fast outflows (UFOs) in active galactic nuclei (AGN) as potential sources of ultra-high-energy cosmic rays (UHECRs). We focus on cosmic-ray nuclei, an aspect not explored previously. These large-scale, mildly-relativistic outflows, characterised by velocities up to half the speed of light, are a common feature of AGN. We study the cosmic-ray spectrum and maximum energy attainable in these environments with 3D CRPropa simulations and apply our method to 86 observed UFOs. Iron nuclei can be accelerated up to $\sim10^{20}\,$eV at the wind-termination shock in some UFOs, but the escaping flux is strongly attenuated due to photonuclear interactions with intense AGN photon fields. The maximum energy of nuclei escaping a typical UFO is limited by photodisintegration to below $\sim 10^{17}\,$eV. However, in the most extreme $5-10\%$ of UFOs, helium (nitrogen) [iron] nuclei can escape with energy exceeding $10^{17.4}$ ($10^{17.8}$) [$10^{18.4}$]\,eV. Protons and neutrons, either primaries or by-products of photodisintegration, escape UFOs with little attenuation, with half of the observed UFOs reaching energies exceeding $10^{18}\,$eV. Thus, UFOs emerge as viable sources of the diffuse cosmic-ray flux between the end of the Galactic cosmic-rays and the highest-energy extragalactic flux. For a few UFOs in our sample, nuclei escape without photodisintegration with energy up to $10^{19.8}\,$eV. This occurs during low-emission states of the AGN, which would make UFOs intermittent sources of UHECR nuclei up to the highest observed energies. The role of UFOs as UHECR sources is testable with neutrino telescopes due to a substantial accompanying flux of PeV neutrinos.

\end{abstract}

\maketitle

\section{Introduction}
    Accreting supermassive black holes in the centres of galaxies, known as \textit{active galactic nuclei} (AGN), are an important source of high-energy radiation and particles in the Universe. Fast outflows are a common feature of AGN \citep[see][for a review]{King:2015caa}, which have been extensively studied in the context of high-energy neutrinos and gamma rays \citep[e.g.][]{Tamborra:2014xia,Lamastra:2016axo,Lamastra:2019zss,Wang:2016zyi,Liu:2017bjr,Padovani:2018hfm}. Recently, \citet{Peretti:2023xqk} (hereafter P23) proposed that protons can be accelerated up to a few exa-electronvolt (EeV) at the shocks resulting from the collision of ultra-fast AGN-driven winds with the interstellar medium. According to the conventional model, where cosmic-ray acceleration depends on magnetic fields that confine the particles, the maximum energy is proportional to the nuclear charge~\citep[cf.\ ``Peters' cycle''][]{Peters:1961}, suggesting that iron-like nuclei could potentially be accelerated to around 100 exa-electronvolts, if they are not rapidly destroyed due to photodisintegration in the source region. In this scenario, \textit{ultra-fast outflows} (UFOs) could serve as a source of the observed flux of \textit{ultra-high-energy cosmic rays} (UHECRs). A stacking analysis of \textsc{Fermi}-LAT data from 11 nearby radio-quiet AGN with ultra-fast outflows ($\vw\gtrsim 0.1c$) has revealed gamma-ray emission from these objects with energies of up to several hundred GeV \citep{Fermi-LAT:2021ibj}, suggesting that particles can be accelerated to high energies in these environments.
    
    Fast ionised outflows from galactic nuclei have long been considered an important feedback mechanism suppressing the star formation rate of their host galaxy~\citep{Silk:1997xw,Crenshaw:2003bh,Kormendy:2013dxa,Tombesi:2015noa}. Ultra-fast outflows constitute a subset of these AGN winds characterised by velocities of up to $\sim0.6c$~\citep{Gianolli:2024jkq}, strong ionisation, and typical distances from the central AGN engine of $10^2-10^4$ Schwarzschild radii ($\lesssim\SI{10}{\parsec}$;~\citealp{Gofford:2015aka,Laha:2020ygb}), suggesting an origin close to the accretion disc. While the mechanism for powering such extreme outflows is still under debate, common interpretations involve winds driven by the radiation of the accretion disc (e.g.~\citealp{Murray:1995ApJ,Ohsuga:2011jk,Jiang:2014tpa,Sadowski:2015hia,Hashizume:2014jva}), or by magnetohydrodynamic effects near the central black hole (e.g.~\citealp{Blandford:1982di,Fukumura:2010ApJ,Fukumura:2017nef,Kraemer:2017rge}). Magnetic reconnection in a magnetically arrested accretion disc and the repeated passage of a compact object through the disc were also proposed~\citep{Sukova:2021thm,Sukova:2023eot}.
    
    The number of active galactic nuclei with observed ultra-fast outflows has increased rapidly over the last years, including both jetted and non-jetted AGN~\citep{Tombesi:2010a,Tombesi:2010b,Tombesi:2015noa,Gofford:2012gw}. The fraction of AGN with observed (ultra-fast) outflows is $\sim50\pm20\%$ for both populations~
    \citep[][see also \citet{Reynolds:1995ve,Crenshaw:2003bh,McKernan:2007MNRAS.379.1359M,Laha:2014faa}]{Tombesi:2010a}, indicating that the presence of a jet and fast wide-angle outflows are not mutually exclusive~\citep{Mestici:2024pjt}.

    The potential role of AGN UFOs as sources of high-energy cosmic rays is linked to one of the mysteries concerning the origin of UHECRs, namely the transition region between Galactic and extragalactic cosmic rays. The identification of an iron ``knee'' feature in the cosmic-ray spectrum around energy $\sim10^{16.8}\,\si{\electronvolt}$~\citep{PhysRevLett.107.171104} likely signals the end of the Galactic cosmic-ray flux. At higher energies, above the ``ankle'' feature of the cosmic ray spectrum at energy $\sim10^{18.7}\,\si{\electronvolt}$ the combined spectrum and composition can be described by a Peters' cycle, attributed to extragalactic sources. This leaves a well-known ``gap'' in flux between Galactic and extragalactic contributions to the spectrum~\citep{Hillas:2005cs,DeDonato:2008wq}. As we show in what follows, UFOs in AGN are an excellent candidate that can account for this spectral component in terms of energetics, maximum energy and chemical composition.
    
    In this paper, we investigate for the first time the maximum energy that cosmic-ray nuclei can obtain via diffusive shock acceleration in UFOs and the potential contribution of UFOs in AGN to the diffuse UHECR and neutrino flux. The assumed structure of the AGN-UFO system is described in \cref{sec:ufo_system}; and in \cref{sec:interactions} we present our treatment of the relevant interaction, escape, and acceleration processes. The results for a representative benchmark UFO and for a sample of $86$ observed UFOs are presented in \cref{sec:results}. We discuss our findings and possible caveats in \cref{sec:discussion}, and conclude in \cref{sec:summary} that UFOs are viable sources of the observed UHECR flux below the ankle of the cosmic-ray spectrum, providing an astrophysical explanation that fills the ``gap'' region of the cosmic-ray spectrum between the Galactic and high-energy extragalactic flux components. 

\section{The AGN-UFO environment}\label{sec:ufo_system}
    \subsection{Geometry of the shocks}
\begin{figure}
        \centering
        \includegraphics[width=\linewidth]{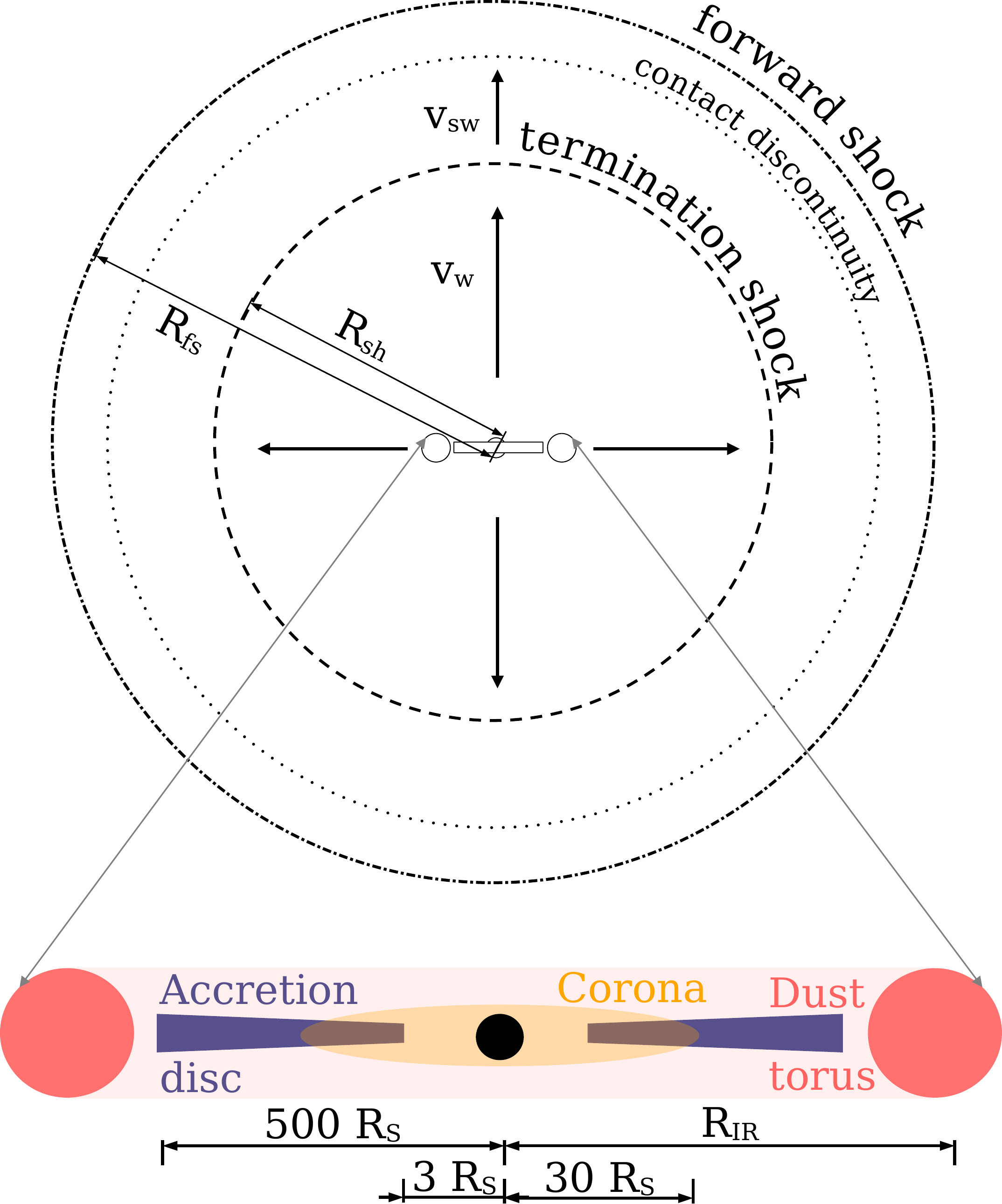}
        \caption{Schematic representation of the UFO shocks (top) and AGN structure (bottom). Distances are not to scale. The torus can be larger than the radius of the wind termination shock in bright AGN.}\label{fig:schematic_agn}
    \end{figure}

    The collision of a supersonic wind with the ambient medium leads to the development of two shocks; a forward shock (FS) that propagates through the interstellar medium (ISM), and an inward-oriented wind termination shock (SH) that separates the unshocked fast and cold wind from the heated shocked wind~\citep{Faucher-Giguere:2012pyj}. The shocked wind and the shocked ambient medium are separated by a contact discontinuity; see \cref{fig:schematic_agn}. The cooling timescale of the shocked wind is similar to the age of the UFO, and is therefore approximately adiabatic, while the shocked ISM cools rapidly and forms a thin shell of cold, dense material close to the forward shock ($R_\text{cd}\simeq\rfs$;~\citealp{Faucher-Giguere:2012pyj,Morlino:2021xpo,Peretti:2021yhc,Peretti:2023xqk}). Initially, the wind expands freely into the ISM; however, after the mass of swept-up matter becomes dynamically relevant the outflow enters the deceleration phase where the radii of the shocks of the energy-conserving outflow are given by~\citep{Weaver+77,Koo:1992ApJa,Koo:1992ApJb,Faucher-Giguere:2012pyj,Morlino:2021xpo,Peretti:2021yhc,Peretti:2023xqk}
    \begin{align}\label{eq:R_sh}
     R_\text{sh} \sim \SI{23}{\parsec} &\left(\frac{t_\text{age}}{\si{\mega\year}}\right)^{2/5} \left(\frac{L_\text{kin}}{10^{38}\,\si{\erg\per\second}}\right)^{3/10} \nonumber\\
            &\times\left(\frac{n_\text{ISM}}{\si{\per\centi\meter\cubed}}\right)^{-3/10} \left(\frac{\vw}{10^3\,\si{\kilo\meter\per\second}}\right)^{-1/2}
    \end{align}
    and
    \begin{align}\label{eq:R_fs}
        R_\text{fs} \sim \SI{76}{\parsec} \left(\frac{t_\text{age}}{\si{\mega\year}}\right)^{3/5} \left(\frac{L_\text{kin}}{10^{38}\,\si{\erg\per\second}}\right)^{1/5} \left(\frac{n_\text{ISM}}{\si{\per\centi\meter\cubed}}\right)^{-1/5}
    \end{align}
    respectively, with $t_\text{age}$ the age of the outflow, $L_\text{kin}=\dot{M}_\text{w}\,\vw^2/2$ the kinetic energy of the wind, $\dot{M}_\text{w}$ the mass outflow rate, $n_\text{ISM}$ the (constant) ambient matter density, and $\vw$ the terminal velocity of the unshocked wind. These expressions are obtained by assuming a balance between the pressure of the hot, shocked wind in the bubble and the ram pressure of the inflowing material at the two shocks. If the wind is driven by radiation from the accretion disc, a close correlation is expected between the wind velocity and the AGN accretion rate~\citep{Reynolds:2012ph,Giustini:2019she}.

    Observations of local AGN indicate large opening angles of the wind of $\Omega/4\pi = 0.4-0.6$~\citep{Tombesi:2010a,Gofford:2012gw}. The termination shock provides an attractive site for the acceleration of cosmic rays up to ultra-high energies since the upstream escape of the particles is strongly suppressed because of the bubble-like geometry. The confinement efficiency is reduced for low covering factors of the wind; however, the pressure of the fast wind generally prevents the cosmic rays from diffusing far upstream.
    
    Because it is oriented inward, toward the fast wind, the termination shock maintains a large velocity gradient over time~\citep{Faucher-Giguere:2012pyj} whereas the velocity of the forward shock with respect to the ISM decreases until it becomes subsonic. This suggests that the wind termination shock can persist on longer timescales in approximate stationary conditions and that the available energy budget for particle acceleration is likely larger than for the forward shock. In the following, we consider only the wind termination shock as a site of cosmic-ray acceleration, referring to the inner region filled with the fast unshocked wind as the upstream and the region between the wind shock and the forward shock as the downstream.

    \subsection{Photon fields}\label{sec:photon_fields}
\begin{figure}
        \centering
        \includegraphics[width=\linewidth]{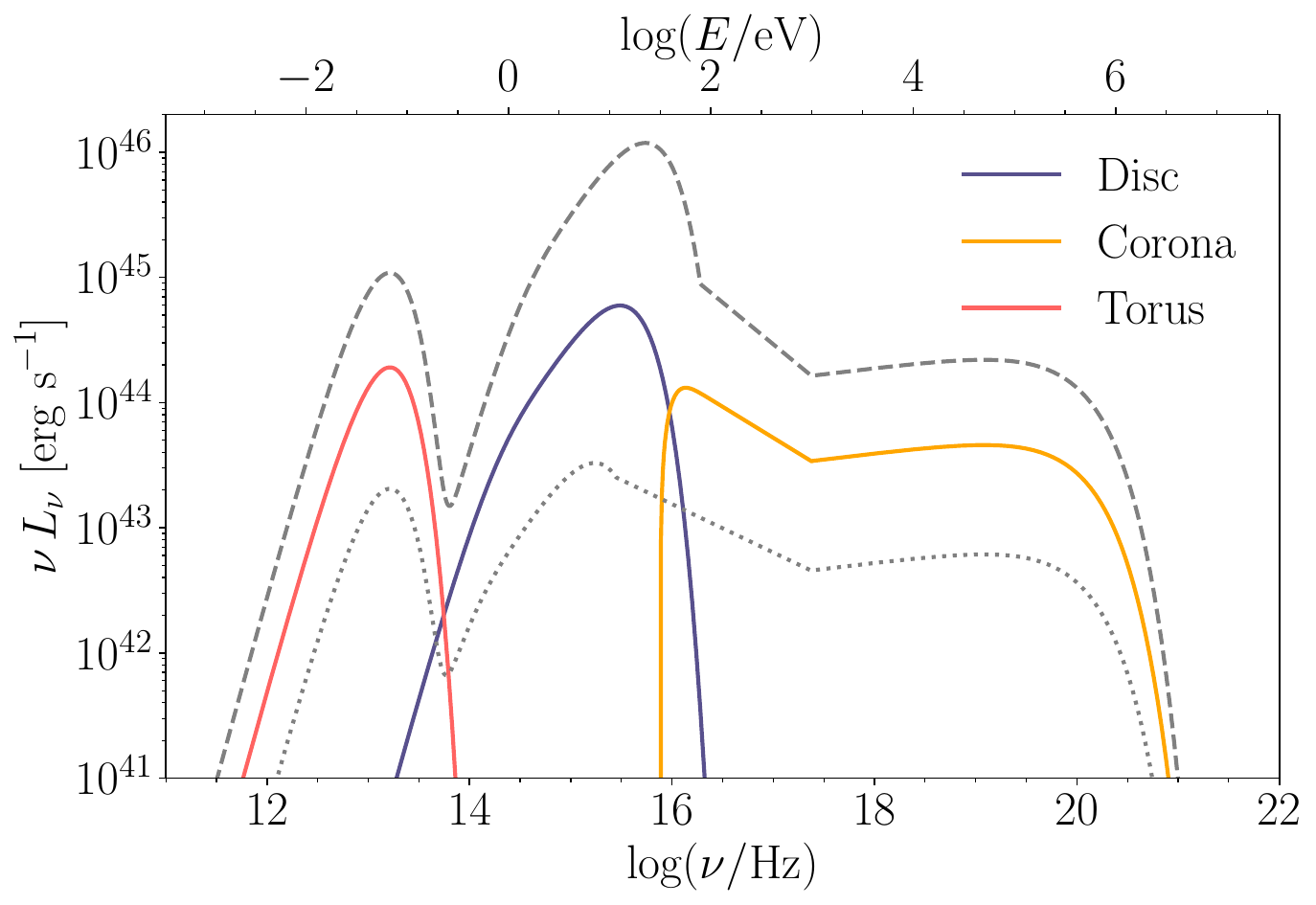}
        \caption{Spectral luminosity of the AGN photon fields for the benchmark scenario ($\lbol=10^{45}\,\si{\erg\per\second}$). The individual components are modelled as greybody (dust torus; red), multicolour greybody (accretion disc; blue), and broken power law (corona; orange) emission. They are normalised according to observed luminosity scaling factors. We also show the spectrum for an otherwise identical AGN with bolometric luminosity of $10^{44}\,\si{\erg\per\second}$ (dotted) and $10^{46}\,\si{\erg\per\second}$ (dashed) respectively.
        }\label{fig:photon_fields_benchmark}
    \end{figure}

    We use observed correlations of the AGN luminosity in different bands, summarised as luminosity scaling factors~\citep{Marconi:2004MNRAS,Mullaney:2011iq,Lusso:2009nq,Lusso:2012yv,Duras:2020A&A}, to normalise the luminosity of the accretion disc, corona, and dust torus relative to the bolometric luminosity of the AGN. The spectrum is modelled as a combination of re-normalised blackbody emission from the disc and torus, and a broken power law for the corona (see \cref{apx:photon_fields}). We assume a thin spherical shell distribution for the dust ``torus''. The effects of a possible thin ring-like distribution of the dust are discussed in \cref{apx:dust_distribution}. For each component (disc, corona, torus), we assume a $1/R^2$ decrease of the photon density at radii larger than the corresponding outer radius and a constant density when $R < R_i$ ($i\in$\{disc, corona, torus\}). In \cref{fig:photon_fields_benchmark} we show the external photon fields for the benchmark UFO as discussed in \cref{sec:results_benchmark}; see \cref{apx:photon_fields} for details on the derivation and the associated photon density. The synchrotron radiation produced by relativistic electrons is subdominant to the external photon fields, and the broad-line region is negligible compared to the other fields at distances from the AGN relevant for typical UFOs.

\section{Acceleration and escape of high-energy cosmic rays}\label{sec:interactions}
    We determine the maximum energy of the cosmic-ray nuclei by comparing the confinement time in the accelerator to the relevant interaction timescales in the AGN environment. At ultra-high energies, the relevant interactions of the cosmic-ray protons and nuclei are Bethe-Heitler (electron-positron) pair production~\citep{Bethe:1934za}, photopion production, and photodisintegration for cosmic-ray nuclei only. Proton-proton and nucleus-proton interactions are not relevant at the assumed matter density in the downstream region. We employ an effective two-zone model, which distinguishes between the acceleration stage near the wind termination shock and the subsequent escape toward the outer boundary of the UFO environment, assumed to coincide with the forward shock. Acceleration is treated with a simple semi-analytical model (\cref{sec:timescales}) to obtain the maximum energy, while suppression of the escaping UHECR flux is computed with 3D \textsc{CRPropa}~\citep{CRPropa2:2013,AlvesBatista:2016vpy,AlvesBatista:2022vem} Monte-Carlo simulations (\cref{sec:escape_ufo}).

    \subsection{Acceleration at the termination shock}\label{sec:timescales}
    We assume that particles are accelerated with an $E^{-2}$ power-law spectrum, as predicted by diffusive shock acceleration~\citep[see e.g.][]{Blandford1}, and an exponential suppression above the maximum achievable energy in agreement with acceleration models at wind termination shocks~\citep[][also P23]{Morlino:2021xpo,Peretti:2021yhc,Mukhopadhyay:2023kel}. We estimate the maximum energy by comparing the acceleration timescale to the escape and the various energy loss timescales in the UFO. We treat the external photon fields as isotropic in the wind frame.

    \subsubsection{Acceleration and escape at the shock}
    The acceleration timescale is approximately $\tau_\text{acc} \approx sD_1(p,B_1)/\vw^2$ (P23), where $s=4$ is the spectral index (strong shock), and the upstream diffusion coefficient $D_1(p,B_1)$ depends on the upstream magnetic field $B_1$ and the cosmic-ray momentum $p$. The magnetic field can be derived as $B_1(R) = \sqrt{2\mu_0\,U_{B_1}}$ from the upstream magnetic energy density $U_{B_1}(R) = \epsilon_B\,\rho_1(R)\,\vw^2$ and the mass density $\rho_1=\dot{M}_\text{w}/(4\,\pi\,R^2\,\vw)$. The acceleration timescale depends on the wind velocity $\vw$, the distance from the central object $R$, the mass outflow rate $\dot{M}_\text{w}$, and the fraction of the total energy that goes into the magnetic fields $\epsilon_B$. The diffusion coefficient is discussed below.

    The advection timescale can be estimated as
    \begin{equation}
        \tau_\text{adv} = \frac{R_\text{esc} - \rsh}{\langle v_\text{sw} \rangle} = \frac{4}{\vw}\frac{\rfs}{3}\left[\left(\frac{\rfs}{\rsh}\right)^2 - \frac{\rsh}{\rfs}\right]\,,
    \end{equation}
    where the final expression is obtained by volume averaging $\langle v_\text{sw} \rangle$ between $\rsh$ and $\rfs$ under the assumption that the contact discontinuity is a thin shell compared to the size of the entire downstream region, i.e.\ $R_\text{esc}\approx R_\text{fs}$. The downstream velocity profile of the shocked wind is given by $v_\text{sw}(R)=v_\text{sw}(\rsh)(\rsh/R)^2$ for ($\rsh\leq R\leq\rfs$), with $v_\text{sw}(\rsh)=\vw/4$ for a strong shock. The advection timescale is independent of the cosmic-ray energy.
    
    The diffusion time can be written as (P23)\footnote{To obtain agreement with the diffusion timescale obtained natively with \textsc{CRPropa} we multiply the diffusion coefficient in this expression by six to account for the three-dimensional nature of the model, cf.\ \citet{Globus:2007bi}.}
    \begin{equation}\label{eq:t_diff}
        \tau_\text{diff}(E) = \frac{(R_\text{esc} - R_\text{sh})^2}{D_2(E)}\,,
    \end{equation}
    again with $R_\text{esc}\approx\rfs$, and $D_2$ the downstream diffusion coefficient. Assuming a Kolmogorov turbulence spectrum ($\delta=5/3$), the diffusion coefficient can be written as
    \begin{equation}\label{eq:diffusion_coefficient}
        D(E) = \frac{c\,l_\text{c}}{6\pi}\left[\left(\frac{E}{E_\text{diff}}\right)^{2-\delta} + \frac{1}{2}\left(\frac{E}{E_\text{diff}}\right) + \frac{2}{3}\left(\frac{E}{E_\text{diff}}\right)^2\right]\,.
    \end{equation}
    Here, $l_\text{c}$ is the coherence length of the magnetic field and $E_\text{diff}$ is the characteristic diffusion energy, i.e. $2\pi r_L(E_\text{diff},Z,B)=l_\text{c}$ \citep{Harari:2002dy,Globus:2007bi,Muzio:2021zud}. The diffusion timescale decreases rapidly for larger energies due to the absence of resonant scattering modes, and the particles enter the regime of ``small pitch-angle scattering''. Under the assumption of a constant magnetic field in the downstream region (see \cref{sec:escape_ufo}), the diffusion coefficient $D_2$ and the timescale $\tau_\text{diff}$ are independent of the distance from the termination shock. The total escape time from the UFO is
    \begin{equation}\label{eq:t_esc}
        \tau_\text{esc}(E) = \left(\frac{1}{\tau_\text{adv}} + \frac{1}{\tau_\text{diff}(E)}\right)^{-1} + \tau_\text{free}\,,
    \end{equation}
    where $\tau_\text{free}\approx(\rfs-\rsh)/c$ is the minimum free-streaming escape time for cosmic rays that are emitted perpendicular to the wind termination shock away from the centre (see \citealp{Globus:2007bi,Muzio:2021zud}).
    
    \subsubsection{Pair production}
    For (Bethe-Heitler) pair production, we use the analytical approximation by \citet{Chodorowski:1992} which is based on the standard solution of \citet{Blumenthal:1970nn}. The energy loss timescale of a particle with charge $Z$ and Lorentz factor $\gamma$ is given by~\citep{Chodorowski:1992,Dermer:2009zz}
    \begin{equation}
        \tau_{BH}^{-1}(\gamma) = \alpha_f r_e^2 c Z^2 m_e c^2 \int_2^\infty\dif\epsilon\,n_\text{ph}\left(\frac{\epsilon}{2\gamma}\right)\frac{\varphi(\epsilon)}{\epsilon^2}\,,
    \end{equation}
    for any isotropic photon field $n_\text{ph}(\epsilon)$ with $\epsilon=E_\gamma / m_ec^2$, and the function $\varphi(\epsilon)$ as in \citet{Chodorowski:1992} (Eq.\ 3.12 onward). For any nucleus with an atomic number $Z$ and a mass number $A$, $\tau_{BH}^{-1}(A,Z)(E) = (Z^2/A)\,\tau_{BH}^{-1}(p)(E/A)$. Capture of the produced electrons by the nucleus can lead to a reduction in the net charge of the cosmic rays. This effect may become relevant for ultra-heavy nuclei ($Z>26$) in dense photon fields~\citep{Esmaeili:2024pwh}.

    \subsubsection{Photopion production}\label{sec:interactions_py}
    The energy loss timescale of protons due to photopion production is obtained by integrating the cross section and inelasticity over all photon energies \citep[e.g.][]{Stecker:1968uc,Dermer:2009zz}
    \begin{equation}\label{eq:interaction_rate}
        \tau_{p\gamma }^{-1}(\gamma_p)\cong\frac{c}{2\gamma_p^2}\int_0^\infty\dif\epsilon~\frac{n_\text{ph}(\epsilon)}{\epsilon^2}\int_0^{2\gamma_p\epsilon}\dif\epsilon'~\epsilon'\,\sigma_{p\gamma}(\epsilon')K_{p\gamma}(\epsilon')\,,
    \end{equation}
    where $\epsilon' = \gamma_p\epsilon(1-\beta_p\cos\theta)$ is the photon energy in the proton-rest-frame, $\gamma_p\approx E_p/m_p c^2$ the Lorentz factor of the proton, and $-1\leq\cos\theta\leq1$ corresponds to the angle between the momenta of the proton and photon in the lab frame. In the ultra-relativistic limit, where $\beta\approx1$, the possible values of $\epsilon'$ for a given photon energy in the lab frame $\epsilon$ are $0\leq\epsilon'\leq2\gamma_p\epsilon$. We take the tabulated cross sections from \textsc{CRPropa}~\citep{AlvesBatista:2016vpy}, and obtain the inelasticity $K_{p\gamma}(\epsilon')$ from simulations with \textsc{Sophia}~\citep{Mucke:1999yb}. The interaction rate of heavier nuclei scales non-trivially with their mass $A$ and charge $Z$; however, it can be approximated as
    \begin{equation}
        \lambda_{A\gamma}(E)\approx0.85\times\left[Z^\zeta\, \lambda_{p\gamma}\left(\frac{E}{A}\right) + (A-Z)^\zeta\,\lambda_{n\gamma}\left(\frac{E}{A}\right)\right]\,,
    \end{equation}
    where $\zeta=2/3$ for $A\leq8$ and $\zeta=1$ otherwise~\citep{CRPropa2:2013}. The interaction rates for protons $\lambda_{p\gamma}$ and neutrons $\lambda_{n\gamma}$ can be calculated with \cref{eq:interaction_rate} by setting the inelasticity to $K(\epsilon_r)\equiv1$.

    \subsubsection{Photodisintegration}
    We use \textsc{CRPropa} to generate the interaction rates and branching ratios for the photodisintegration of cosmic-ray nuclei. For heavy nuclei with $A\geq12$, this is based on precomputed cross sections from \textsc{TALYS 1.8}~\citep{Koning:2005ezu}. At lower masses, the nuclear models included in \textsc{TALYS} are not reliable and the respective cross sections were sourced from multiple references, which are listed in~\citet{CRPropa2:2013}. The energy loss rate due to photodisintegration is then obtained by convolving the interaction rate with the effective probability of losing the nucleus during an interaction which we take to be proportional to the relative mass loss $\dif A/A$. For $\dif A/A\to0$ this implies that a single interaction has minimal impact while for $\dif A/A\to1$ the nucleus is completely destroyed during an interaction. The average mass loss per photodisintegration event is calculated by summing over the contributions of all possible branching ratios and their respective probability.

    \subsection{Modelling the escape from the UFO environment}\label{sec:escape_ufo}
    \begin{figure}
        \centering
        \includegraphics[width=\linewidth]{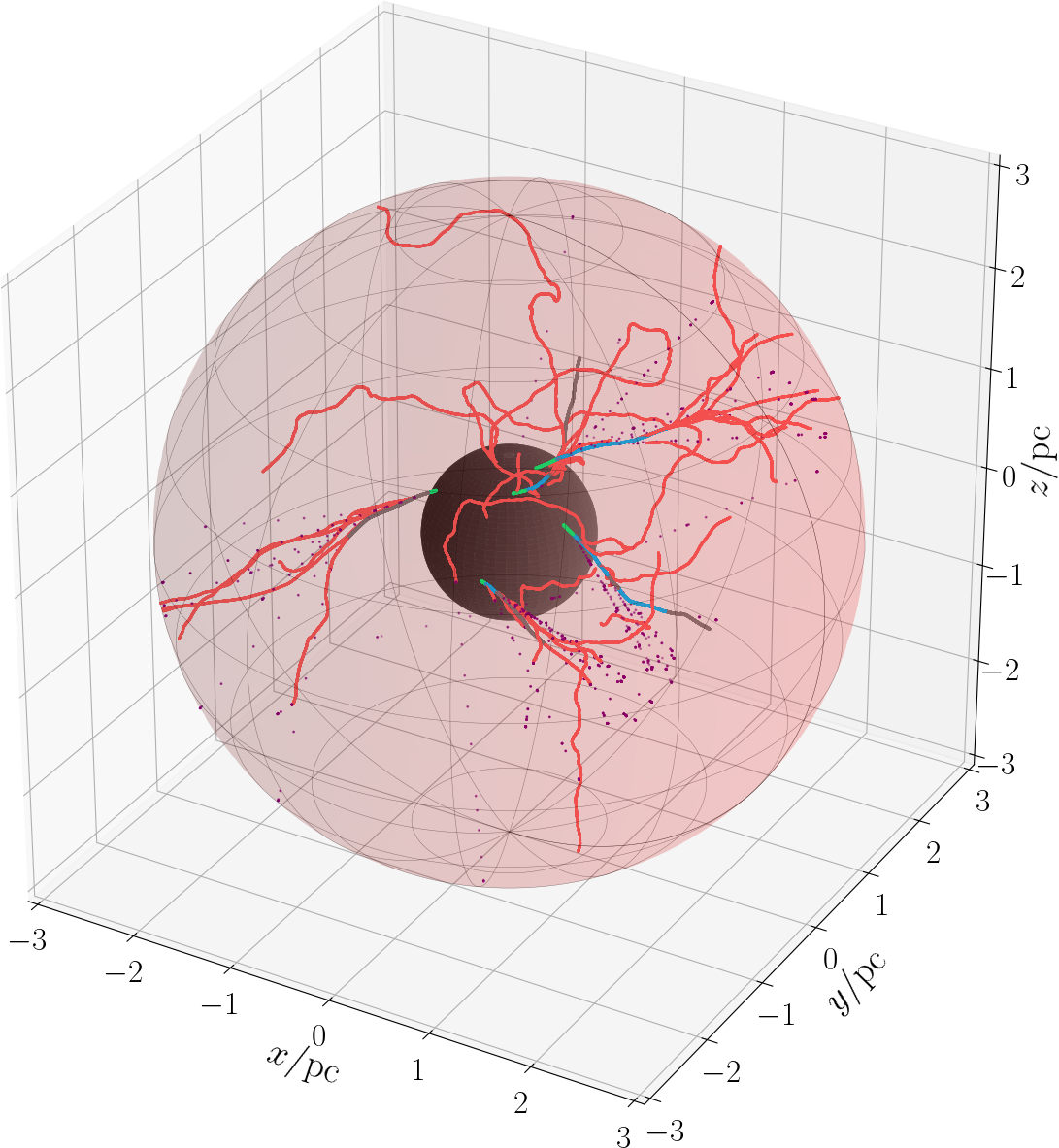}
        \caption{Illustration of the 3D \textsc{CRPropa} simulations performed to estimate the suppression of the cosmic-ray flux injected at the wind termination shock (black shell) until escape at the forward shock (red shell). We show the example of five primary nitrogen nuclei (green) injected with $10^{20}\,\si{\electronvolt}$. Produced secondaries include helium (grey), protons (red), and neutrons (purple, dotted). Intermediate nuclei between nitrogen and helium are indicated in blue.}\label{fig:cr_tracks}
    \end{figure}
    We model the escape of cosmic rays after acceleration at the wind termination shock with 3D Monte-Carlo simulations utilising \textsc{CRPropa} (see \cref{fig:cr_tracks}), where the cosmic rays are injected uniformly on a spherical shell corresponding to the termination shock according to the power-law spectrum with exponential suppression at the maximum energy as obtained from our semi-analytical approximation (see \cref{sec:timescales}). They are propagated in discrete steps with the Boris push method~\citep{Boris:1970} until they leave the UFO system through the forward shock at $\rfs$, in which case they are recorded as successfully escaped, or until they have lost enough energy due to interactions to fall below $E_\text{CR}<10^{15.5}\,\si{\electronvolt}$, in which case they are removed from the simulation. We also remove cosmic rays that return to the termination shock to maintain self-consistency of the injection spectrum. The weight of the remaining cosmic rays and secondary particles is adjusted to account for this leakage at the inner boundary. We inject particles in the simulation at a fiducial distance from the shock in order to minimise shock re-crossing. We assume as a reference one Larmor radius of protons at around $10^{18.7}\,\si{\electronvolt}$, which corresponds roughly to $\SI{0.05}{\parsec}$. We verified that the extent of this shift, while making the simulation more efficient, does not modify our result. This can be expected as, in the nearest neighbourhood of the shock, advection is dominating and the spectral shape is unmodified.
    
    Depending on their energy, the wind velocity, and the magnetic field strength, the transport is dominated by advection with the outflow (at low energy), diffusion (at intermediate energy) or quasi-ballistic free-streaming escape when the Larmor radius significantly exceeds the maximum coherence length of the magnetic field. As the wind launching region is typically much smaller than the size of the system and the wind quickly reaches its terminal wind speed, we assume a constant upstream velocity $\vw$. The velocity is reduced by a factor of four at the termination shock (strong shock) and decreases $\propto1/R^2$ in the downstream. This is implemented as the \texttt{SphericalAdvectionShock} module in \textsc{CRPropa}\footnote{We estimate the shock width as $\lambda=\vw/\omega_\text{p}\sim\mathcal{O}(\text{km})$, where $\omega_\text{p}$ is the proton plasma frequency. This corresponds to a thin shock for typical UFOs.}. The upstream magnetic field is
    \begin{equation}
        B(R) = \sqrt{2\mu_0U_B} = \left(2\mu_0\,\epsilon_B\,\frac{\dot{M}_\text{w}}{4\pi\,R^2}\,\vw\right)^{1/2} \propto R^{-1}\,,
    \end{equation}
    where $\epsilon_B$ is the fraction of the wind kinetic energy that is converted into magnetic field, and $\dot{M}_\text{w}$ is the mass outflow rate. At the shock, the field is compressed along two of the three spatial dimensions, resulting in an amplification of the average field strength by a factor of $\sqrt{11}$ (see e.g.\ P23, \citealp{Marcowith:2010pn}). For typical UFOs, the velocity of the shocked wind is less than the Alfv\'en velocity in the downstream, motivating the assumption of a constant magnetic field in this region. The magnetic field is simulated in \textsc{CRPropa} on a 3D grid with $\SI{5e-4}{\parsec}$ separation between the grid points. The typical field in the shocked wind is $\mathcal{O}(\SI{0.1}{\gauss})$, resulting in strong deflections of the cosmic rays and quasi-diffusive propagation.
    
    We include energy losses due to Bethe-Heitler pair production, photopion production, and photodisintegration by the photon fields of the accretion disc, corona and dust torus, and nuclear decay. We have modified the \textsc{CRPropa} code to simulate a radial scaling of the energy density of the tabulated photon fields. In addition, we have integrated the advection process into the standard Boris-Push particle propagation algorithm in \textsc{CRPropa} and included both the radial dependence of the advection field and the magnetic field. The modified version is available at \href{https://github.com/ehlertdo/CRPropa3/tree/outflows}{github.com/ehlertdo/CRPropa3}. We have verified that the escape spectra derived from the simulations for spatially constant photon fields agree well with simple analytical estimates (see \cref{apx:escape_analytical}).
    
    Cooling via hadronic interactions with the shocked ambient medium (SAM) is not included. If the density of the ISM is similar to ordinary galaxies ($\sim1\,\si{\per\centi\meter\cubed}$) then interactions in the SAM are negligible. If, however, the density is large ($n_\text{ISM}\sim10^3-10^4\,\si{\per\centi\meter\cubed}$)~\citep{Liu:2017bjr,Peretti:2023xqk} they may become relevant depending on the magnetic field in the shocked wind. For the fiducial ISM density and the same coherence length of the magnetic field as in the shocked wind (see \cref{tab:parameters_benchmark}), the time scale for proton-proton/nucleus interactions in the SAM is larger than the typical diffusive escape time for $B^\text{SAM}_\text{rms}\,\sqrt{l_c^\text{SAM}}\lesssim\SI{10}{\gauss\parsec}^{1/2}\,/\,Z$ at $E_\text{CR}\gtrsim10^{17}\,\si{\electronvolt}$. If the magnetic field in the shocked ambient medium is comparable to the field in the shocked wind then this condition is fulfilled, within our model, for all known UFOs. However, while the cosmic-ray flux will not be attenuated substantially when crossing the SAM, a significant flux of secondary neutrinos may still be produced (see P23).

    \begin{table}
    \centering
        \caption{Parameters of the AGN/UFO system in our benchmark scenario. The properties of the individual photon fields are described in \cref{apx:photon_fields}.}\label{tab:parameters_benchmark}
        \renewcommand{\arraystretch}{1.3}
        \begin{tabular}{llcl}
            \toprule
            Parameter               & Description           & Benchmark     \\ \hline
            \textbf{AGN}            &                       &               \\
            $L_\text{bol}$\,[\si{\erg\per\second}] & bolometric luminosity & $10^{45}$     \\
            $L_\text{X}$\,[\si{\erg\per\second}]   & $2-10\,\si{\kilo\electronvolt}$ luminosity & $10^{43.8}$     \\
            $M\,[M_\odot]$          & black-hole mass       & $10^8$        \\
            $\eta_\text{rad}$       & radiation efficiency  & $0.1$         \\
            \textbf{UFO}            &                       &               \\
            $\dot{M}_\text{w}~[M_\odot\,\si{\per\year}]$ & mass outflow rate  & $0.1$ \\
            $n_\text{ISM}\,[\si{\per\centi\meter\cubed}]$ & ambient matter density    & $10^4$   \\
            $\vw/c$                 & terminal wind velocity & $0.2$        \\
            $t_\text{age}$ [yr]     & age of the UFO        & $1000$        \\
            
            \textbf{Magnetic}       &                       &               \\
            \textbf{Field}          &                       &               \\
            $\epsilon_B$            & magnetic energy fraction  & $0.05$    \\
            $B_2$ [mG]              & downstream field strength & $85$      \\
            $l_\text{c}$ [pc]       & coherence length      & $0.01$        \\
            $\delta$                & turbulence index      & $5/3$         \\
            \bottomrule
        \end{tabular}
    \end{table}

\section{Results}\label{sec:results}
    \subsection{Benchmark UFO}\label{sec:results_benchmark}
    \subsubsection{UFO model definition}
    The parameters of our benchmark scenario are summarised in \cref{tab:parameters_benchmark}. They were chosen to represent an average UFO with respect to the population of observed UFOs (see \cref{sec:ufo_list}). We assume a black-hole mass $M_\text{BH}=10^8\,M_\odot$, bolometric luminosity $L_\text{bol}=10^{45}\,\si{\erg\per\second}$ ($8\%\,L_\text{Edd}$), mass outflow rate $\dot{M}_\text{w}=0.1\,M_\odot\si{\per\year}$, and terminal wind velocity $\vw=0.2\,c$. In addition, we assume the standard value for the radiation efficiency of the accretion disc of $\eta_\text{rad}=0.1$~\citep{Dermer:2009zz,Yu:2002sq,Elvis:2002ApJ,Marconi:2004MNRAS,Kato:2008bhad}, and an ambient matter density of $n_\text{ISM}=10^4\,\si{\per\centi\meter\cubed}$ (P23). The typical lifetime of ultra-fast outflows is not well constrained at present~\citep{Matzeu:2023A&A}. Here and in the following we consider young UFOs with $t_\text{age}=10^3\,\si{\year}$, corresponding to the minimum time required to ensure that the steady state is reached where the evolution of both shocks is described by \cref{eq:R_sh,eq:R_fs}. The magnetic field is assumed to be turbulent with a coherence length of $l_c=\SI{0.01}{\parsec}$, as motivated by models of the disc wind \citep[e.g.][]{Murray:1995ApJ}, turbulence cascade spectral index $\delta=5/3$ (Kolmogorov turbulence), and energy fraction relative to the ram pressure of $\epsilon_B=0.05$. We have confirmed that the maximum energy of the escaping cosmic rays is not strongly sensitive to the choice of $\epsilon_B$. The nominal value results in a magnetic field strength of $B_2\approx\SI{85}{\milli\gauss}$ and an energy density of $U_\text{B,2}\approx\SI{3.5e-4}{\erg\per\centi\meter\cubed}$ at the wind termination shock. To obtain the desired coherence length in our \textsc{CRPropa} simulations we set the maximum and minimum length scale of the turbulence cascade to \SI{0.052}{\parsec} and \SI{0.001}{\parsec} respectively \citep[see][]{Harari:2002dy}. Although the latter is larger than the Larmor radius of cosmic rays with rigidity of $E/Z\lesssim\SI{0.1}{\exa\volt}$ (for $B_2=\SI{85}{\milli\gauss}$), the simulation remains accurate at lower rigidities because of the dominance of advection over diffusion in this regime. Our results change only minimally for a Kraichnan turbulence spectrum ($\delta=3/2)$.
    
    Based on the above parameters, we obtain an accretion rate of $\dot{M}=L_\text{disc} / \eta_\text{rad}\,c^2\approx0.21\,M_\odot\si{\per\year}$, and wind kinetic energy of $L_\text{kin}\approx10^{44}\,\si{\erg\per\second}$. The shocks are located at $\rsh\approx\SI{0.8}{\parsec}$ and $\rfs\approx\SI{3.1}{\parsec}$ respectively. The spectral luminosity of the external photon fields is shown in \cref{fig:photon_fields_benchmark}; see \cref{apx:photon_fields} for details.
    
    Our model is similar to the benchmark scenario of P23; however, we have adjusted several parameters to provide a better representation of the average UFO/AGN of our sample. We have modified the bolometric luminosity ($10^{45.53}\,\si{\erg\per\second}\to10^{45}\,\si{\erg\per\second}$), the temperature of the dust torus ($\SI{70}{\kelvin}\to\SI{200}{\kelvin}$) to improve compatibility with the template of \citet{Mullaney:2011iq}, and the spectral index of the magnetic turbulence cascade ($3/2\to5/3$, i.e. Kraichnan $\to$ Kolmogorov). In addition, we have used updated luminosity scaling factors of the different AGN components and a more realistic distance scaling of the photon field densities. We have verified that the predicted timescales for photopion interactions, acceleration, and escape of protons are in good agreement with the results of P23 (Fig.\ 2 in their paper) if identical parameters, luminosity scaling factors, and distance scaling are used.

    \subsubsection{Maximum energy at acceleration}
    \begin{figure}
        \centering
        \includegraphics[width=\linewidth]{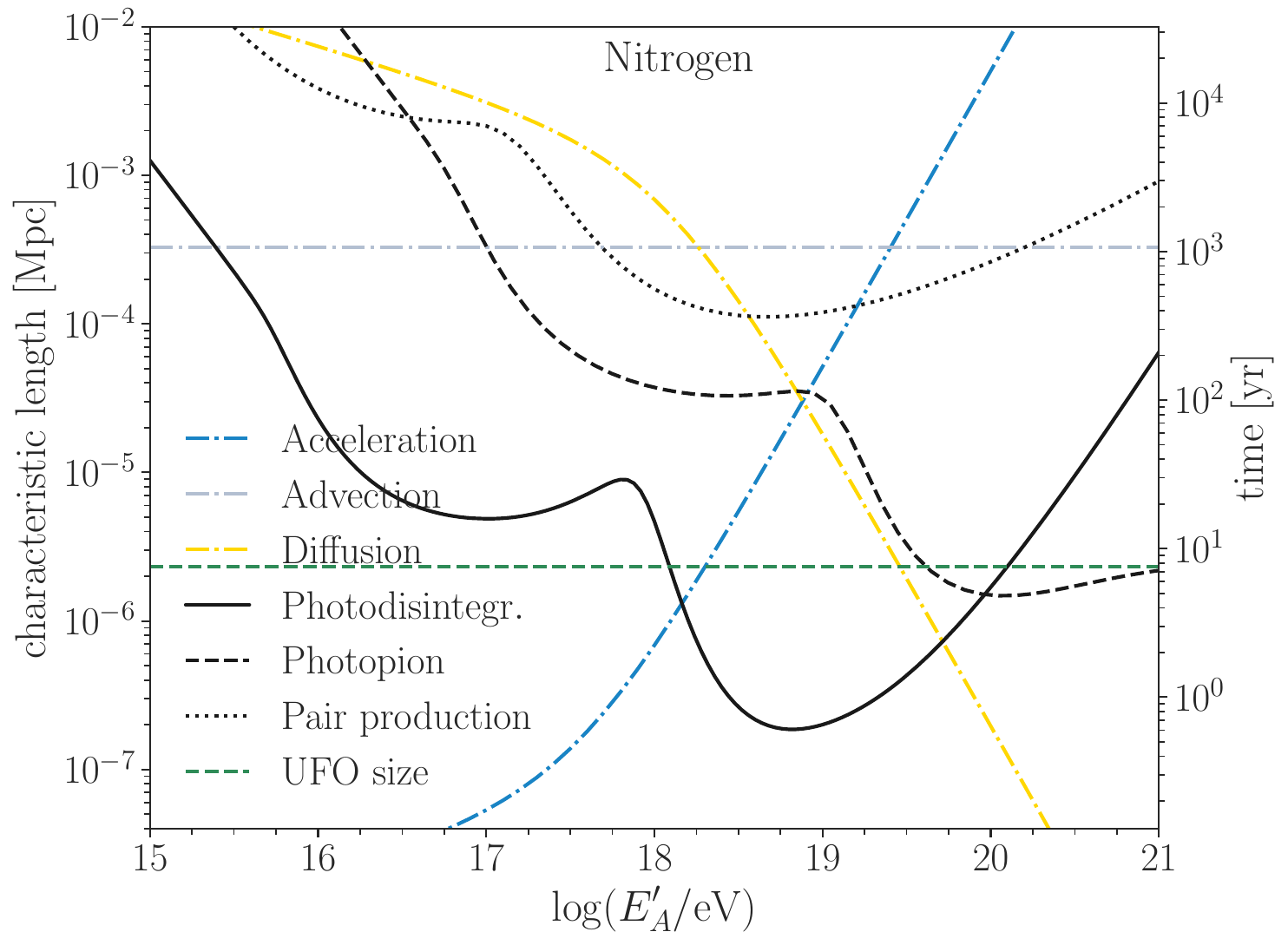}
        \includegraphics[width=\linewidth]{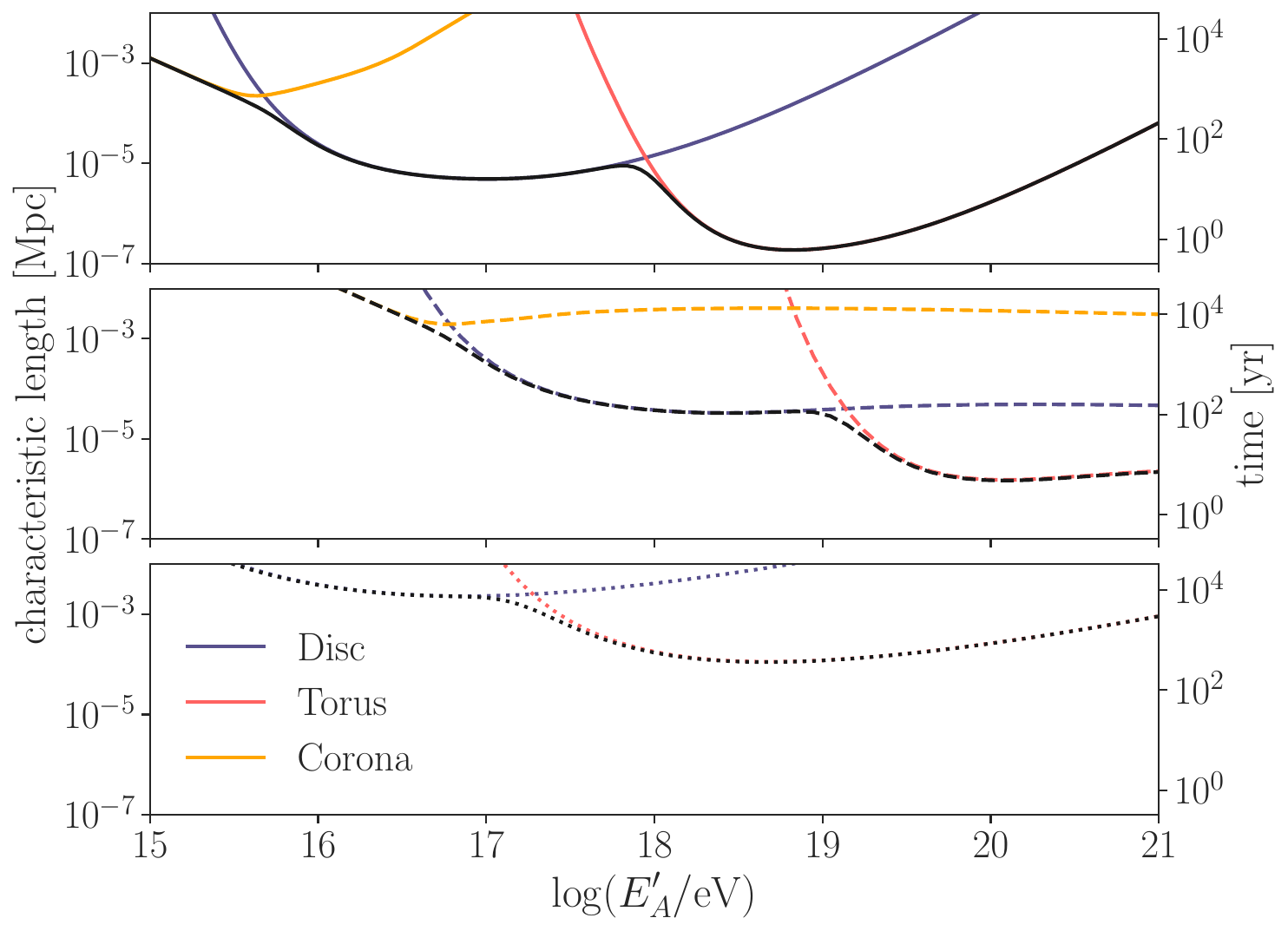}
        \caption{Top panel: Energy loss length and characteristic timescales of nitrogen nuclei in the shocked wind at distance $\rsh + \delta R$ from the central engine for our benchmark AGN. Bottom three panels: Individual contribution of the accretion disc, dust torus, and corona photon field to the energy loss length for photodisintegration, photopion, and pair production (top, middle, bottom). See \cref{apx:results_benchmark} for protons, helium, silicon, and iron nuclei.}\label{fig:results_benchmark_nitrogen}
    \end{figure}

    We use the semi-analytical approach discussed in \cref{sec:timescales} to estimate the maximum energy cosmic rays can be accelerated to at the wind termination shock of the benchmark UFO. This is done for five representative cosmic-ray species ($^1$p, $^4$He, $^{14}$N, $^{28}$Si, $^{56}$Fe) by comparing the relevant timescales for acceleration, interaction, and escape. The timescales are evaluated in the shocked wind downstream of the wind termination shock at $R=\rsh+\delta R$ where the particles spend most of their time, since they are quickly advected back to the shock whenever they enter the upstream region.

    The maximum energy of cosmic-ray nuclei is limited by photodisintegration due to the infrared field of the dust torus (see \cref{fig:results_benchmark_nitrogen} for nitrogen). The maximum energy per nucleon is similar for all nuclei at approximately $0.1-0.15\,A\,\si{\exa\electronvolt}$. This is consistent with the threshold energy for photodisintegration scaling with the energy per nucleon, i.e.\ $E_\text{th}\propto A$. The maximum energy of protons is $\SI{0.9}{\exa\electronvolt}$, comparable to the $\mathcal{O}(\SI{1}{\exa\electronvolt})$ found in P23. It is limited by the characteristic escape timescale from the shock region; however, the energy loss timescale of photopion production due to the infrared field of the dust torus is comparable (see \cref{apx:results_benchmark}).
    
    Electron-positron and photopion production are generally subdominant at UHE for all species except protons. The energy losses of nuclei are dominated by photodisintegration on the dust torus photon field and the disc field (at lower energies). Above approximately\ $A\,(3-5)\,\si{\exa\electronvolt}$, photopion losses exceed photodisintegration; however, strong suppression of the flux due to photodisintegration at lower energies prevents cosmic rays from reaching this regime.

    \subsubsection{Escaping cosmic-ray flux}\label{sec:cr_source_spectrum}
    \begin{figure}
        \centering
        \includegraphics[width=\linewidth]{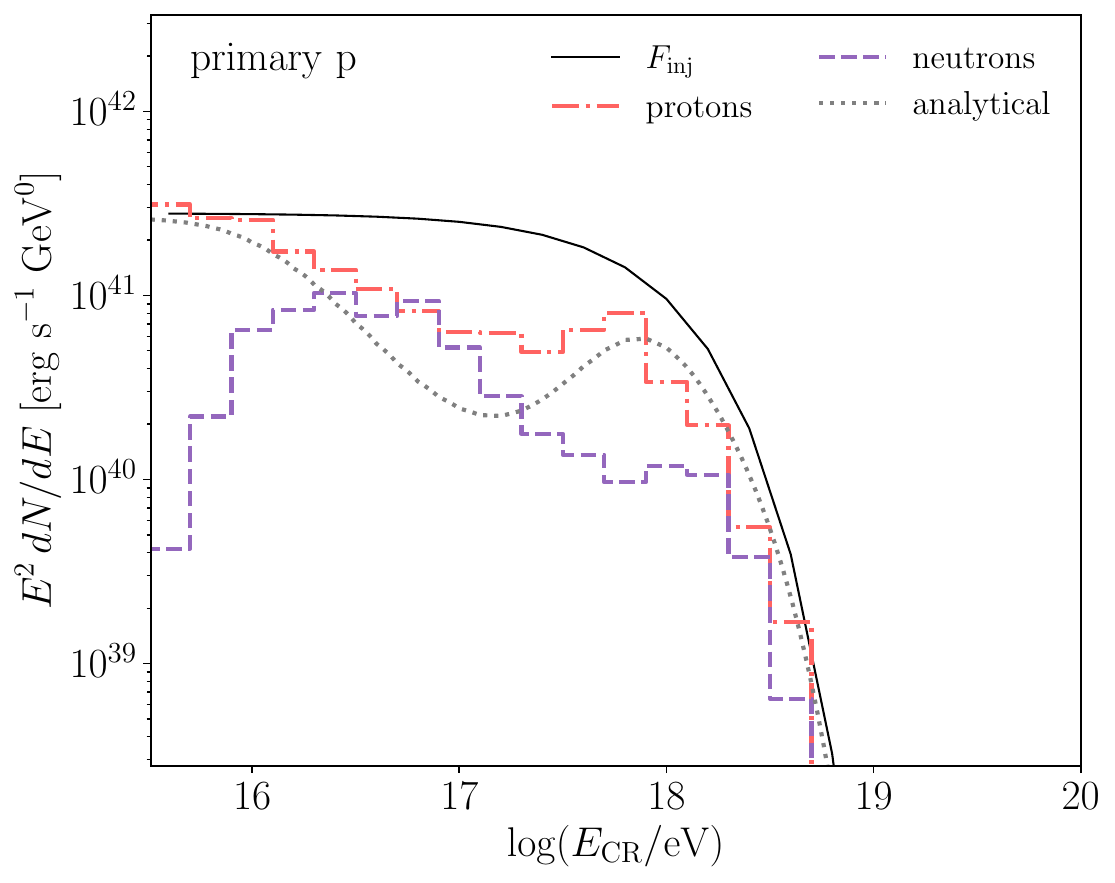}
        \includegraphics[width=\linewidth]{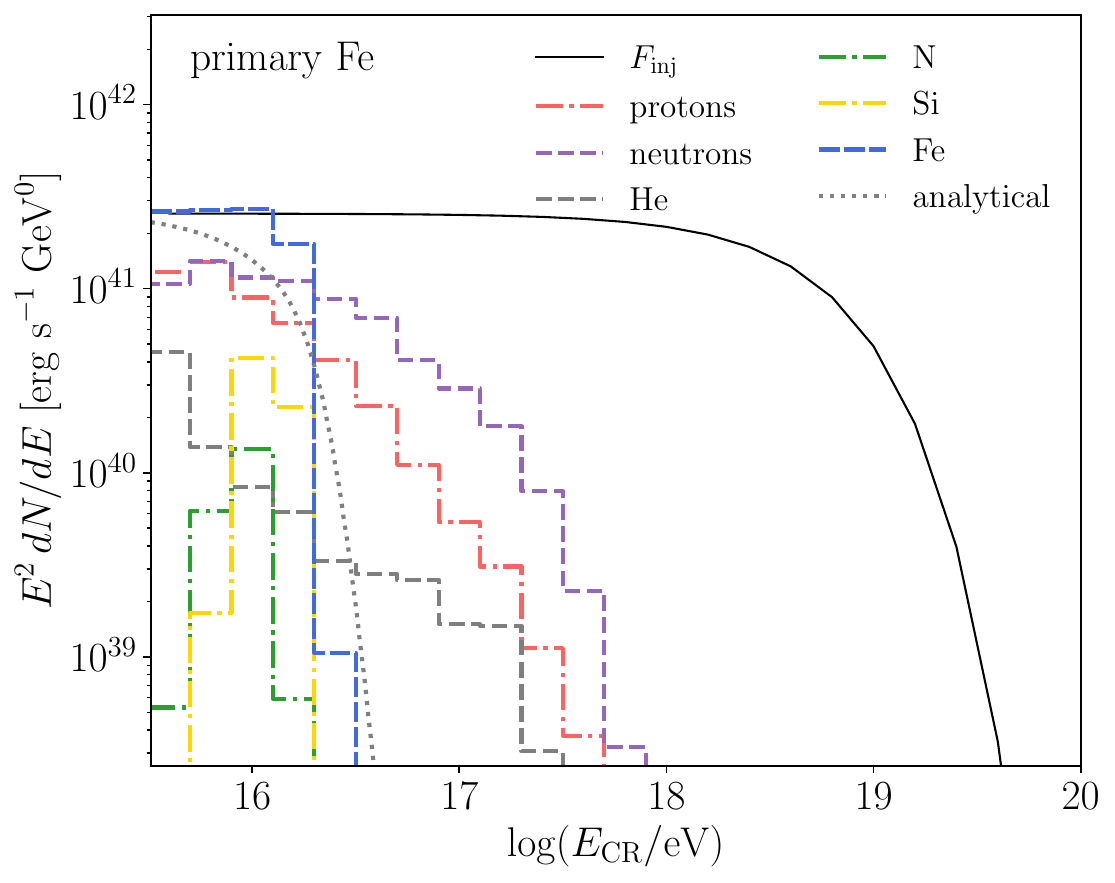}
        \caption{Escaping spectrum at the forward shock for injection of primary protons (top) and Fe-56 (bottom) at the wind termination shock. The lines correspond to the distribution of escaping cosmic rays in the mass range $[A_\text{min},\,A_\text{max}]$. The semi-analytical estimate of the surviving primaries (p / Fe), without production of secondaries and for constant photon fields, is indicated by the dotted line.}\label{fig:benchmark_escape_spectra}
    \end{figure}
    We derive the injected and escaping cosmic-ray spectrum for the benchmark UFO. The injection spectrum is assumed to follow a $E^{-2}$ distribution characteristic for diffusive shock acceleration with an exponential cut-off at the previously derived maximum energy, i.e.
    \begin{equation}\label{eq:cr_flux_inj}
        Q_\text{inj}(E) = Q_0\,\left(\frac{E}{E_0}\right)^{-2}\,\exp\left(-\frac{E}{\emax(A,Z)}\right)\,.
    \end{equation}
    The normalisation factor $Q_0$ $[\si{\per\electronvolt\per\second}]$ is determined by the total cosmic-ray luminosity of the source
    \begin{equation}
        L_\text{CR} = \int_{E_\text{min}}^\infty\,\dif E \left[E\, Q_\text{inj}(E)\right],~~E_\text{min}=\SI{1}{\giga\electronvolt}\,.
    \end{equation}
    Following P23, we assume that $\eta_\text{CR}=5\%$ of the kinetic power of the wind is converted into hadronic cosmic rays. This represents a conservative estimate compared to the $\sim10\%$ or more suggested by observations of supernova remnants~\citep{Helder:2009fm}.

    The escaping flux of all cosmic-ray nuclei is strongly suppressed above $\sim10^{16}\,\si{\electronvolt}$ by interactions with photons from the disc and torus fields, as illustrated in \cref{fig:benchmark_escape_spectra}. This includes secondary nuclei produced in the interactions; however, the nuclear cascade results in a significant flux of sub-EeV secondary protons, which escape the UFO environment because of a reduced interaction probability and intermediate conversion through the photopion channel to neutrons, which are not constrained to the system by magnetic deflections. A similar disintegration of primary nuclei by the external photon fields of the AGN is expected in the jets of high-luminosity flat-spectrum radio quasars~\citep{Rodrigues:2017fmu,Rodrigues:2020pli}. The impact of the external photon fields in the AGN environment was also previously studied by ~\citet{Murase:2014foa,Murase:2019vdl} in the context of astrophysical neutrinos.

    For injection of iron at the termination shock, $40\%$ of the injected luminosity is lost in the form of low-energy cosmic rays, electron-positron pairs, and neutrinos. An additional $31\%$ escapes the UFO in the form of protons and neutrons, while $25\%$ escapes as iron-like nuclei ($A\in[29,56]$). The remaining intermediate mass groups (helium-, nitrogen-, silicon-like) carry away at most $2\%$ of the injected luminosity respectively.
    Primary protons benefit from a shorter confinement time due to their larger rigidity ($E/Z$) at a given energy and are therefore less attenuated, typically experiencing only a few interactions. The flux of escaping high-energy primary protons carries $73\%$ of the original injected luminosity.
    
    We conclude that the benchmark scenario -- representing an average UFO -- does not provide a suitable environment to produce heavy UHECRs with energies above the ankle. Even if the flux was not strongly attenuated by the intense AGN photon fields, the modest maximum energies achievable through diffusive shock acceleration at the wind termination shock disfavour average UFOs as the sources of the observed UHECRs. The case of ``extreme'' UFOs, with properties more conducive to accelerating and releasing UHECRs, is discussed in the following section.

    \subsection{Application to observed UFOs}\label{sec:ufo_list}
    \subsubsection{The UFO sample}
    \begin{figure}
        \centering
        \includegraphics[width=\linewidth]{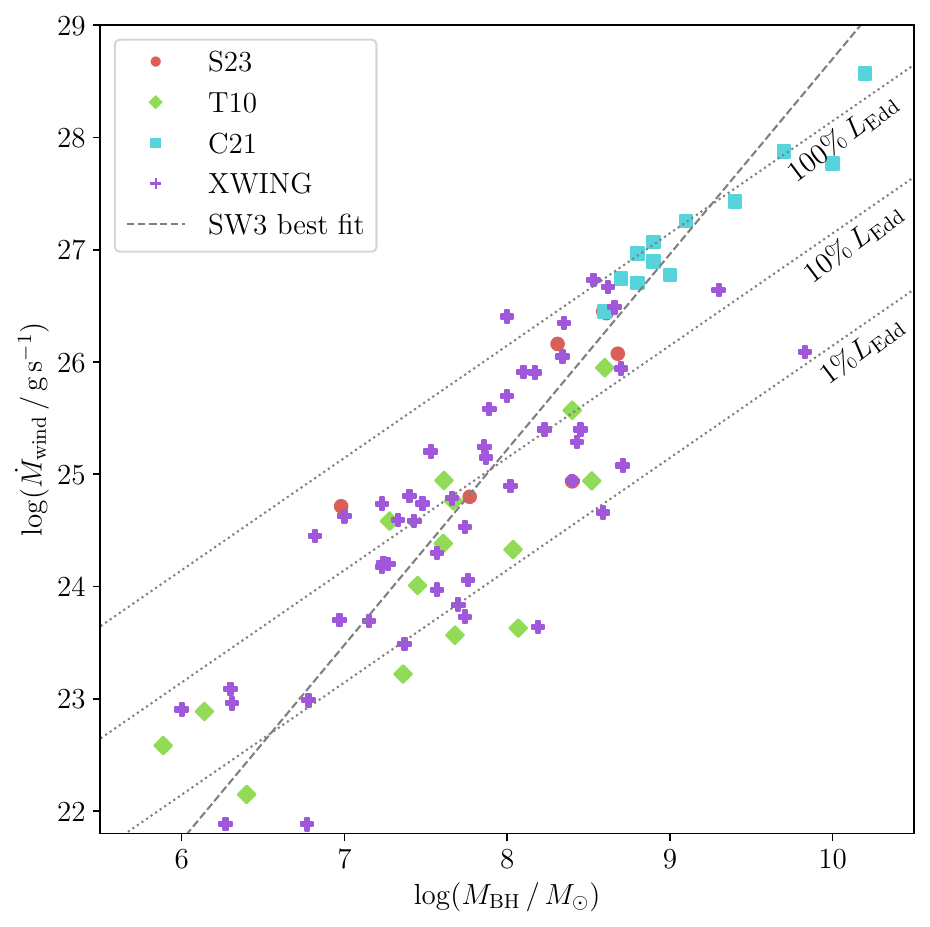}
        \caption{Inferred mass outflow rates of the sample of observed UFOs. The input parameters for objects in the T10~\citep{Tombesi:2010a}, S23~\citep{Matzeu:2023A&A} and C21~\citep{Chartas:2021ApJ} samples were taken from ~\citet{Gianolli:2024jkq} (SW3), and XWING-sources from ~\citet{Yamada:2024lss}. Uncertainties are omitted for clarity.}\label{fig:results_ufo_outflow}
    \end{figure}
    We apply the above framework to a sample of $86$ observed ultra-fast outflows, composed of $34$ UFOs from the SUBWAYS III sample of \citet{Gianolli:2024jkq} (in the following ``SW3''), and $52$ additional UFOs from the XWING sample of \citet{Yamada:2024lss}. The SW3 sample consists of three subsamples taken from previous papers; \citet{Tombesi:2010a} (``T10''), \citet{Chartas:2021ApJ} (``C21''), and \citet{Matzeu:2023A&A} (``S23''). The XWING data set includes outflows with lower velocities (``warm absorbers'' and ``low-ionisation parameter'' UFOs or ``LIPs''). Out of 573 observations of such outflows in 132 unique AGN, we identified 96 objects where at least a single observation was tagged as an UFO (93 objects) or LIP (28 objects). We include all objects in our analysis where all relevant parameters are provided to calculate the mass outflow rate ($M_\text{BH},\,\vw$; see \cref{apx:outflow_rate}). This includes 84 XWING objects\footnote{The existence of outflows in six of these AGN is uncertain (1ES\,1927+654, IRAS\,04416+1215, NGC\,1068, NGC\,6240, PG\,0844+349, PG\,1202+281). See discussion in \citet{Yamada:2024lss}.} of which 52 are not part of the SW3 sample\footnote{Including 16 low-velocity UFOs with $0.03\lesssim\vw/c\lesssim0.1$.}. We have identified some uncertainties in the derivation of the mass outflow rate in SW3\footnote{The outflow rates provided in the supplementary material were derived from the launching radii inferred via the observed column density, in contrast to the approach based on the minimum escape radius discussed in the paper.}. Here, we re-derive the outflow rates using the outflow velocity, column density, and black-hole mass provided in SW3, and the procedure outlined in \cref{apx:outflow_rate}. This results in a conservative estimate of the mass outflow rate. The same procedure is applied to UFOs in the XWING sample. For XWING UFOs with multiple observations of an ultra-fast outflow, we take the representative outflow velocity as the average of all individual observations, and the launching radius and mass outflow rate as the log-average. The list of UFOs used in our study and associated parameters can be found in \cref{apx:ufo_list}. The estimated mass outflow rate $\dot{M}_\text{w}$ is shown for all UFOs in \cref{fig:results_ufo_outflow} as a function of the mass of the AGN.
    
    For most objects, we find mass-outflow rates corresponding to between $1\%$ and $100\%$ of the Eddington luminosity (for $\eta_\text{rad}=0.1$). However, the outflow rate exceeds the mass accretion rate derived from the bolometric luminosity for around half of the UFOs. These outflows are likely fully or partially accelerated by magnetic driving~\citep{Fukumura:2010ApJ,Kraemer:2017rge}, and may be of an intermittent nature~\citep{Belloni:1997fk,Cappi:2009ji,Luminari:2020qus,Gianolli:2024jkq}; see \cref{sec:limit_mass_outflow_rate}.
    Alternatively, the outflow can extract a significant fraction of the mass from the accretion flow before it reaches the inner part of the accretion disc~\citep{Parker:2017wnh,Honig:2019ApJ,Laurenti:2020ftw}. The velocities used here and typically quoted for observed UFOs in the literature refer to the observed outflow velocities along the line of sight. For reasonable values of the observation angle, the predicted mass outflow rate is consistent within a factor of a few~\citep{Krongold:2007ApJ}.
    
    \subsubsection{Distribution of maximum energy at acceleration}
    \begin{figure*}
        \centering
        \includegraphics[width=\linewidth]{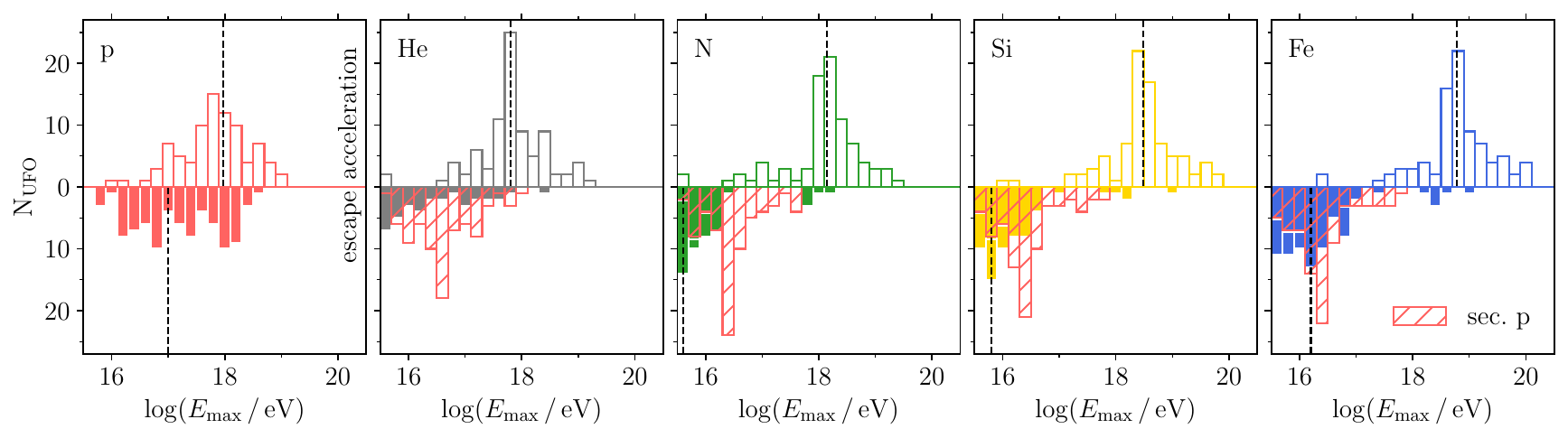}
        \caption{Distribution of maximum energies over the UFO sample for the injection of different primary cosmic-ray species; at acceleration (un-filled bars) and after escape (filled bars). The red hatched bars in panels 2-5 indicate the maximum energy of secondary protons. The maximum energy of the benchmark UFO for each species is marked by the black dashed line (below the considered energy range if no line present).}\label{fig:results_sample_Emax_distribution}
    \end{figure*}
    \begin{figure*}
        \centering
        \includegraphics[width=\linewidth]{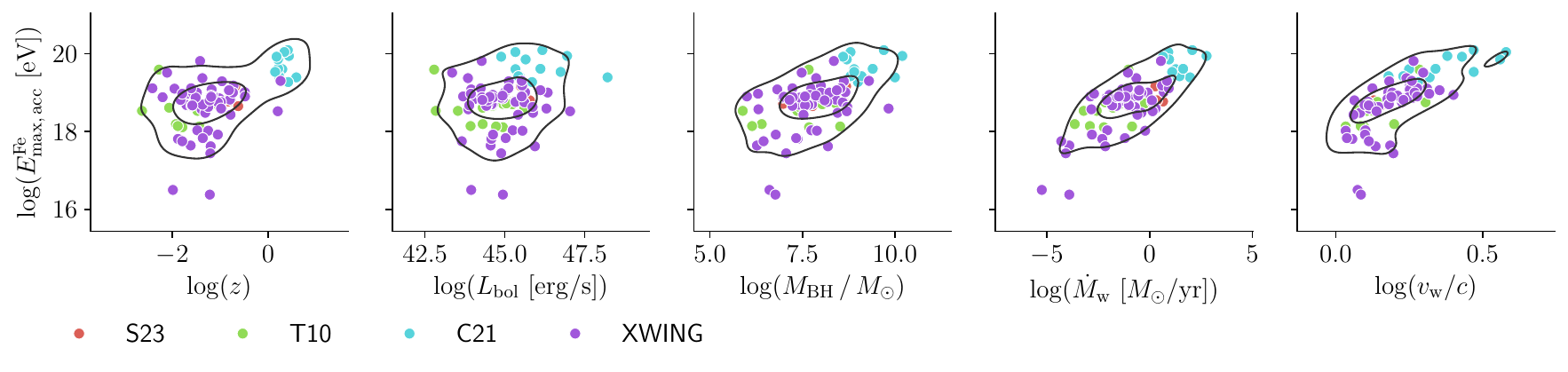}
        \caption{Correlation between the predicted maximum energy of iron nuclei at acceleration (before escape) and key parameters of the UFO/AGN system. Concentric contours indicate the $40\%$ and $86\%$ containment interval of the smoothed bivariate source density distribution for the nominal choice of parameters (corresponding to the one- and two-sigma intervals at 2 dof).}\label{fig:Emax_line_benchmark_acc}
    \end{figure*}
    For most investigated outflows, the maximum cosmic-ray energy after acceleration scales approximately with the mass number,  similar to the results for our benchmark UFO model (see \cref{sec:results_benchmark}). The maximum energy of protons is between a factor of $2$ (for NGC\,7582 and NGC\,4051) and $35$ (for IRAS\,17020+4544) times higher than the characteristic $\emax/A$ of the iron nuclei. The large variance is related to the strength of the ambient photon fields as these are generally more effective in limiting the maximum energy of heavier cosmic rays. The distribution of maximum energies at acceleration obtained for the sample of UFOs is shown in \cref{fig:results_sample_Emax_distribution} (unfilled, top histograms). We confirm that the previously discussed benchmark scenario represents an average UFO in terms of the ability to accelerate UHECRs.
    
    Previous fits to the Auger data have shown that the energy spectrum can be fitted by sources with a maximum energy of at least $\sim10^{19}\,\si{\electronvolt}$~\citep{AlvesBatista:2018zui,Heinze:2019jou,PierreAuger:2022atd,Ehlert:2023btz}. This is satisfied by $33\%$ (28) of the UFOs in our sample for iron, $17\%$ (15) for silicon, $7\%$ (6) for nitrogen, and $3\%$ (3) for helium. No outflow reaches $10^{19}\,\si{\electronvolt}$ energy for protons. The UFOs with the highest maximum energy at acceleration for iron nuclei are listed in \cref{tab:sample_best_sources_acc}. We observe a strong correlation (Spearman $p<10^{-5}$) of the maximum energy at acceleration (for primary iron) with the terminal wind velocity (Spearman $\rho=0.80$), mass outflow rate ($\rho=0.67$), and mass of the AGN ($\rho=0.51$); see \cref{fig:Emax_line_benchmark_acc}. There is a weak, marginally significant correlation with the bolometric luminosity ($\rho=0.18$, $p=0.09$), and a spurious correlation with the redshift ($\rho=0.44$, $p\approx10^{-5}$) due to the bias introduced by the C21 sample. The results are qualitatively similar for the other cosmic-ray nuclei. We have confirmed that the uncertainty of the observed wind velocity has a limited impact on the predicted maximum energy. The caveats of the estimated mass outflow rate are discussed in \cref{sec:limit_mass_outflow_rate}. Comparable correlations are found for primary protons, although the maximum energy is most strongly correlated with the mass outflow rate in this case ($\rho=0.78$), and the correlation with the bolometric luminosity is more pronounced ($\rho=0.35$, $p=0.001$). We conclude that protons and cosmic-ray nuclei require similar conditions to allow acceleration to high energies. The majority of the highest-energy UFOs are part of the C21 sample -- a consequence of objects in this sample exhibiting larger mass outflow rates and higher outflow velocities than UFOs from other studies. The C21 UFOs also have significantly larger black hole masses and redshift than the UFOs in the other samples \citep[cf.\ \cref{fig:Emax_line_benchmark_acc}, and][Fig.\ 2]{Yamada:2024lss}.

    \begin{table}
        \caption{All investigated UFOs with a maximum energy at acceleration of at least $10^{19.7}\,\si{\electronvolt}$ for iron nuclei. Energies are in electronvolt. See \cref{apx:ufo_list} for the full list.}\label{tab:sample_best_sources_acc}
        \begin{tabular}{lccc}
            \toprule
            Name            & $\log E_\text{max}^\text{acc}(\mathrm{p})$ & $\log E_\text{max}^\text{acc}(\mathrm{N})$  & $\log E_\text{max}^\text{acc}(\mathrm{Fe})$ \\  \midrule
            Mrk-273         & 18.6  & 19.1  & 19.8  \\
            SDSS-J1128+2402 & 18.9  & 19.1  & 19.8  \\
            HS-0810+2554    & 18.8  & 19.2  & 19.9  \\
            HS-1700+6416    & 18.9  & 19.2  & 19.9  \\
            SDSS-J1029+2623 & 18.9  & 19.3  & 20.0  \\
            SDSS-J1442+4055 & 18.9  & 19.4  & 20.1  \\ \bottomrule
        \end{tabular}
    \end{table}

    \subsubsection{Escaping cosmic-ray flux for the UFO sample}
    For most UFOs in our sample, the escaping flux of cosmic-ray nuclei is strongly suppressed at energies where interactions with the photon fields of the accretion disc and dust torus are relevant. The optical-UV field of the accretion disc and the IR field of the dust torus efficiently disintegrate the nuclei and suppress the flux above approximately $\sim A\times10^{16}\,\si{\electronvolt}$ for the majority of outflows. Ultra-fast outflows that are good UHECR accelerators due to their fast outflow velocities are especially prone to this suppression as a consequence of a strong correlation between the outflow rate of the UFO and the bolometric luminosity and mass of the associated AGN. In more luminous AGN, the photon fields in the shocked wind are generally much stronger and photodisintegration losses are more severe as a result. We define the maximum energy of the escaping flux as the energy at and beyond which the flux is suppressed by a factor of $1/e~(\approx37\%)$ or more compared to the injected $F\sim E^{-2}$ flux. For most UFOs, with strong external photon fields, the cutoff is sharp and the precise choice of the suppression factor is not important as all reasonable choices result in a similar estimate for the maximum energy. The spectrum of the escaping flux can show a bump at high energies for UFOs with weak photon fields. The distribution of maximum energies at escape for all cosmic-ray species is shown in \cref{fig:results_sample_Emax_distribution}, and the outflows with the highest energies are listed in \cref{tab:sample_best_sources_esc}.

    \begin{table}
        \caption{Candidate UFOs with the largest maximum energy after escape for primary protons (including the conversion of neutrons after escape) and iron nuclei. We also include the energy where the escaping flux is suppressed by a factor of $10$. Energies are in electronvolt. The escape energies are binned in steps of $\dif\log E=0.2$ to minimise statistical fluctuations. See \cref{apx:ufo_list} for the full list of all investigated outflows.}\label{tab:sample_best_sources_esc}
        \begin{tabular}{lcccc}
            \toprule
            Name            & Sample & $\log E_\text{max}^\text{acc}$  & $\log E_\text{max}^\text{esc}$  & $\log E_\text{sup,10}^\text{esc}$ \\  \midrule
            \textbf{Protons}  &       &      &       &       \\
            SDSS\,J0904+1512  & XWING & 18.5 & 18.2  & 18.6  \\
            NGC\,6240         & XWING & 18.2 & 18.2  & 18.4  \\
            NGC\,2992 	      & XWING & 18.2 & 18.2  & 18.6  \\
            NGC\,7582 	      & T10   & 18.2 & 18.2  & 18.4  \\
            MCG\,03-58-007    & XWING & 18.2 & 18.2  & 18.4  \\
            SDSS\,J1128+2402  & C21   & 18.9 & 18.2  & 18.8  \\
            SDSS\,J1442+4055  & C21   & 18.9 & 18.2  & 18.6  \\
            MG\,J0414+0534    & C21   & 18.5 & 18.2  & 18.4  \\
            SDSS\,J1529+1038  & C21   & 18.6 & 18.2  & 18.8  \\
            SDSS\,J1029+2623  & C21   & 18.9 & 18.4  & 19.0  \\ 
            IRAS\,13349+2438  & XWING & 18.4 & 18.4  & 18.4  \\
            HS\,0810+2554 	  & C21   & 18.8 & 18.4  & 19.0  \\
            Mrk\,273 	      & XWING & 18.6 & 18.6  & 18.8  \\ \midrule
            \textbf{Iron}     &       &      &       &       \\
            HS\,0810+2554 	  & C21   & 19.9 & 17.4  & 18.2  \\
            IRAS\,13224-3809  & XWING & 19.1 & 18.2  & 18.4  \\
            Mrk\,231 	      & XWING & 19.4 & 18.4  & 18.4  \\
            Mrk\,273 	      & XWING & 19.8 & 18.4  & 18.4  \\
            NGC\,4051 	      & T10   & 18.5 & 18.4  & 18.8  \\
            NGC\,2992 	      & XWING & 19.5 & 18.6  & 19.8  \\
            NGC\,7582 	      & T10   & 19.6 & 19.0  & 19.8  \\ \bottomrule
        \end{tabular}
    \end{table}

    \begin{figure}
        \centering
        \includegraphics[width=\linewidth]{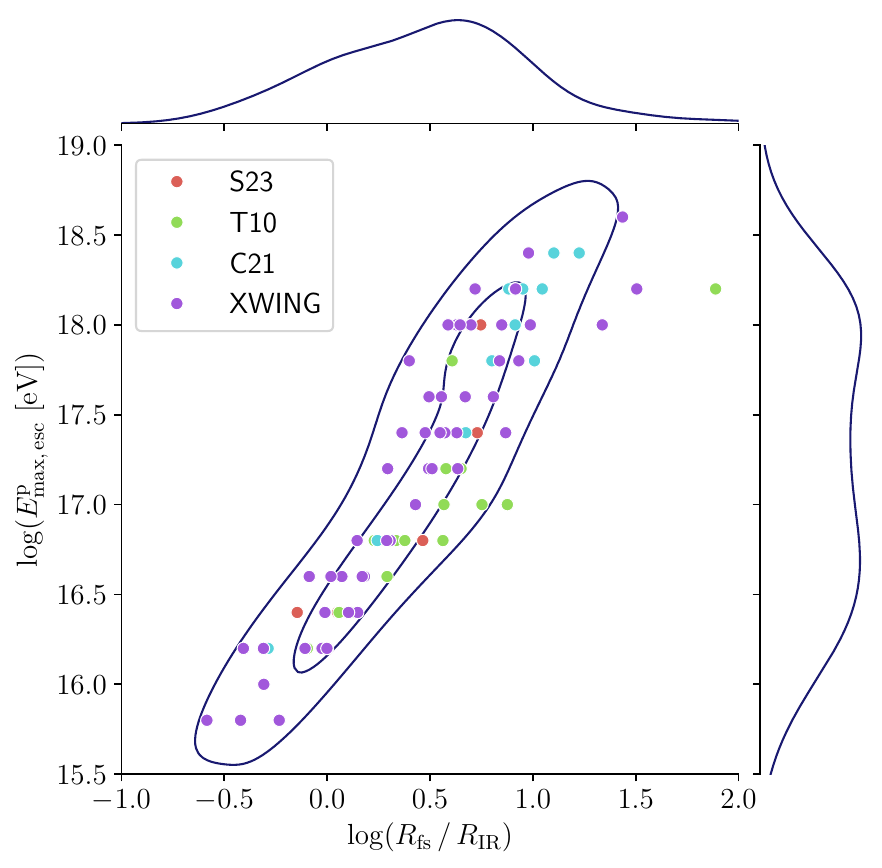}
        \includegraphics[width=\linewidth]{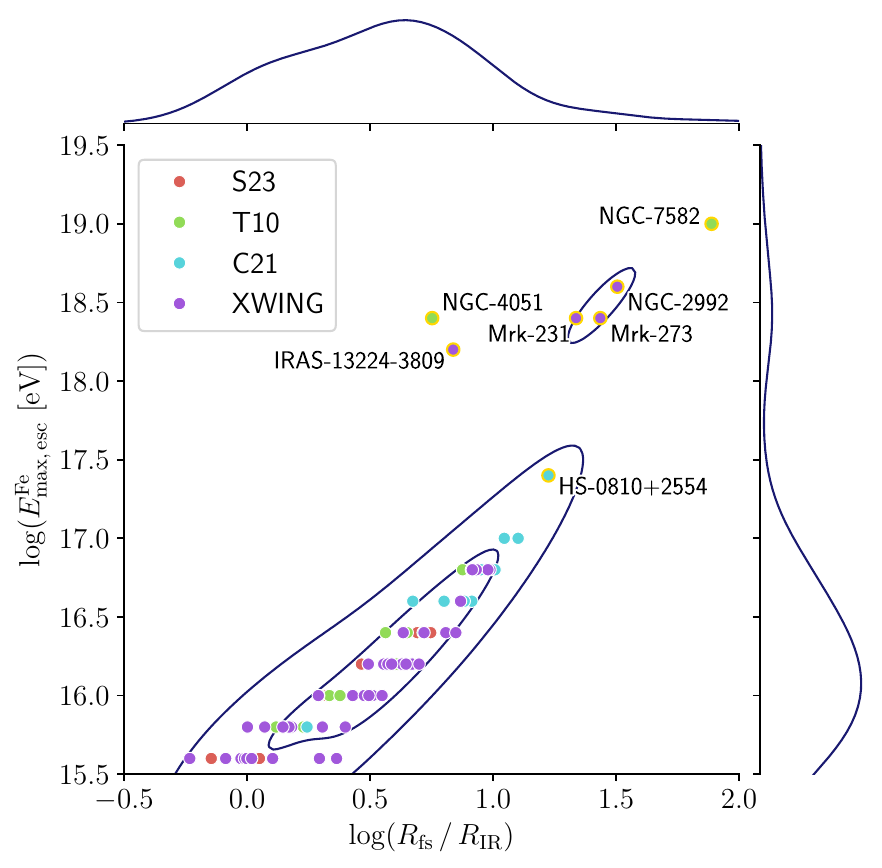}
        \caption{Maximum energy of primary protons (top) and iron nuclei (bottom) after escape from the UFO environment as function of the ratio of the radius of the forward shock to the radius of the dust torus. Discrete levels in $\emax$ are due to the numerical binning process. Objects with $\emax<10^{15.5}\,\si{\electronvolt}$, the lower limit of our simulations, are omitted. Contours indicate the $40\%$ and $86\%$ containment interval of the smoothed bivariate source density distribution for the nominal choice of parameters (one- and two-sigma intervals at 2 dof).}\label{fig:Emax_radius_correlation}
    \end{figure}

    For the nominal parameters, only the low-luminosity active galactic nucleus NGC\,7582 reaches a maximum energy of $10^{19}\,\si{\electronvolt}$ for iron nuclei. This is due to weak photon fields because of the low overall luminosity, and a dust torus radius significantly smaller than the radius of the wind termination shock which further reduces the photon density in the downstream region. NGC\,7582 is the lowest-luminosity AGN in our sample and at the same time also one of the most nearby at a distance of approximately $\SI{23}{\mega\parsec}$. We observe a strong correlation between redshift and AGN bolometric luminosity (Spearman $\rho=0.79$, $p<10^{-5}$) in our UFO sample. This reflects a systematic bias in the available UFO surveys towards AGN with higher luminosity and suggests the existence of a population of undiscovered, similarly promising UFOs at higher redshift.

    The maximum energy of the escaping iron nuclei is correlated with the mass outflow rate (Spearman $\rho=0.49$, $p<10^{-5}$), AGN mass ($\rho=0.37$, $p=0.0004$), and wind velocity ($\rho=0.30$, $p=0.005$), and marginally anti-correlated with the bolometric luminosity ($\rho=-0.19$, $p=0.09$). Similar correlations are found for other cosmic-ray nuclei ($A>1$). Compared with the correlation of the maximum energy at acceleration with these parameters, this represents a strong reduction of both correlation strength and significance for all parameters. The UFOs with the highest maximum energy of iron nuclei are unremarkable in terms of AGN mass, outflow rate, and wind velocity. However, they are associated with some of the least luminous AGN in our sample. For primary protons, where the escaping flux is generally less suppressed, the correlations observed at acceleration are qualitatively preserved at escape at high significance and strong, but reduced, correlation strength; wind velocity ($\rho=0.55$, $p<10^{-5}$), outflow rate ($\rho=0.55$, $p<10^{-5}$), and black-hole mass ($\rho=0.39$, $p=10^{-4}$). There is no significant correlation with the bolometric luminosity. This emphasises that different conditions are required for UFOs to be good sources of ultra-high-energy protons and nuclei.

    \begin{figure*}
        \centering
        \includegraphics[width=\linewidth]{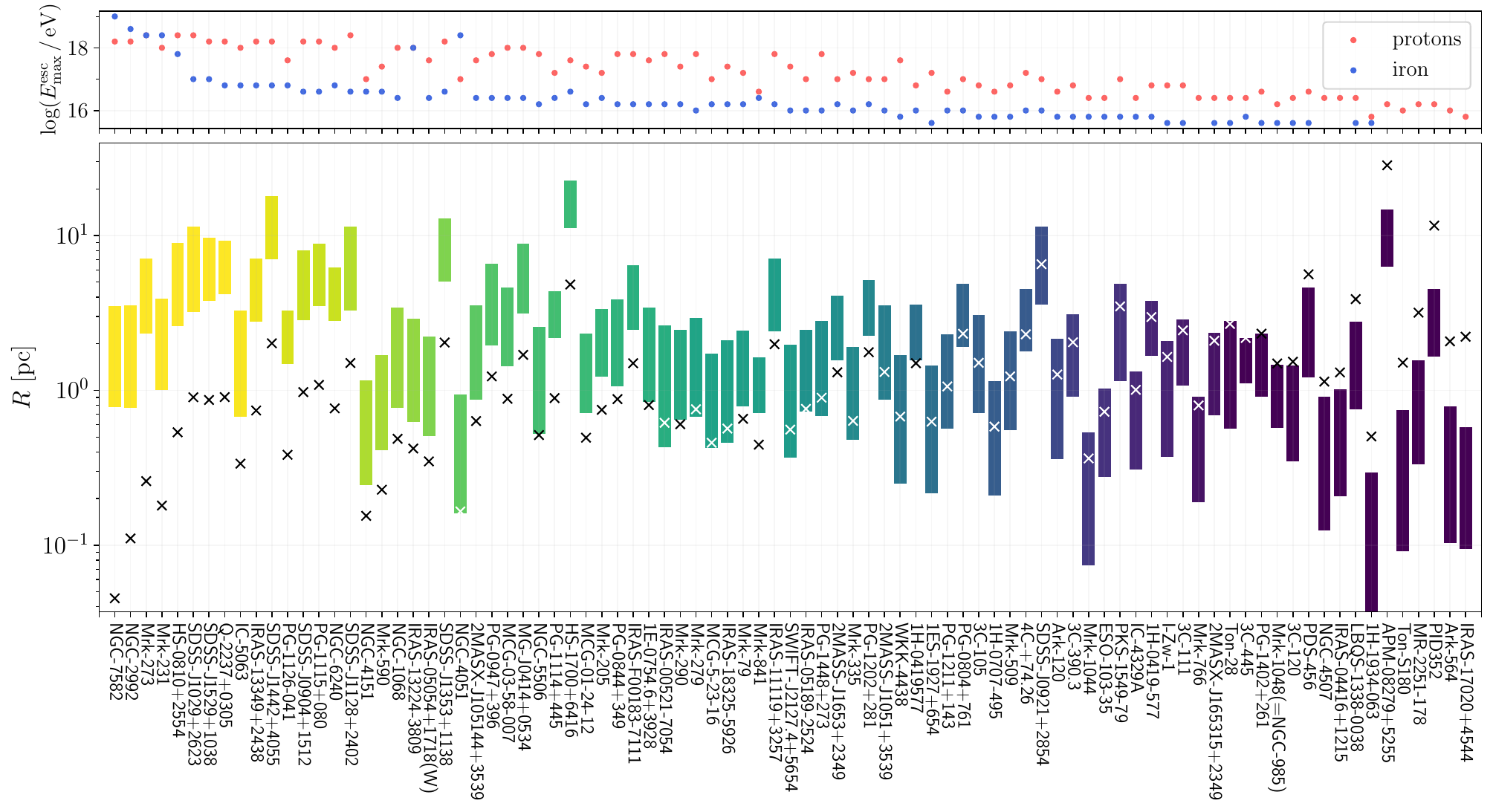}
        \caption{Main panel: Characteristic radii of the considered UFOs; $R\ir$: ``x'' sign. The bar indicates the region between the termination shock $\rsh$ (lower boundary) and the forward shock $\rfs$ (upper boundary), with the colour corresponding to $\log(\rfs/R\ir)$ -- green-yellow: $\rfs\gg R\ir$, purple-blue: $\rfs \lesssim R\ir$. Top panel: Maximum energy of the escaping flux for primary protons (red) and iron nuclei (blue) for each UFO. Outflows with a larger radial separation between the shocked wind bubble and the dust ``torus'' typically allow the escape of heavy cosmic-ray nuclei with larger energy (cf.\ \cref{fig:Emax_radius_correlation}).
        }\label{fig:ufo_radii}
    \end{figure*}

    \begin{figure}
        \centering
        \includegraphics[width=0.85\linewidth]{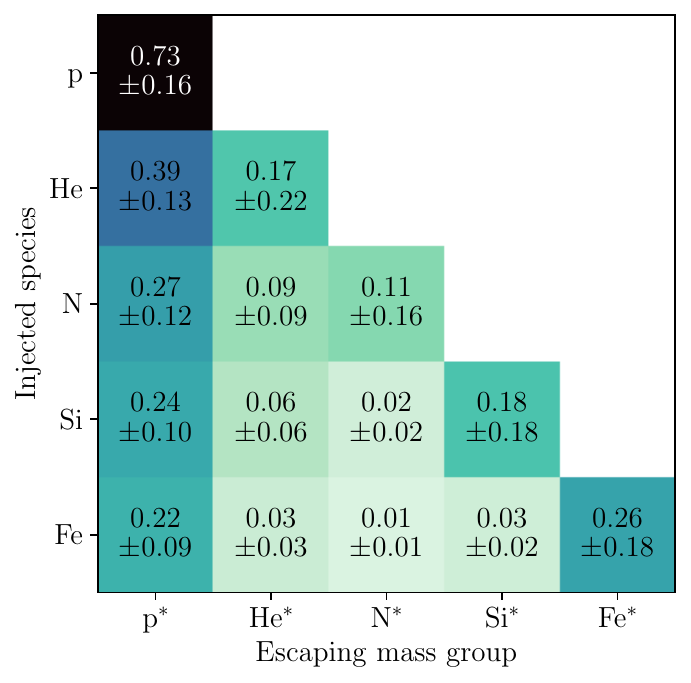}
        \caption{Average fraction (incl.\ $68\%$ uncertainty) of the injected luminosity above $10^{15.5}\,\si{\electronvolt}$ contained in each escaping mass group for a given species of primary cosmic rays. The mass groups correspond to ($[A_\text{min},\,A_\text{max}]$), $\text{p}^*=[1,\,1],~\text{He}^*=[2,\,4],~\text{N}^*=[5,\,14],~\text{Si}^*=[15,\,28],~\text{Fe}^*=[29,\,56]$. The amount missing to unity, summing all contribution for a given primary species, gives the relative luminosity transferred to neutrinos, low-energy cosmic rays, and electrons.}\label{fig:results_sample_lumi_fractions}
    \end{figure}    

    We identify the distance between the dust torus and the two shocks of the fast wind, particularly the forward shock $\rfs$, as the most important parameter that defines the maximum energy of the escaping cosmic-ray nuclei ($\rfs/R\ir$: $\rho=0.97$, $p<10^{-5}$); see \cref{fig:Emax_radius_correlation}. The same behaviour is observed for UHE protons ($\rho=0.91$, $p<10^{-5}$). For larger separations, the $1/R^2$ decrease of the photon density results in weak ambient photon fields in the region between two shocks where cosmic rays spend the majority of their time. This allows more cosmic rays to escape without interaction and reduces the suppression of the escaping flux. The characteristic radii ($R\ir,\,\rsh,\,\rfs$) of all the UFOs in our sample are shown in \cref{fig:ufo_radii}. Visual inspection confirms NGC\,7582 as the most favourable source of UHECRs since it exhibits the largest distance between the dust torus and the UFO shocked-wind region.

    In addition to NGC\,7582, we find a second population, comprising five objects (NGC\,4051, Mrk\,231, Mrk\,273, NGC\,2992, IRAS\,13224-3809), with a maximum energy of the escaping iron nuclei of $1-4\,\si{\exa\electronvolt}$. While there is a significant attenuation of the cosmic-ray nuclei in these UFOs, they form a class of sources clearly distinct from the main population where the maximum energy does not exceed $\sim10^{17}\,\si{\electronvolt}$. All of these sources exhibit a comparatively large separation between the dust torus and the characteristic radius of the shocks in the outflow (cf.\ \cref{fig:Emax_radius_correlation}, \cref{fig:ufo_radii}), resulting in weak photon fields in the downstream region and a mild suppression of the cosmic-ray flux due to interactions. Here, the maximum energy of the iron nuclei is not limited by interactions with the disc field but only by the torus field, and the maximum energy per nucleon is therefore comparable to the maximum energy of the protons in each outflow respectively.

    The spectrum of escaping nuclei does not always terminate rapidly when interactions with the disc or torus field become energetically viable. The interactions with the disc field can suppress the cosmic-ray flux by more than a factor of $1/e$ but not lead to a full attenuation. In this case, a brief flux recovery at higher energies can occur, before interactions with the torus field become relevant, or the escaping flux can be consistently non-negligible above $\emax$ but below the $e$-fold suppression threshold. In either case, the escaping spectrum cannot be understood as a simple power law with (exponential) cutoff and the nominal maximum energy, estimated as the point of $e$-fold flux suppression, does not capture the true maximum energy of the source. To adequately describe these UFOs, we define the energy of 10-fold flux suppression $E_\text{sup,10}$ as a second estimate of the maximum energy. For UFOs with a sharp spectral cutoff, such as the benchmark scenario, the value will be in close agreement with the nominal maximum energy $\emax$. We identify eight outflows with $\emax^\text{esc} < 10^{18}\,\si{\electronvolt}$ but $E_\text{sup,10}^\text{esc} \geq10^{18}\,\si{\electronvolt}$ for iron nuclei; NGC\,4151, HS\,0810+2554, 1ES\,1927+654, 1H\,0707-495, IC\,5063, IRAS\,05054+1718(W), Mrk\,590, and NGC\,1068.

    For primary protons, we observe a similar correlation of the maximum energy with $\rfs/R\ir$ (see \cref{fig:Emax_radius_correlation}, top) but without the significant outliers observed for nuclei. We find that a quarter of observed UFOs can emit a proton flux with $e$-fold maximum energy of $10^{18}\,\si{\electronvolt}$ or more; however, only four of these can reach up to $10^{18.4}\,\si{\electronvolt}$ (SDSS\,J1029+2623, HS\,0810+2554, IRAS\,13349+2438, Mrk\,273). When considering the 10-fold flux suppression, to capture protons in the high-energy tail of the flux cutoff, $60\%$ ($15\%$) of outflows can reach an energy of $10^{18}\,\si{\electronvolt}$ ($10^{18.6}\,\si{\electronvolt}$), with a maximum of $10^{19}\,\si{\electronvolt}$ found in two outflows (see \cref{tab:sample_best_sources_esc}). This suggests that the UFO population can provide an extragalactic flux of protons up to the ankle, especially when including the secondary protons produced during the disintegration of heavier primary cosmic rays in the source.

    On average, a significant fraction of the luminosity injected into high-energy cosmic rays at the wind termination shock is transferred to low-energy cosmic rays, electron-positron pairs, and neutrinos. In addition, a fraction of the remaining luminosity that eventually escapes the UFO in (ultra-)high-energy cosmic-rays is contained in secondary cosmic-rays with lower mass that are produced in the nuclear cascade; however, a majority of the luminosity in escaping cosmic rays is typically carried away by nuclei in the same mass group as the original primaries and secondary protons. The intermediate mass groups are only sparsely populated; see \cref{fig:results_sample_lumi_fractions}. The luminosity contained in secondary protons (incl.\ neutrons) is typically larger than that of the surviving nuclei in the same mass group as the injected species. In addition, this flux of light secondaries typically extends to higher energies than the heavy primary where most of the surviving luminosity is concentrated at energies just above the simulation threshold ($10^{15.5}\,\si{\electronvolt}$). The spectral shape of the secondary protons is generally not a simple power law with an exponential cutoff. We define the maximum energy of the secondary protons analogous to the maximum energy of the primaries as the energy beyond which the flux is suppressed by at least an $e$-fold compared to injected, primary, flux (see \cref{fig:results_sample_Emax_distribution}, hatched histogram).

    The average UFO gamma-ray spectrum in our model is lower than derived in a stacking analysis of eight UFOs by \citet{Fermi-LAT:2021ibj}, assuming cascading of the initially high-energy photons to sub-TeV energies. This suggests that an additional gamma-ray production mechanism, such as leptonic interactions or proton-proton/nucleus interactions in the shocked ambient medium may be required to match the observations, and that the UHECR from UFOs do not significantly contribute to the observed diffuse gamma-ray background.

    \subsection{The contribution of AGN ultra-fast outflows to the diffuse UHECR and neutrino flux}\label{sec:results_population_spectrum}
    \begin{figure*}
        \centering
        \includegraphics[width=0.48\linewidth]{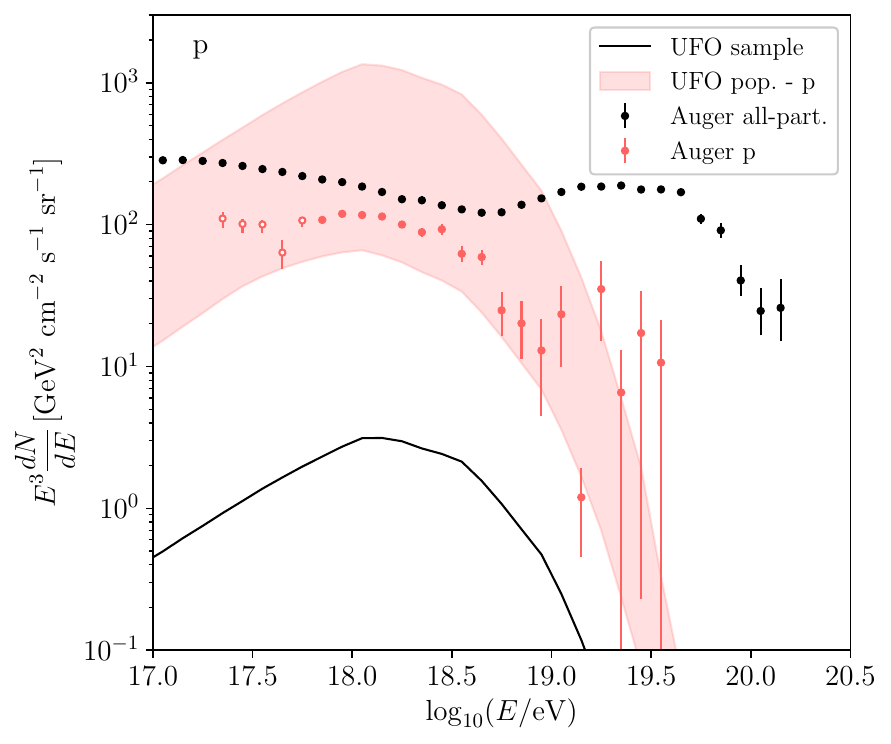}
        \includegraphics[width=0.48\linewidth]{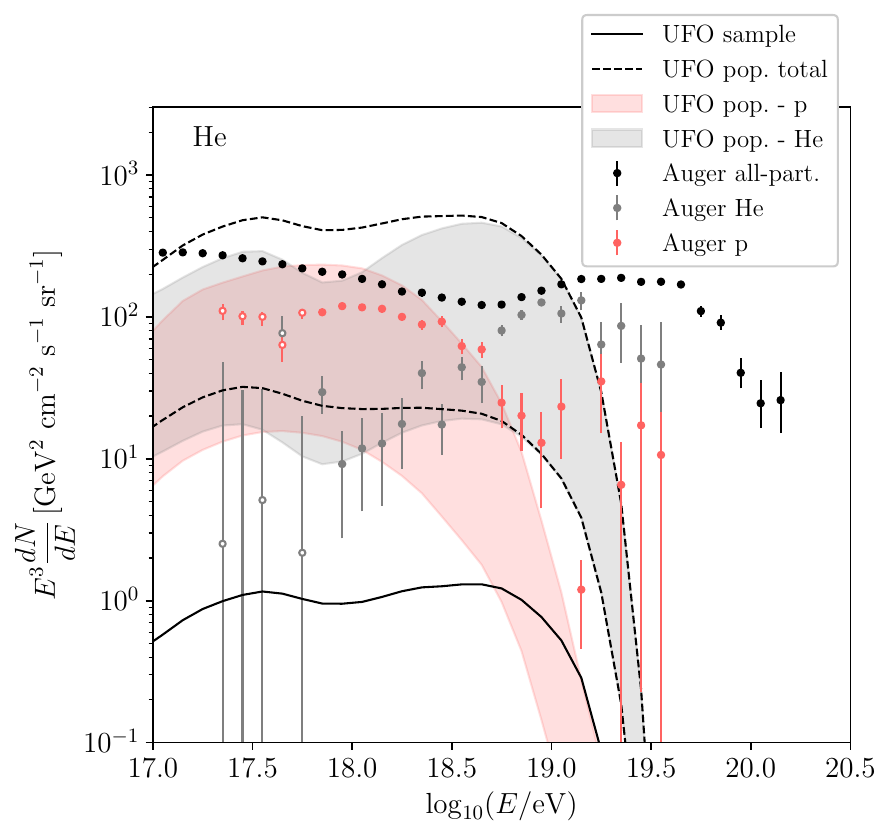}
        \includegraphics[width=0.48\linewidth]{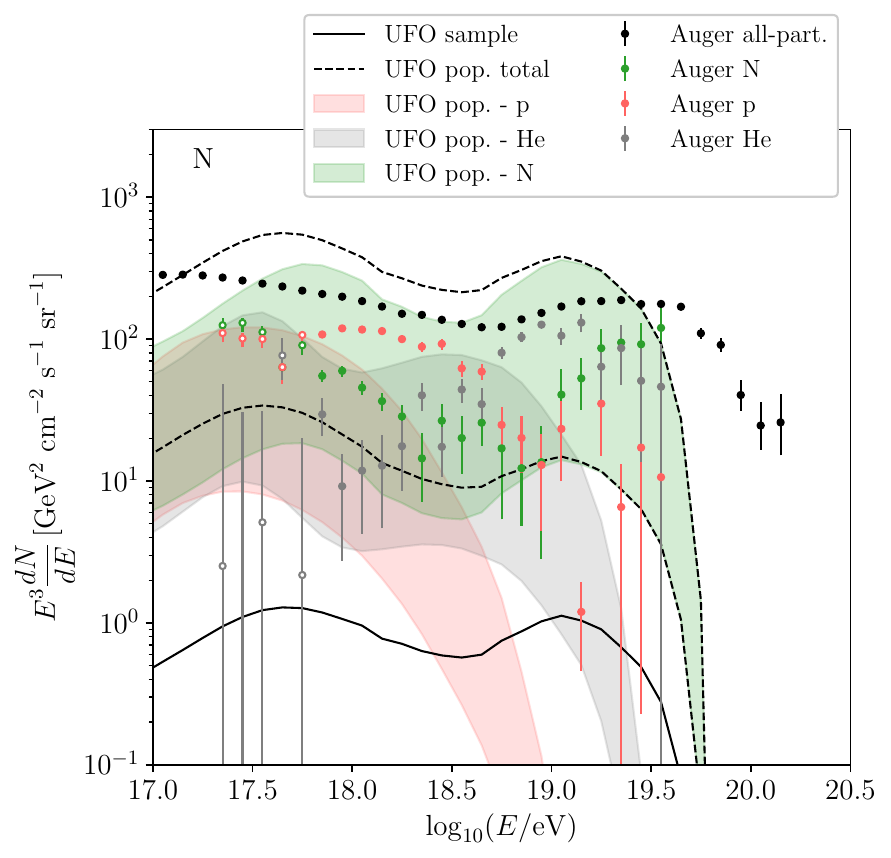}
        \includegraphics[width=0.48\linewidth]{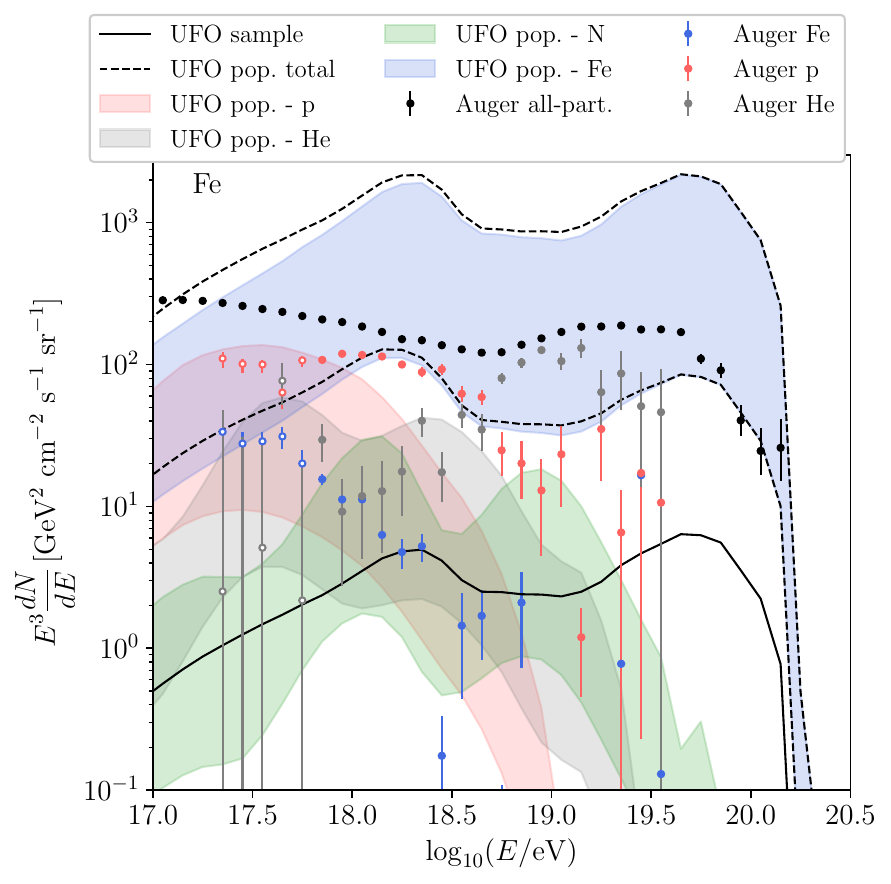}
        \caption{Expected cosmic-ray flux at Earth for injection of pure primary protons, helium, nitrogen, and iron nuclei; from all investigated UFOs (black, solid), and from the predicted UFO population up to $z=1.5$. The flux is bracketed by assuming a constant emissivity (lower limit) and number density (upper limit) of the UFO-hosting AGN at all redshifts. We also show the observed diffuse all-particle flux~\citep{PierreAuger:2014gko,PierreAuger:2021hun} and the flux per mass group from \citet{Bellido:2017cgf} (open circles) and \citet{Tkachenko:2021bja} (closed circles) assuming \textsc{EPOS-LHC}~\citep{Pierog:2015} for the hadronic interactions.}\label{fig:population_flux}
    \end{figure*}
    \begin{figure}
        \centering
        \includegraphics[width=\linewidth]{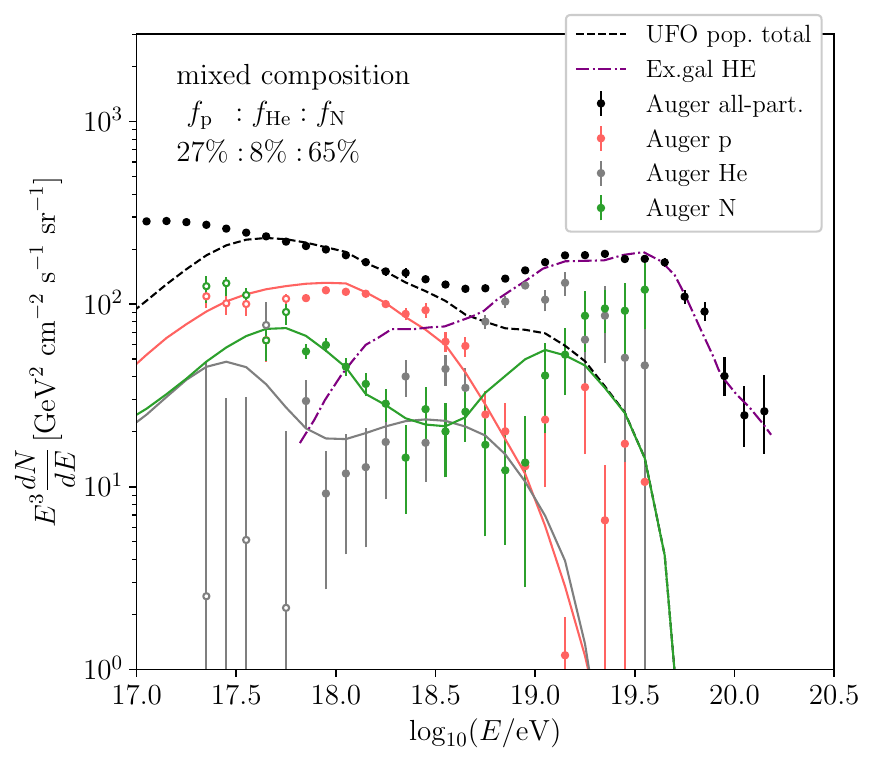}
        \caption{Expected cosmic-ray flux at Earth from the population of UFOs for a rescaled, injected composition of $f_\text{p}:f_\text{He}:f_\text{N} = 27\%:8\%:65\%$ (solid lines). The extragalactic, high-energy component of Scenario B of~\citet{PierreAuger:2022atd} is also shown with a purple dot-dashed line.}\label{fig:population_flux_combined}
    \end{figure}
    
    We show the expected cosmic-ray spectrum at Earth in \cref{fig:population_flux}, for the investigated UFOs and extrapolated to the entire population including undiscovered UFOs. As the composition of the accelerated cosmic rays is unknown we show the results for injection of protons, helium, nitrogen, and iron separately. A combined spectrum for injection fractions of $f_\text{p}:\,27\%,f_\text{He}:\,8\%,~f_\text{N}:65\%$, which gives a good description of the observed flux below the ankle, is shown in \cref{fig:population_flux_combined}.

    We provide two estimates of the completeness of our sample by comparing the sample emissivity and number density with the total (Compton-thin) AGN luminosity function~\citep{Ueda:2014tma}. The expected flux of UHECRs from UFOs, shown in~\cref{fig:population_flux}, is bracketed by the limits of these two approaches. We investigate redshift shells of width $\dif z=0.05$. In the ``local'' Universe ($z<0.05$), the AGN in our sample capture approximately $7.7\%$ of the X-ray emissivity and $0.3\%$ of the comoving number density of UFO AGN ($\sim50\%$ of all AGN). This applies to the interval of $2-10\,\si{\kilo\electronvolt}$ X-ray luminosities covered by AGN in our sample, i.e. $L_\text{X}\approx[10^{41.5},\,10^{46.3}]\,\si{\erg\per\second}$. We note the bias of our sample towards medium-/high-luminosity AGN in the local Universe;~\citep[cf.\ ][]{Gianolli:2024jkq}. Because of rapidly decreasing completeness at larger redshift, we estimate the contribution to the observed UHECR flux from the population of AGN UFOs by extrapolating from the observed local UFOs. In the local redshift shell out to $z=0.05$ we consider the individual contributions of the investigated UFOs (38 at $z<0.05$) and rescale their predicted flux by the completeness stated above. For the more distant redshift shells, we assume that the unobserved UFOs are well represented by the observed local sample. For each redshift shell except the local shell, we assume that the total cosmic-ray spectrum of the existing UFOs can be described by the ``effective'' stacked contribution of the local ($z<0.05$) sources but with the normalisation rescaled so that the X-ray emissivity and the number density of the associated AGN match the completeness observed in the local Universe. Since both the X-ray emissivity~\citep{Ueda:2014tma} and comoving volume increase for the more distant redshift shells this results in an enhancement of the total UHECR flux produced by more distant sources.
    
    We model the interactions during extragalactic propagation to Earth with \textsc{CRPropa} simulations, assuming ballistic transport and including the production of secondary cosmic rays and neutrinos. For local UFOs, we model the propagation individually considering their distance based on the observed redshift. For the effective population at larger redshift, we assume, for simplicity, that all of the flux is produced by a single source at a distance corresponding to the centre of the respective redshift shell. The flux at Earth is dominated by nearby sources; however, the contribution of distant sources to the sub-ankle protons and nuclei can be significant. Secondary cosmic rays from interactions with the cosmic microwave background and the extragalactic background light provide a subdominant contribution to the sub-ankle protons.

    In \cref{fig:population_flux}, we compare the predicted cosmic-ray flux at Earth for the population of UFOs with the diffuse UHECR flux observed by the Auger Observatory~\citep{PierreAuger:2014gko,Bellido:2017cgf,Tkachenko:2021bja,PierreAuger:2021hun}, using \textsc{EPOS-LHC}~\citep{Pierog:2015} to reconstruct the individual mass groups. We illustrate four separate scenarios, with an injection of a pure flux of cosmic rays at the acceleration shock, composed of protons, helium, nitrogen and iron nuclei, respectively. The population of UFOs can, in principle, explain the entire observed spectrum of UHECRs within the estimated uncertainties. However, as we discuss below, providing a good fit to the observed UHE nuclei above the ankle is challenging due to the predicted spectral shape of the flux of nitrogen and iron nuclei. In addition, the expected flux of cosmic-ray nuclei above the ankle is significantly lower when excluding NGC\,7582 from the analysis (see \cref{apx:no_ngc7582}). We highlight that the UHECR spectra of the UFO population presented here, including the spectra of individual mass groups, are based on a limited number of investigated sources. The true spectral shapes may be different.
    
    \subsubsection{Protons}
    The shape and normalisation of the observed proton spectrum is reproduced remarkably well by the expected protons from the UFOs. Interestingly, the maximum energy of the escaping protons is very similar (a few EeV) for the known UFOs which provide the largest contributions to the total expected flux. Sources with lower maximum energy are typically also less luminous in terms of the escaping proton flux and do not significantly affect the shape of the high-energy proton flux at Earth. As a result, UFOs emerge as viable sources of the observed protons below the ``ankle'' of the cosmic-ray spectrum relatively robust to modelling details and population uncertainties.
    
    However, this requires the injection of primary protons in the sources, as secondary protons from the disintegration of heavier nuclei cannot explain the high-energy tail of the observed proton spectrum due to the limited maximum energy of the primary nuclei within the sample of studied UFOs. The secondary protons can provide a significant contribution to the expected flux below $\sim10^{18.5}\,\si{\electronvolt}$, similar to the model of \citet{Unger:2015laa}, and can account for all the sub-ankle protons below $\sim10^{18}\,\si{\electronvolt}$.

     \subsubsection{Nuclei}
    The predicted nitrogen flux from the population of UFOs exhibits good agreement with the diffuse flux of nitrogen-like nuclei derived from observations, both for the normalisation and the spectral shape. However, the observed nitrogen above $\sim10^{19}\,\si{\electronvolt}$ is difficult to explain because of the limited maximum energy of nitrogen nuclei escaping the UFOs. The sub-ankle nitrogen is more robust and does not rely on the outsized contribution of a single source (NGC\,7582, see~\cref{apx:no_ngc7582}). The dip in the observed flux between $10^{18}-10^{19}\,\si{\electronvolt}$ -- approximately coincident with the ankle of the spectrum -- is well reproduced, even when excluding NGC\,7582 from the analysis. In our model, the dip is caused by the attenuation of the cosmic-ray flux in the source environment by photodisintegration due to the infrared field of the dust torus. The peak energy of this suppression for nitrogen nuclei coincides nicely with the energy of the ankle in the UHECR spectrum for the assumed dust temperature of $T_\text{IR}=\SI{200}{\kelvin}$.

    When interpreting the composition of the observed diffuse UHECRs with \textsc{EPOS-LHC} little helium is required up to the ankle. This observed flux can easily be explained by the expected flux from AGN UFOs if the injection fraction of helium at the sources is sufficiently low. In contrast, the substantial flux of helium nuclei above the ankle cannot be explained by UFOs without over-predicting the helium abundance at low energies. In addition, as for nitrogen nuclei, UFOs are challenged as sources of the observed flux of helium nuclei above $\sim10^{19}\,\si{\electronvolt}$ due to a limited maximum energy. A larger flux of sub-ankle helium is required when analysing the observed flux with \textsc{Sibyll 2.3}~\citep{Riehn:2017mfm}, see \cref{apx:no_ngc7582}. This flux cannot be explained easily by helium from UFOs without exceeding the sub-ankle protons around $\SI{0.5}{\exa\electronvolt}$ due to the production of secondary protons, which are in addition to the required primary protons (see above).

    For iron nuclei, UFOs can supply the observed sub-ankle iron derived with \textsc{EPOS-LHC} if the injection fraction is sufficiently low (few percent), although they may exceed the required low iron flux above the ankle. Better agreement can be reached when excluding NGC\,7582, which reduces the high-energy peak of the predicted iron flux, or when using \textsc{Sibyll 2.3} to interpret the observed flux. The flux of sub-ankle iron nuclei may instead be provided by a second, high-energy Galactic component~\citep{Thoudam:2016syr,PierreAuger:2022atd}.

    \subsubsection{Mixed injection}
    Considering the predicted spectra of the different cosmic-ray species discussed above, we conclude that AGN UFOs are an excellent candidate to explain the transition region between the end of the spectrum of Galactic cosmic-ray sources and the high-energy component of extragalactic sources typically used to explain the observed flux above the ankle. A contribution up to the highest observed energies may be possible, especially for iron nuclei. See \cref{sec:discussion_variable_luminosity,sec:discussion_compB} for more discussion.
    
    In \cref{fig:population_flux_combined}, we show the combined flux at Earth expected from the population of UFOs assuming that a fraction of $f_\text{p}:f_\text{He}:f_\text{N}=27\%:8\%:65\%$ of the total cosmic-ray luminosity of the UFOs is transferred to protons, helium and nitrogen respectively. These luminosity injection fractions provide a good description of the observed total diffuse UHECR flux and composition below the ankle. We indicate the additional, high-energy extragalactic component of Scenario B from \citet{PierreAuger:2022atd} at higher energies. In combination with the flux from the UFOs, we obtain a good qualitative description of the cosmic-ray spectrum across the ankle. The same mixed-composition scenario but without NGC\,7582 and for \textsc{Sibyll 2.3} is discussed in \cref{apx:no_ngc7582}.

    \subsubsection{Neutrinos}
    If UHECRs are accelerated in UFOs we expect a significant high-energy neutrino flux (see also P23) given their interactions with the photon fields in the vicinity of the AGN. Additional neutrinos are produced during the extragalactic propagation of the escaping cosmic rays. Here, we present the neutrinos that must accompany the UHECR signal if UFOs are major sources of the sub-ankle UHECRs. 
    
\begin{figure}
        \centering
        \includegraphics[width=\linewidth]{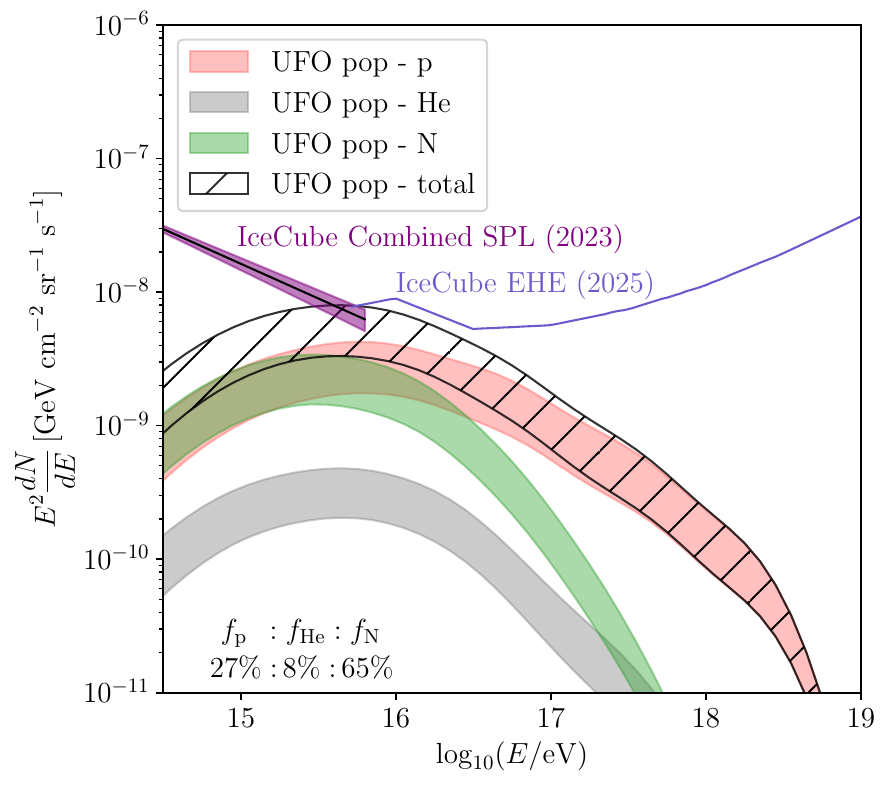}
        \caption{Expected all-flavour neutrino flux from the UFO population if they produce a major fraction of the UHECR diffuse flux below the ankle (see  Fig.\ref{fig:population_flux_combined}); for a typical outflow age of $\SI{1000}{\years}$. The bands bracket the uncertainty of the UFO redshift evolution; AGN-like~\citep{Ueda:2014tma} (upper limit) and no redshift evolution (lower limit). We also show the measured IceCube diffuse, single power law (SPL) neutrino flux~\citep{Naab:2023xcz} and the upper limits from the EHE analysis~\citep{IceCube:2023luu}.\label{fig:neutrinos}}
    \end{figure}

    Fig~\ref{fig:neutrinos} shows the all-flavour neutrino flux from the UFO population, estimated in the same way as the cosmic-ray flux in Fig~\ref{fig:population_flux_combined} and including neutrinos from the extragalactic propagation which we calculated with \textsc{CRPropa} out to $z = 1.5$\footnote{We have checked using the analytical approach of 
    \citet{Dermer:2009zz} -- see their Eq.\ 4.54 -- and assuming that the redshift evolution of UFOs follows the star formation rate of~\citet{Y_ksel_2008} that integrating out to redshift $z =5$ increases the neutrino intensity by less than $1\%$.}. Here, we do not show the entire range of possible UHECR emissivities shown in \cref{fig:population_flux}. Instead, we choose a UHECR emissivity sufficient to match the UHECR observations below the ankle as in \cref{fig:population_flux_combined}.
    If the majority of observed UHECRs in the transition region are produced by UFOs, there is a guaranteed accompanying neutrino flux that peaks at $\sim\SI{5}{\peta\electronvolt}$ and can, depending on the redshift evolution of the UFO population, provide a significant contribution to the observed, diffuse neutrino flux above $\SI{1}{\peta\electronvolt}$. A large flux at ultra-high energy ($E_\nu\gtrsim10^{17}\,\si{\electronvolt}$) is not expected because of the limited maximum energy of most UFOs. The diffuse UFO neutrino signal is consistent with the neutrino excess in the direction of non-jetted AGN reported in~\citet{IceCube:2021pgw}. In \cref{apx:neutrinos}, we present an astrophysically motivated UHECR composition scenario and discuss the detectability of neutrino point sources from the UFO population. In older outflows, where the wind termination shock and the forward shock are further away from the AGN, the expected neutrino flux may be reduced because of a reduction of in-source interactions.
    
\section{Discussion}\label{sec:discussion}
    \subsection{Maximum energy of cosmic rays}
    We have studied the maximum energy that cosmic-ray nuclei can reach in the termination shocks of mildly-relativistic ultra-fast outflows. Such UFOs are observed in about 50\% of AGN. We found that in typical UFOs, the energy of escaping cosmic-ray nuclei is limited by photodisintegration to about $E \sim10^{16}-10^{17}\,\si{\electronvolt}$. This is a result of confinement by substantial magnetic fields $\mathcal{O}(0.01 - 1~\rm G)$ in the vicinity of the AGN, where the nuclei lose energy via interactions with strong photon fields. The dominant photon fields are those of the accretion disc and the dust torus. The most promising sources, approximately $5-15\%$ of outflows in the investigated sample, reach energies exceeding $\sim10^{18}\,\si{\electronvolt}$ for iron, and one source (NGC\,7582) reaches $10^{19}\,\si{\electronvolt}$; see~\cref{tab:sample_best_sources_esc}.
    Protons experience lower energy losses than nuclei during the escape process, in part due to intermediate conversion to neutrons that escape the UFO easily. Approx.\ $70\%$ of the luminosity injected in protons at the wind termination shock escapes the UFO environment as high-energy protons, whereas for nuclei only $10-30\%$ escapes in the form of high-energy cosmic-ray nuclei with mass comparable to the injected primary. In the 86 investigated UFOs, protons attain the highest energies after escape among the nuclear species studied (p, He, N, Si, Fe) in 81 outflows, despite the maximum energy at acceleration being larger for iron in all investigated outflows. Up to $25\%$ of the investigated UFOs can provide a significant flux of protons with an energy greater than $10^{18}\,\si{\electronvolt}$, with a maximum of $10^{18.4}-10^{18.6}\,\si{\electronvolt}$ in four outflows. The highest-energy protons can reach energies of $10^{18}\,\si{\electronvolt}$ or more (considering $E_\text{sup,10}$) in approximately $60\%$ of the investigated UFOs. Due to strong photodisintegration, most ultra-fast outflows are unlikely to be the sources of the observed UHECR nuclei above the ankle. However, nuclei might escape during low-emission states of the AGN, as we demonstrated for NGC\,7582 which, during its low-luminosity state, could be allowing cosmic rays with energy up to $E \sim 10^{19.8}\,\si{\electronvolt}$ to escape. Primary protons, whose flux is less suppressed, and secondary protons from the spallation of heavier primary nuclei can contribute to the light sub-ankle flux component.

    \subsection{Ultra-fast outflows as intermittent sources of ultra-high-energy nuclei above the ankle}\label{sec:discussion_variable_luminosity}
    We have identified the outflow in NGC\,7582, a type 2 Seyfert galaxy, as the most probable source of UHECR nuclei above the ``ankle'' (energy $10^{18.7}$ eV) in the present sample of observed outflows due to the low luminosity of the associated AGN ($\lbol[L_X]\approx10^{42.8}[10^{41.6}]\,\si{\erg\per\second}$) reported in the SUBWAYS-III sample of \citet{Gianolli:2024jkq} and corroborated by the XWING sample of \citet{Yamada:2024lss} ($L_X\approx10^{41.7}\,\si{\erg\per\second}$). However, archival data indicate a large variability of observed X-ray emission, which could be due to a change in intrinsic luminosity of the AGN~\citep{Piconcelli:2007gt}, or due to the variable obscuration of the central region by a clumpy torus~\citep{Rivers:2015ifo}. The intrinsic X-ray luminosity of the central source can be as low as $\sim10^{41.5}\,\si{\erg\per\second}$~\citep{Piconcelli:2007gt,Rivers:2015ifo}, our nominal value, or as high as $\sim10^{42.6}-10^{43.5}\,\si{\erg\per\second}$~\citep{Goulding:2012tz,Ricci:2015rzl}. Assuming the latter luminosity, the maximum energy per nucleus of the cosmic-ray nuclei at acceleration and escape is reduced to $\lesssim10^{17.3}\,\si{\electronvolt}$ and $<10^{15}\,\si{\electronvolt}$, respectively, and NGC\,7582 becomes unremarkable compared to the bulk of UFOs. The maximum energy of primary protons is affected only marginally. If the dimming of NGC\,7582 is due to increased obscuration of the central region along the line of sight~\citep{Rivers:2015ifo}, the photon fields within the UFO are stronger than anticipated in our analysis and prevent the escape of UHE cosmic-ray nuclei at all times.
    
    Significant luminosity variations were also observed in other AGN with associated UFOs, e.g.\ PDS\,456~\citep{Reeves:2013opa}, 1H\,0707-495~\citep{Hagino:2015hca}, IRAS\,13224-3809~\citep{Jiang:2018ixk}, NGC\,4151~\citep{AstrophysicsGroupCavendishLaboratoryUK:2020fyz}, and PG\,1448+273~\citep{Laurenti:2020ftw}. This suggests that UFOs could be intermittent sources of ultra-high-energy nuclei where the escape of nuclei beyond the ``ankle'' energy is only possible when the associated AGN is in a low state. 
    
    As shown in~\cref{fig:population_flux_combined}, the contribution of UFOs to the flux beyond the ankle is relatively limited in our model, unless sources like NGC\,7582 are more abundant in the UFO population than in our sample. In addition, UFOs could contribute substantially to the highest energies if there is a significant fraction of iron in that energy range, as suggested in~\citet{PierreAuger:2024flk}. Furthermore, here, we have only explored a specific scenario in which UHECRs are accelerated at the wind-termination shock, and our study has been limited by our incomplete knowledge of the UFO luminosity function. A larger sample of UFOs, and different models of UHECR acceleration in UFOs, for example including time dependence and possibly the impact of the forward shock in the early phases of the expansion, are needed to conclusively resolve the possible role of UFOs as sources of the highest-energy extragalactic cosmic-ray component.

    \subsection{Ultra-fast outflows as the sources of Hillas' ``Component B''}\label{sec:discussion_compB}
    The KASCADE experiment has identified an iron knee feature in the cosmic-ray spectrum at energy $\approx10^{16.8}$~eV~\citep{PhysRevLett.107.171104}, which likely marks the maximum energy of the Galactic cosmic-ray component. At higher energies, above the ankle of the UHECR spectrum, the combined spectrum and composition can be fit with a Peters' cycle and are attributed to extragalactic sources. However, in this picture, there is a gap in flux between the Galactic and extragalactic components~\citep{Hillas:2005cs,DeDonato:2008wq}. 
    In the ``gap" region, the composition consists primarily of protons and nitrogen~\citep{Tkachenko:2021bja}. Interpreting the Auger composition measurements with \textsc{Sibyll 2.3} also requires a non-negligible helium component; however, this component is not necessary when using \textsc{EPOS-LHC}. The protons that have a soft spectrum below the ankle have previously been explained to originate in an additional unspecified extragalactic source population~\citep{Aloisio:2009sj}, as the products of photodisintegration in the source environment~\citep{Globus:2015xga, Unger:2015laa}, or as protons from FRI-type jetted AGN~\citep{Giacinti:2015pya}. But for the ``sub-ankle'' nitrogen, there exist no astrophysical models of extragalactic origin. In~\citet{PierreAuger:2022atd}, this component was interpreted as a second lower-energy extragalactic UHECR component, without specifying a specific source class. The older work of~\citet{Hillas:2005cs} explained this part of the spectrum as an iron-rich Galactic ``Component B'' produced by supernovae. However, such an iron-rich composition is disfavoured by the latest UHECR composition measurements. More recently, \citet{Thoudam:2016syr} have proposed that the sub-ankle nitrogen could be produced in supernova explosions of Galactic Wolf-Rayet stars which have a CNO-rich stellar wind. We have shown here that UFOs provide an alternative explanation for the cosmic ray flux in the Galactic-extragalactic transition region and can account for the sub-ankle protons and nitrogen simultaneously.

    Compared to starburst winds, which have also been discussed as possible sources of UHECRs~\citep{Anchordoqui:2018vji,Romero:2018mnb,Condorelli2023}, UFOs seem much more likely to accelerate UHECRs. The former were shown to not have enough power to reach the ankle in \citet{Peretti:2021yhc}. However, this does not constrain starburst galaxies as hosts of UHECR accelerators, as a substantial fraction of star-forming galaxies is also characterised by an AGN~\citep{Gruppioni:2013}. Interestingly, NGC\,4945, the starburst galaxy responsible for $\sim40\%$ of the observed correlation of UHECRS with starburst galaxies~\citep{PierreAuger:2018qvk}, hosts a well-known heavily obscured AGN at its core~\citep{Puccetti:2014pqa}.

    \subsection{The impact of magnetic horizons}
    Astrophysical magnetic fields can lead to a delay of the emitted cosmic rays compared to the electromagnetic emission of the AGN/UFO, resulting in a possible loss of correlation between the different messengers. In the most extreme case, if the time delay for a given UFO exceeds the Hubble time, the cosmic rays of this source will be hidden beyond the magnetic horizon. In addition, the cosmic-ray flux will be suppressed if the effective path length of the cosmic rays becomes comparable to or larger than the typical energy loss length. These effects can result in a hardening of the observed cosmic-ray flux.

    We find that the galactic magnetic fields, both of the typical host galaxy of the UFO and of the Milky Way, do not induce a significant suppression of the cosmic ray flux for $E_\text{CR}\gtrsim10^{17}\,\si{\electronvolt}$ although the time delay of up to $\mathcal{O}(100\,\si{\kilo\years})$ at sub-ankle energies may lead to a loss of time correlation between the cosmic rays and a potential electromagnetic or neutrino counterpart. The cosmic ray flux below the ankle can be suppressed significantly if the AGN/UFO is located in a galaxy cluster; see \citet{Condorelli:2023xkx} for a general discussion. However, most galaxies are not located in large clusters~\citep{Bahcall:1995tf}. Finally, cosmic rays may be confined by the extragalactic magnetic field. If the field strength is close to the current upper limit ($\sim\SI{0.01}{\nano\gauss}$, see \citealp{Jedamzik:2018itu}), the flux of intermediate and heavy cosmic rays can be suppressed by approximately an order of magnitude at $10^{17}\,\si{\electronvolt}$. The suppression is less significant for lighter nuclei and protons and at higher energies where deflections in the magnetic field are reduced. If the extragalactic magnetic field is significantly weaker, compatible with present lower limits (e.g.\ \citealp{Fermi-LAT:2018jdy}), the flux suppression and time delay become small. The spectral hardening induced by the magnetic horizons can limit the viability of ultra-fast outflows as the sources of the cosmic ray flux in the transition region, depending on the strength of the intervening magnetic fields. However, the harder spectrum can make UFOs more attractive sources of the highest-energy cosmic rays.

    \subsection{Uncertainties of the model}
    \subsubsection{Geometry of the dust distribution}
    The maximum energy of the cosmic rays is sensitive to the strength of the infrared field of the hot dust in the AGN, for which we have assumed a spherical distribution with all of the dust concentrated in a thin shell at radius $R\ir$. However, recent high-resolution observations of individual AGN indicate a more complex structure \citep[see e.g.][]{Honig:2019ApJ}, favouring a small scale height with most of the dust close to the plane of the accretion disc~\citep{GRAVITY:2020A&A}.
    
    To study the dependence of our results on the assumed dust geometry, we investigate the opposite limit, namely an equatorial ring of dust. Our previous conclusions are unchanged when this dust distribution is adopted instead of our fiducial spherical model. In the most optimistic case of the dust-ring model (at small polar angles), the maximum energy at acceleration is increased by, on average, $(5\pm5)\%$ for cosmic-ray nuclei and by $(1\pm1)\%$ for protons. In some individual UFOs, we observe an increase in maximum energies of up to $30\%$ and $10\%$, respectively. The maximum energy after escape is reduced by $10-30\%$ on average. However, it remains invariant within numerical binning uncertainties for between $15\%$ (helium) to $60\%$ (iron) of outflows, and increases for up to $10\%$ of outflows both for nuclei and protons; see \cref{apx:dust_distribution}.

    \subsubsection{Uncertainty of the kinetic luminosity}\label{sec:limit_mass_outflow_rate}
    We find a median ratio of kinetic luminosity to bolometric luminosity of $L_\text{kin}/\lbol\approx0.1$ in our UFO sample, consistent with \citet{Fiore:2017A&A}. However, $25\%$ ($15\%$) [$10\%$] of investigated outflows have a predicted luminosity ratio greater than $1$ ($5$) [$10$], leading to a mean value of $\langle L_\text{wind}/\lbol\rangle=3.7^{+0.7}_{-3.7}$. Such super-bolometric outflows were previously noted by \citet{Chartas:2021ApJ} and \citet{Mestici:2024pjt}. For radiation-driven outflows, the luminosity ratio cannot exceed the covering factor of the outflow $f_c\leq1$ (maximum for spherical outflows). Outflows driven by magneto-hydrodynamic effects~\citep[see][]{Fukumura:2010ApJ,Kraemer:2017rge} can exceed this limit. The large number of objects with $L_\text{kin}/\lbol>1$ probably suggests that magnetic driving is dominant or at least contributes to the acceleration of UFOs. Analogous behaviour is well known in AGN jets where the jet luminosity exceeds the accretion luminosity~\citep{Ghisellini_2014}. 
    
    The cosmic-ray flux of the UFO population -- as extrapolated from the sample of known outflows -- can supply only a small fraction of the observed UHE nuclei if the kinetic luminosity of the outflows is limited to at most $\lbol$ by reducing the mass outflow rate; see \cref{apx:limit_outflow_rate}. In contrast, it is still possible to explain the sub-ankle protons. This suggests that our conclusions about UFOs as potential sources of the sub-ankle protons are robust.

\section{Summary and Conclusion}\label{sec:summary}
    We have explored the potential of ultra-fast outflows (UFOs) in AGN to produce ultra-high energy cosmic rays (UHECRs), focusing on acceleration at the wind-termination shock. We obtained the maximum energies of five representative UHECR species, from protons up to iron, by comparing the characteristic timescale for particle acceleration and the most important loss processes. Subsequently, we determined the spectral shape and total normalisation of the cosmic-ray flux escaping the UFO environment using 3D simulations of the interactions of UHECRs in these sources. 
    
    We studied 86 observed UFO systems with known velocities and mass-outflow rates, and show that the maximum acceleration energy is positively correlated with both of these quantities. Of the observed ultra-fast outflows, six objects can accelerate iron nuclei to at least $10^{19.8}\,\si{\electronvolt}$. The maximum energy of nuclei at acceleration is limited by photodisintegration due to the infrared field of the dust torus and the optical-UV field of the accretion disc.
    
    Intense AGN photon fields efficiently prevent the escape of the highest-energy cosmic-ray nuclei from the majority of outflows. In our model, cosmic-ray protons can escape the source with energy up to $\sim10^{18}\,\si{\electronvolt}$ and nuclei with at most $10^{16}-10^{17}\,\si{\electronvolt}$ in the majority of studied UFO. Most of the energy injected into UHE nuclei at acceleration is typically transferred to secondary protons, electron-positron pairs, and neutrinos. In approximately $5-15\%$ of our sample, nitrogen and helium nuclei escape with maximum energy of $10^{17.6}\,\si{\electronvolt}$ or more, making these intriguing candidate sources of the intermediate-mass nuclei in the ``sub-ankle’’ region of the UHECR spectrum. 
    
    The maximum energy of the escaping cosmic rays is most significantly correlated with the separation between the UFO shocks and the dust torus. For a few UFOs, with weak external photon fields, nuclei may escape with comparatively little attenuation, even at the highest energies. The low-luminosity Seyfert galaxy NGC\,7582 provides the most suitable environment for the production of UHECR nuclei within our sample of UFOs. The weak photon fields are insufficient to suppress the flux of escaping cosmic rays, resulting in an iron flux with energy up to $10^{19.8}\,\si{\electronvolt}$. The lack of flux-complete surveys of ultra-fast outflows implies the existence of a large population of hitherto undiscovered UFOs, especially in low-luminosity AGN, which are disproportionally under-represented in our sample. Some of these could provide conditions comparable to NGC\,7582 for the acceleration and escape of UHE nuclei up to the highest observed energies.
    
    Protons are less attenuated because of the absence of photodisintegration and because of the conversion to neutrons (in photopion interactions), which are not confined by the strong magnetic fields in the UFO. This allows $\sim\,$half of the observed UFOs to produce a proton flux that reaches $10^{18}\,\si{\electronvolt}$ or more. The escaping protons retain most of the injected luminosity, and the escaping flux inherits the correlation of the maximum acceleration energy with the wind velocity and outflow rate. Primary protons accelerated at the wind termination shock, and secondary protons from the disintegration of heavier nuclei can comfortably provide the observed proton flux in the transition region below the ankle.
    
    All in all, we have demonstrated that ultra-fast outflows are viable sources of the observed ``sub-ankle'' UHECRs. As such, they provide an excellent astrophysical explanation for the observed diffuse cosmic-ray flux between the Galactic iron knee and the extragalactic UHECRs above the ankle of the cosmic-ray spectrum in terms of energetics, spectral shape, and chemical composition. 
    This ``gap'' in the Galactic-extragalactic transition region is a long-standing open question~\citep{Hillas:2005cs,DeDonato:2008wq}. To our knowledge, ultra-fast outflows of AGN are the first extragalactic source class shown to be suitable for this component while accounting for the CNO-type nuclei observed by Auger in this energy range. 
        
    If the observed ``sub-ankle'' UHECRs are produced by ultra-fast outflows in AGN, they must be accompanied by a guaranteed flux of neutrinos at energy $\gtrsim\,$few PeV, providing an important multimessenger signature for the acceleration of cosmic rays in these objects. The expected neutrino signal is independent of the neutrino emission at $\sim 10-30\,\si{\tera\electronvolt}$ from the cores of AGN such as NGC\,1068~\citep{IceCube:2022der} and should contribute substantially to the diffuse neutrino flux at $\gtrsim\,$few PeV energy. 
      
    In this work, we have explored a specific scenario in which UHECRs are accelerated in the wind-termination shock of the UFO. Our study has been limited by our incomplete knowledge of the UFO population and luminosity function. Better understanding of the latter and different models of UHECR acceleration in UFOs, for example, including intermittency, time dependence and the possible role of the forward shock, are needed to elucidate the possible role of UFOs as sources of the highest-energy extragalactic cosmic-ray flux.

\section*{Acknowledgements}
    We thank Michael Kachelriess for help with photopion inelasticity calculations with \textsc{SOPHIA}, Egor Podlesny for useful discussions on electron-positron pair production, Michael Unger for advice on \textsc{CRPropa3-data} and multiple helpful discussions, and Marco Muzio and Arjen van Vliet for feedback on the paper. We also thank Francesco Tombesi, Alessandra Lamastra and Antonio Condorelli for useful discussions. Finally, we thank the authors of the SUBWAYS-III and XWING papers, especially Vittoria Gianolli, Claudio Ricci and Satoshi Yamada, for their help with understanding the details of their compiled UFO samples.
    EP was supported by Agence Nationale de la Recherche (grant ANR-21-CE31-0028).

\section*{Software}
    Part of the interaction rates in the sources and the entire propagation from the sources to Earth were calculated with \textsc{CRPropa\,3}~\citep{CRPropa2:2013,AlvesBatista:2016vpy,AlvesBatista:2022vem}, which is publicly available from \href{https://crpropa.desy.de}{crpropa.desy.de}. The version adapted to model the UFO environment with radially dependent photon fields, magnetic field, and advection field can be retrieved from \href{https://github.com/ehlertdo/CRPropa3/tree/outflows}{github.com/ehlertdo/CRPropa3}.

\section*{Data availability}
    A machine-readable version of \cref{tab:ufo_list}, as well as the supplementary data files, for (1) the predicted cosmic-ray and neutrino flux of the entire UFO sample and population and (2) the predicted neutrino flux of all observed ($z<1$) UFOs, are available at \citet{ehlert_2025_zenodo}. Additional files may be made available upon reasonable request.

\bibliographystyle{mnras}
\bibliography{bibliography}

\clearpage
\appendix
\section{Diffuse UHECR Flux Contributions from the AGN UFO Population - Model Variations}\label{apx:no_ngc7582}

\begin{figure}
    \centering
    \includegraphics[width=\linewidth]{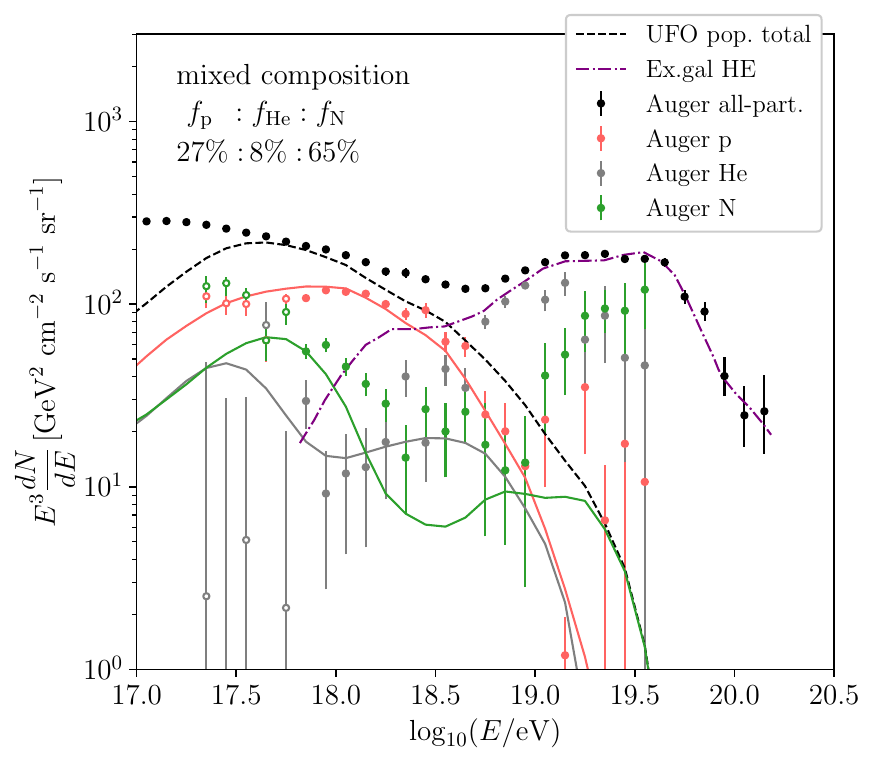}
    \caption{Same as~\cref{fig:population_flux_combined} but without NGC\,7582.\label{fig:no_ngc7582}}
\end{figure}

\begin{figure}
    \centering
    \includegraphics[width=\linewidth]{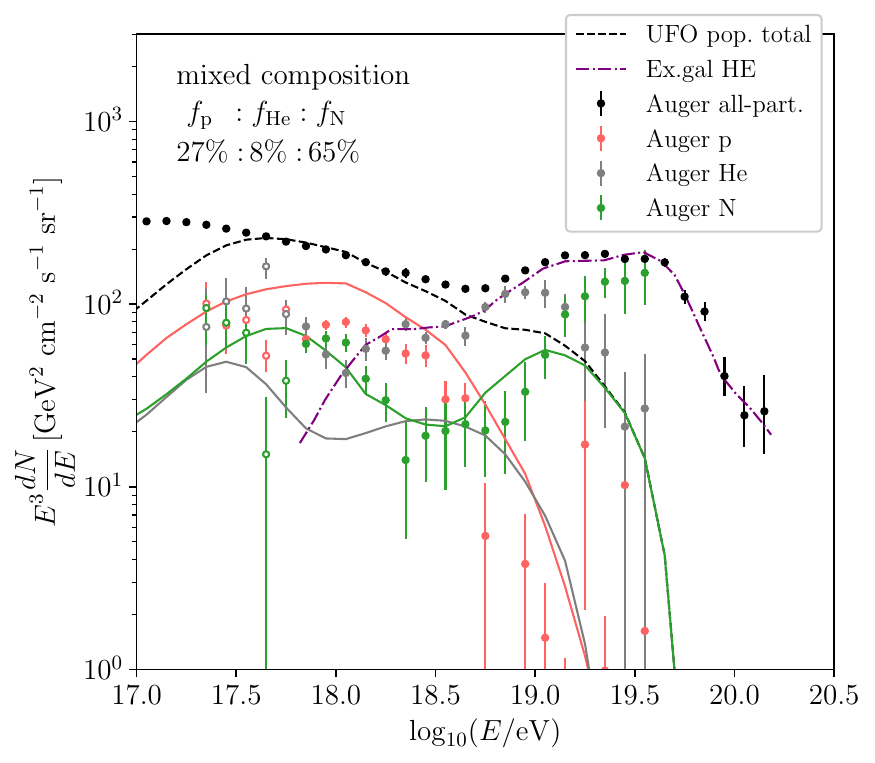}
    \caption{Same as~\cref{fig:population_flux_combined} but using \textsc{Sibyll 2.3} instead of \textsc{EPOS-LHC}. \label{fig:Sibyll}}
\end{figure}

\paragraph*{Diffuse UHECR flux in the absence of a nearby source of UHE nuclei}
In the absence of more detailed statistics or physical understanding of the prevalence of UFOs in AGN of different luminosities, the prevalence of NGC\,7582-like systems is difficult to quantify. On the one hand, such low-luminosity AGN are abundant locally, and from the point of view of the AGN luminosity, NGC\,7582, at a distance of 21.2~Mpc, is not unusually nearby. On the other hand, the combination of powerful-enough UFO and low-enough AGN luminosity is unique in our sample of 86 sources. To demonstrate the effect of the presence of NGC\,7582 in our sample, in \cref{fig:no_ngc7582} we show an extreme scenario without this source in the local, $z < 0.05$, UFO population. The contribution of NGC\,7582-like sources at larger redshift is still considered as before. The injected fractions are the same as those shown in \cref{fig:population_flux_combined}. In such a case, the contribution of UFOs to the highest energy part of the spectrum is likely very limited because increasing the nitrogen or iron contribution would lead to overshooting the total flux at lower energies even with a pure nitrogen or iron injection. On the other hand, the sub-ankle contribution of UFOs to the diffuse UHECR flux is largely insensitive to the exclusion of NGC\,7582.
\paragraph*{Sibyll 2.3}
\cref{fig:Sibyll} shows the contribution of the UFO population to the observed diffuse UHECR flux using the composition fractions obtained with \textsc{Sibyll 2.3} instead of~\textsc{EPOS-LHC}. In this case, the predicted flux of nitrogen-like nuclei exhibits better agreement with the observations, and the observed proton flux can already be explained by a slightly lower proton flux. On the other hand, \textsc{Sibyll 2.3} requires a substantial helium component, which can only be matched at the expense of the primary protons and nitrogen. A significant injection of primary helium at the sources results in an over-prediction of the sub-ankle proton flux around $\SI{0.5}{\exa\electronvolt}$ due to the associated secondary protons. However, the injection of primary protons is necessary to explain the high-energy tail of the observed proton flux, which cannot be explained by secondary protons within our analysis.

\section{Neutrinos from the UFO population with an astrophysically motivated injection composition}\label{apx:neutrinos}

In this work, we have investigated the acceleration and escape of protons and nuclei from UFO environments. The injection of nitrogen is motivated by the observed UHECR composition, but our model does not address the origin of nitrogen in the AGN environment. One of the interesting sources in our sample is NGC\,6240. It is a merging, dual AGN that exhibits a highly supersolar metallicity in its central region $Z/Z_{\odot} \sim 2 - 10$, meaning that heavy elements constitute $4 - 20\%$ of the matter by mass~\citep{Puccetti:2015nba}. Such conditions are promising for UHECR production sites with respect to the injected composition, but we do not explore this topic further here. 

\begin{figure}
    \centering
    \includegraphics[width=\linewidth]{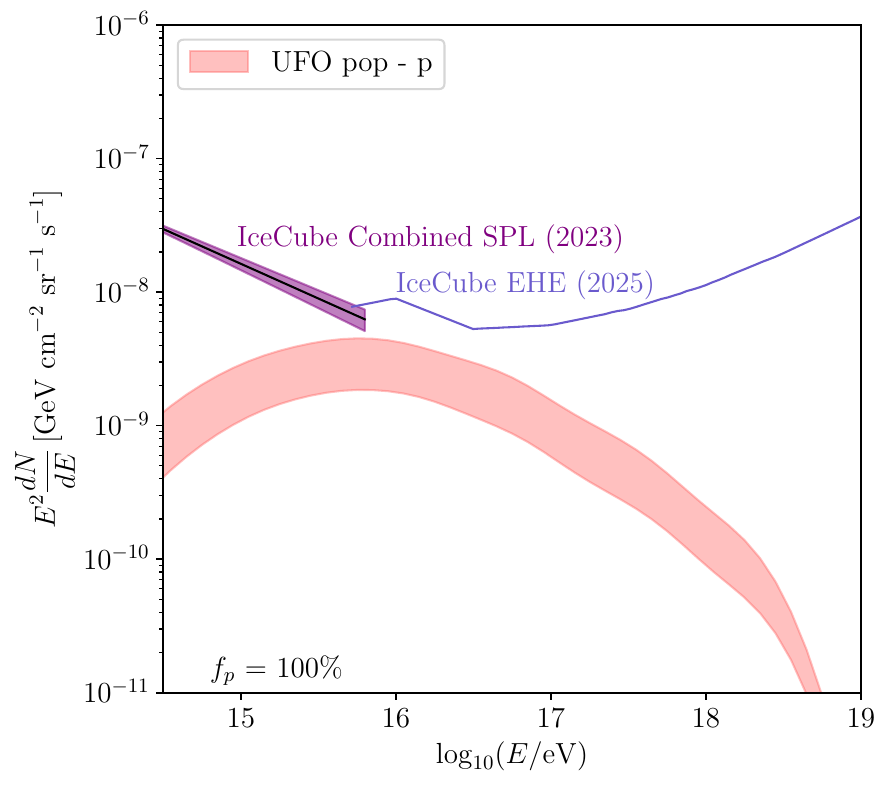} 
    \caption{Same as Fig.~\ref{fig:neutrinos} except that here we model the expected neutrino flux from the UFO population assuming that the UHECRs produced in UFOs produce only the protons measured with Auger.\label{fig:neutrinosapx}}
\end{figure}

In Fig.~\ref{fig:neutrinosapx} we show the expected diffuse neutrino intensity from the UFO population in an alternate scenario, where the ``sub-ankle'' UHE proton flux is entirely supplied by primary protons from UFOs and no other nuclear species are accelerated in UFOs. We see that if the UHECR protons measured by Auger are produced by AGN UFOs, there is a guaranteed accompanying flux of neutrinos with a peak intensity at energy of a few PeV, similar to the baseline scenario shown in~\cref{fig:neutrinos}. 

\begin{figure}
    \centering
    \includegraphics[width=\linewidth]{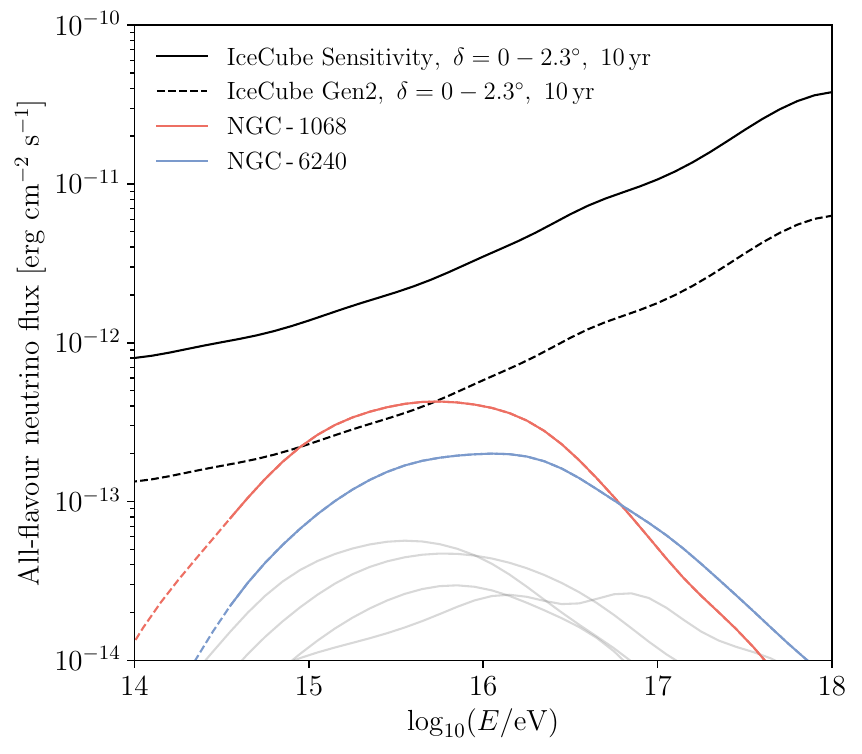}
    \caption{Predicted all-flavour neutrino flux from the tentative observations of ultra-fast outflows in NGC\,1068 and NGC\,6240 (see~\citealp{Yamada:2024lss} and references therein)\label{fig:neutrino_point_sources}, assuming pure-proton cosmic rays. The differential sensitivity of the IceCube throughgoing muon analysis with effective area $A_{\rm eff}(E_{\nu}, \delta)$ at declination $\delta$, which is $\sim 3E_{\nu}/[A_{\rm eff}(E_{\nu},\delta)\, t_\text{det} \ln 10]$, is shown with a black solid line~\citep{IceCube:2016tpw}. For IceCube-Gen2 (black-dashed line), we estimate the sensitivity assuming a six-fold increase in effective area with respect to that of IceCube. The sensitivity of IceCube toward the UFOs with lower neutrino flux (grey lines) is less favourable due to source declinations not near the local horizon.}
\end{figure}

We have also investigated the sensitivity of neutrino telescopes to neutrino point sources from the studied UFO sample. Our most promising neutrino point sources are NGC\,1068, and NGC\,6240. Both of these are obscured AGN and the UFO observations should therefore be considered as uncertain (see Sec 2.2 of~\citealp{Yamada:2024lss}). These sources are located at declination $\delta \approx 0^{\circ}, 2^{\circ}$, respectively, and are at an ideal location in the sky for IceCube, and IceCube-Gen2. We find that if these UFOs are real, the sources would be detectable with IceCube-Gen2 with peak energy at $\sim 10^{16}\,\si{\electronvolt}$, see \cref{fig:neutrino_point_sources}. Such a neutrino signal should be expected in addition to the lower-energy neutrinos from the core of NGC\,1068~\citep{IceCube:2022der} as, generally, the launching radius of the UFO is thought to be outside the corona of the AGN. 

\section{AGN photon fields}\label{apx:photon_fields}
    We simulate the AGN photon spectrum as a combination of black-body (dust torus), multi-colour back-body (disc) and (broken-) power law (corona) emission \citep[see][]{Ghisellini:2009wa,Marconi:2004MNRAS}. The individual luminosities are then re-normalised relative to the bolometric luminosity of the AGN using observational luminosity scaling factors to connect the different wavelength bands~\citep{Duras:2020A&A,Mullaney:2011iq}. The bolometric luminosity here refers to the total intrinsic luminosity, whereas the integral of the entire spectral energy distribution corresponds to the total observed luminosity of the AGN.

\begin{figure}
        \centering
        \includegraphics[width=\linewidth]{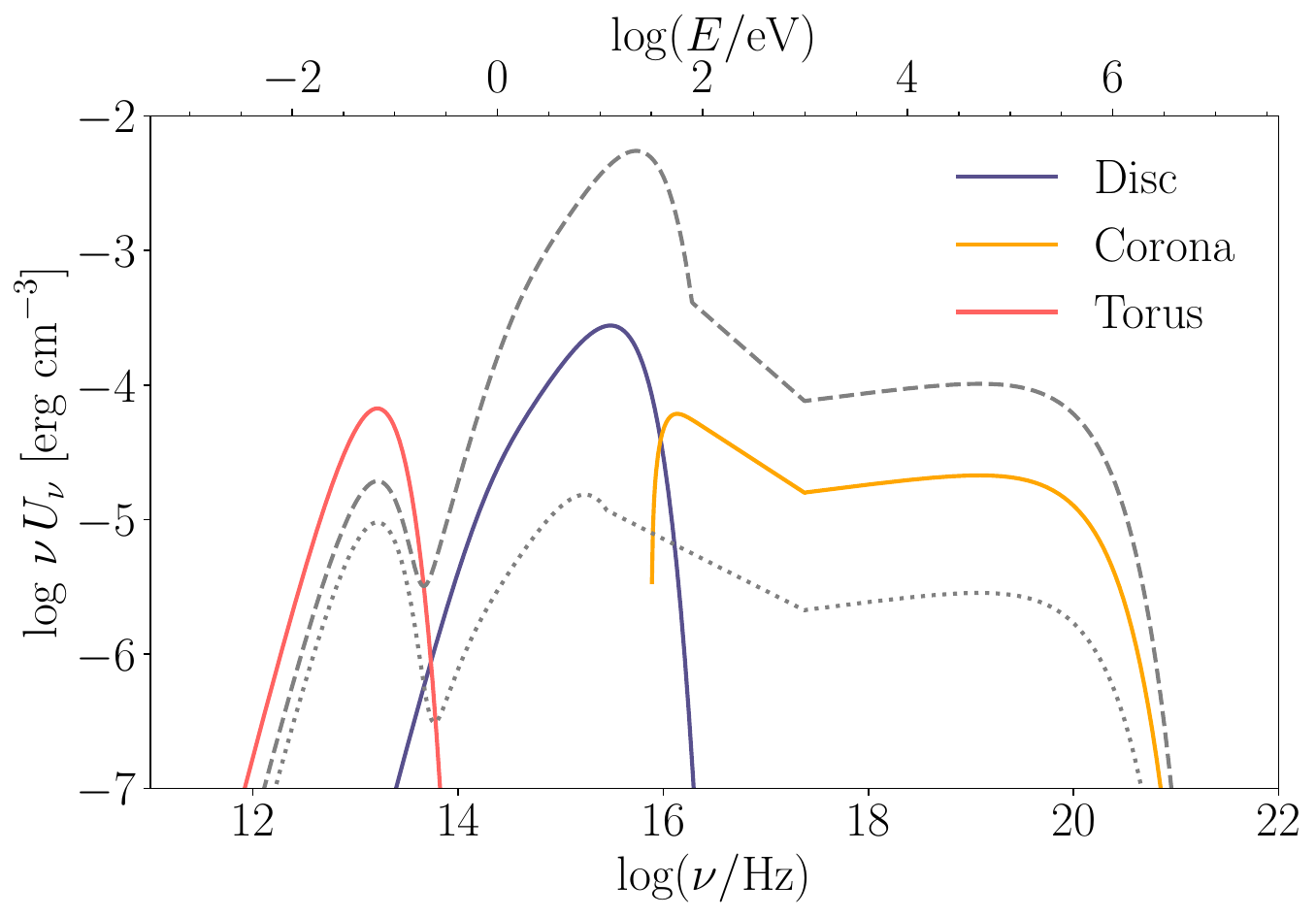}
        \caption{Photon field energy density at distance $R_\text{sh}$ for the benchmark scenario ($\lbol=10^{45}\,\si{\erg\per\second}$), and for bolometric luminosities of $10^{44}\,\si{\erg\per\second}$ (dotted) and $10^{46}\,\si{\erg\per\second}$ (dashed). In the latter case, the photon density of the IR torus field at the location of the wind termination shock is lower despite the larger luminosity due to the larger size of the torus. which results in a stronger ``dilution'' of the field.}\label{fig:photon_density_benchmark}
    \end{figure}

    For all photon fields, we assume a $1/R^2$ decrease of the energy density at radii larger than the outer radius of the emitting structure (disc, corona, torus), and a constant density inside. We consider the emission to be isotropic in all directions, i.e.\ not depending on the viewing angle of the AGN. In \cref{apx:dust_distribution}, we discuss the effect of considering a different geometry. The energy density in the benchmark scenario at the wind termination shock is shown in \cref{fig:photon_density_benchmark}.
    
    \subsection{Accretion disc}    
    In the standard picture, the black hole is surrounded by an accretion disc wherein matter moves inward while the potential energy is radiated away and the angular momentum is transported outward~\citep{Shakura:1972te}. The accretion disc exhibits a radial temperature gradient~\citep{Dermer:2009zz,Ghisellini:1996fm},
    \begin{equation}\label{eq:T_disk}
        T^4 = \frac{3 R_\text{S} L\disc}{16\pi\eta\sigma_\text{SB}R^3} \left[1 - \left(\frac{3 R_\text{S}}{R}\right)^{1/2}\right]\,.
    \end{equation}
    Here, $R_\text{S}$ is the Schwarzschild radius of the black hole, $R$ the distance from the AGN, $L\disc$ the luminosity of the disc\footnote{We encounter a circular dependency since a derivation of the temperature requires knowledge of the luminosity of the disc which, in turn, requires prior knowledge of the temperature. We overcome this problem by assuming, for this particular calculation only, that $L\disc=0.5\,\lbol$.}, $\eta$ the accretion efficiency, and $\sigma_\text{SB}$ the Stefan-Boltzmann constant.

    The total radiation field produced by the disc can be understood as a superposition of black-body spectra produced by discrete disc annuli, each with a width $\dif R$ and the temperature given by \cref{eq:T_disk}. The total monochromatic disc luminosity [$\si{\erg\per\second\per\hertz}$] is then
    \begin{equation}
        L\disc(\nu) = \int_{R_\text{in}}^{R_\text{out}}\dif R~L_\text{BB}(\nu,R)\,,
    \end{equation}
    where $R_\text{in}=3 R_\text{S}$~\citep{Dermer:2009zz} ($R_\text{out}=500\,R_\text{S}$\footnote{The location of the outer edge of the disc is an ongoing matter of discussion. However, the luminosity contribution of large radii, $R>\mathcal{O}(100\,R_S)$, is negligible due to the low temperature of the disc at these distances.}) is the inner (outer) radius of the accretion disc, and the monochromatic luminosity of each annulus is 
    \begin{align}
        L_\text{BB}(\nu,R) &= \dif A\dif\Omega\, \left[B_\nu(T(R))\cos\theta\right] \\
                            &\text{with } B_\nu = \frac{2h\nu^3}{c^2} \left[e^{-\frac{h\nu}{k_\text{B}T(R)}}\right]^{-1}. \nonumber
    \end{align}
    Including both hemispheres of the AGN system, the geometry of the disc is given by 
    \begin{align}
        &\dif A = 2 \left(2\pi\dif R\, R - \pi \dif R^2\right)\,,~\text{and}\\
        &\dif\Omega\cos\theta = \int_0^{2\pi}\dif\phi \int_0^{\pi/2}\dif\theta\cos\theta\sin\theta = \pi\,.
    \end{align}
    We normalise the luminosity of the accretion disc using the scaling factor of \citet{Duras:2020A&A} to convert the bolometric luminosity of the AGN to the monochromatic luminosity at the nominal frequency of the B band, i.e. $\nu_B L_{\nu_B} / \lbol = 5.13\pm0.10$.

    \subsection{Corona}
    Active galactic nuclei commonly feature a non-thermal spectrum of X-ray photons with energies up to $\mathcal{O}(\SI{100}{\kilo\electronvolt})$. The precise origin of this emission is poorly understood but is typically associated with a hot ``corona'' located above the inner accretion disc~\citep[e.g.][]{Haardt:1991tp,Haardt:1993nj,Merloni:1999pe}. The corona reprocesses a fraction of the photons emitted by the accretion disc to higher energies via inverse Compton scattering with a population of hot, high-energy electrons. From this process, a power-law distribution of photon energies is expected. In addition, for a significant number of AGN a soft X-ray excess below $\SI{2}{\kilo\electronvolt}$ has been observed~\citep{Gierlinski:2003ti,Waddell:2023rek}, the origin of which is uncertain but could be associated with a secondary warm corona, reflection of corona photons by the disc~\citep{Ross:1999rf,Ross:2005dm,Done:2011at,Petrucci:2017niz,Petrucci:2020cda,Ballantyne:2024lyy}, or absorption features~\citep[e.g.][]{Gierlinski:2003ti}.

    We use a model-independent approach and describe the hard X-ray (hot) corona emission as a power law distribution with exponential cutoff at high energies
    \begin{equation}\label{eq:L_corona}
        L_\text{corona}(\nu)\propto
        \begin{cases}
            \left(\frac{E}{E_\text{break}}\right)^{-b} &E_\text{break}\leq E \leq E_\text{crit}, \\
            \left(\frac{E}{E_\text{break}}\right)^{-b} e^{-E/E_\text{crit}} &E > E_\text{crit},
        \end{cases}
    \end{equation}
    with $b=0.9,\,E_\text{break}=\SI{1}{\kilo\electronvolt}$ and $E_\text{crit}=\SI{500}{\kilo\electronvolt}$~\citep{Marconi:2004MNRAS}. For a given AGN luminosity we normalise the coronal spectrum such that the integral in the 2-10~keV band equals the X-ray luminosity, $\lhx$, of the AGN. Subsequently, we model the soft X-ray regime by connecting the normalised disc and hot-corona spectra between $\SI{25}{\electronvolt}$ and $\SI{1}{\kilo\electronvolt}$ with a power law. This provides an approximate description of the soft X-ray excess.

    \subsection{Dust torus}
    Surrounding the inner disc/corona structure is a region containing warm dust that is heated by the emission of the accretion disc, with a temperature inversely correlated with the distance from the central object. The distribution may be clumpy and extend close to the black hole~\citep{Nenkova:2008uk,Nenkova:2008um,Schartmann:2008qb,Mullaney:2011iq}; however, the inner edge is bounded by the dust sublimation temperature $\sim\SI{1500}{\kelvin}$~\citep{Barvainis:1987ApJ}. The distribution of this dust is not well understood; however, a torus-like geometry is typically assumed in support of models for AGN unification~\citep{Antonucci:1993sg}. The radial extend of this torus is generally unconstrained except for a few sources, but it can be large, resulting in a non-trivial profile of the temperature and surface brightness as a function of the distance from the central source~\citep{Burtscher:2013aza,Kishimoto:2011hz}. However, the common picture of an extended dust torus has recently been challenged by high-resolution imaging of NGC\,1068 by GRAVITY/VLTI where a much thinner ring of hot dust was identified~\citep{GRAVITY:2020A&A}.
    
    The characteristic distance of the inner boundary of the torus is given by the sublimation radius~\citep{Ghisellini:2009wa,Nenkova:2008um,Barvainis:1987ApJ}
    \begin{equation}\label{eq:R_IR}
        R\ir = 2.5\times10^{18}\left(\frac{L\disc}{10^{45}\,\si{\erg\per\second}}\right)^{1/2}\si{\centi\meter}.
    \end{equation}
    For typical luminosities, this corresponds to a radius of $0.1-10\,\si{\parsec}$. The above relation between disc luminosity and torus size represents an approximation, and the true inner radius can deviate significantly from this simple prescription due to a clumpy structure or other environmental factors~\citep{Burtscher:2013aza}.

    We model the emission of the dust torus as a black-body spectrum with constant temperature. An upper limit on the temperature of the dust is provided by the dust sublimation temperature of $1000-2000\,\si{\kelvin}$ depending on the dust species~\citep{Baskin:2018MNRAS}; however, observations indicate that the true temperature can be significantly lower, down to a few hundred Kelvin or less \citep[e.g.][]{Lopez-Rodriguez:2018ApJ,Kishimoto:2011hz,Burtscher:2013aza,Rosas:2021zbx}. We choose $T\ir=\SI{200}{\kelvin}$ in our analysis as this corresponds to the temperature where a pure black-body spectrum is in closest agreement with the parametric template of \citet{Mullaney:2011iq}. The IR spectrum is normalised relative to the hard X-ray luminosity $\lhx$ by using the appropriate luminosity scaling factor~\citep{Mullaney:2011iq}.
    
    To avoid an angular dependence of the infrared photon field, we assume, by default, that all the dust is concentrated in a thin shell at the radius $R\ir$ \citep[as in][]{Ghisellini:2009wa}. In addition, we assume that the torus is uniformly illuminated by the accretion disc, i.e.\ there is no angular dependence of the photon field due to the disc not being spherically symmetric. If instead a torodial shape is assumed, cosmic rays with larger polar angle will experience a weaker photon field relative to those closer to the planar torus. A dust distribution with a flat ring-like shape is studied in \cref{apx:dust_distribution}.

\section{Geometry of the dust distribution}\label{apx:dust_distribution}
    \begin{figure*}
        \centering
        \includegraphics[width=\linewidth]{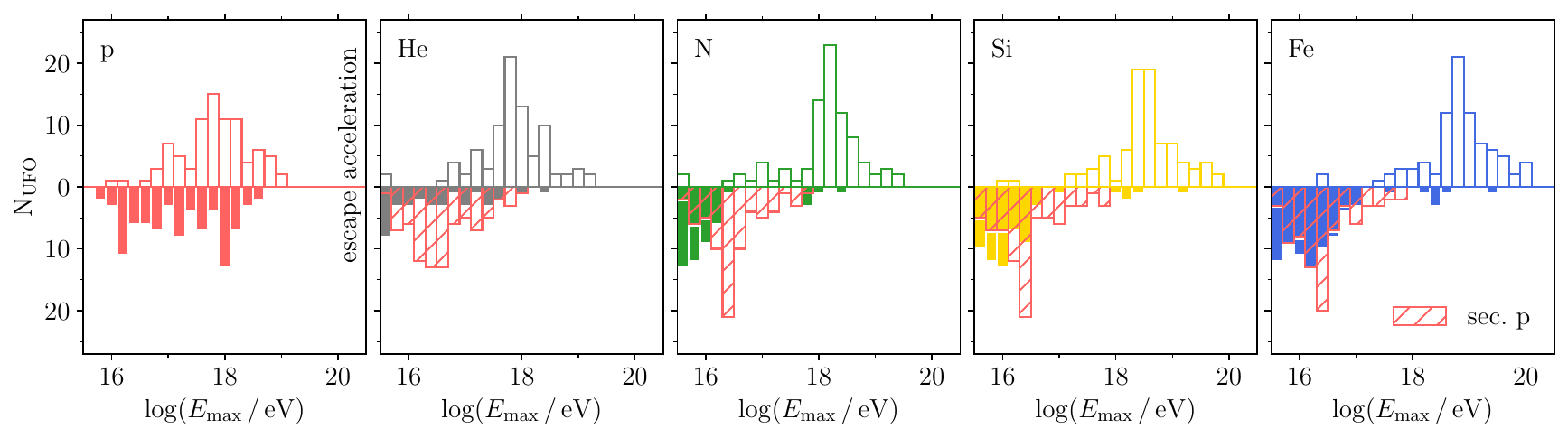}
        \caption{Same as \cref{fig:results_sample_Emax_distribution} but with the infrared-emitting dust distributed in a thin ring co-planar to the accretion disc.}\label{fig:results_sample_Emax_distribution_dustRing}
    \end{figure*}    
    
    By default, we have assumed a spherical dust distribution, with all material concentrated in an infinitesimally thin shell at radius $R\ir$, and heated to a constant temperature of $T\ir=\SI{200}{\kelvin}$; see \cref{sec:photon_fields}). The exact shape of the dust ``torus'' is not well understood. Recent observations indicate a complex, multi-component dust geometry (see e.g.\ \citealp{Honig:2019ApJ}). A thin (scale height $h/r < 0.14$), disc-like dust distribution is favoured in the case of NGC\,1068 by recent high-resolution GRAVITY observations~\citep{GRAVITY:2020A&A}. 
    
    Here we consider, as the opposite limit to our default spherical-shell scenario, the case where the IR emission is produced by dust located in a thin ring co-planar to the accretion disc, i.e.\ $h/r\to0$. We assume that the luminosity of the dusty ring is described by the same scaling factor relative to the bolometric luminosity as before~\citep{Mullaney:2011iq}.

    For this dust configuration, the position of any point along the dust ring is specified by the distance from the AGN, $R\ir$, and the azimuthal angle $\phi_2$ in the $x-y$ plane. The distance of an arbitrary point $p$, expressed in terms of the distance $R$ from the AGN, the azimuthal angle $\phi$, and the polar angle $\theta$ relative to the $z$-axis, is then given by
    \begin{equation}
       d^2 = R^2 + R\ir^2 - 2\, R R\ir \sin\theta \cos(\phi-\phi_2)\,, 
    \end{equation}
   which reduces to \ $d^2 = R^2 + R\ir^2$ for $\theta = 0$ (on the $z$ axis). Assuming isotropic emission, the contribution to the photon energy density at any point $p$ by a particular ring element at distance $d(p,R\ir,\phi_2)$ is $U\ir = L\ir/(4\pi\,d^2\,c)$. Noting the symmetry of the problem, we can redefine $\phi-\phi_2 = \mu$, and integrate the contribution of all ring elements, as \begin{align}
        U\ir    &= \int_0^{-\pi}\dif\mu\left(\frac{L\ir}{4\pi d^2(R\ir,\mu)\,c}\right) \\
                &= - 2\tan^{-1}\left[\tan\left(\frac{\mu}{2}\right)\frac{a}{b}\right]\frac{1}{b}~\Bigg|_{\mu=0}^{-\pi}\times\frac{L\ir}{4\pi\,c}\\
                \text{with }    &a = R^2 + 2RR\ir\sin\theta + R\ir^2 \nonumber \\
                \text{and }     &b = \sqrt{R^4 + 2R^2R\ir^2\cos(2\theta) + R\ir^4}\,. \nonumber
    \end{align}

    For small radii ($R\ll R\ir$), the photon density in the ring model matches the photon field of a homogeneously emitting shell. For large radii ($R\gg R\ir$) both scenarios converge to the $1/R^2$ behaviour expected for point sources. Only when both radii are comparable ($R\sim R\ir$) can the geometry of the dust distribution have an appreciable impact on the expected photon density. However, when the radius of the dust ``torus'' is significantly smaller than the radius of the wind termination shock ($\rsh\gtrsim2R\ir$: $\sim20\%$ of UFOs), or when the forward shock is much closer to the AGN than the dust ``torus'' ($\rfs\lesssim2R\ir$: $\sim6\%$ of UFOs) the expected difference between a shell-like and ring-like dust distribution is expected to be negligible.

    Under the ring model, the photon density is reduced at small polar angles and increased at large polar angles near the accretion disc plane compared to the shell model. For polar angles less than $45\,\si{\deg}$, the photon density predicted by the ring model is smaller compared to the shell model at all radii (by up to $50\%$ at $R\ir$ for $\theta=\SI{0}{\deg}$), while for angles above $60\,\si{\deg}$ densities are higher (up to a factor of 2 at $R\ir$ for $\theta=\SI{75}{\deg}$) in the ring model. At $\theta=\SI{90}{\deg}$, the photon density diverges as $R$ approaches the radius of the dust ring, a result of our assumption of infinitesimal thickness. In more realistic scenarios, where $\dif R\ir>0$, the density would reach a constant value once the radius becomes comparable to $R\ir$.

    As discussed in \cref{sec:escape_ufo}, the $\mathcal{O}(0.1\,\si{\gauss})$ magnetic field in the shocked wind causes quasi-diffusive motion, leading to long confinement times. Consequently, cosmic rays sample the downstream region over various polar angles, regardless of their initial injection angle (see \cref{fig:cr_tracks}). Even at higher energies, where the escape is quasi-ballistic, escape trajectories are in general not radial, and cosmic rays still sample a range of polar angles. Furthermore, since cosmic-ray nuclei have a long path length, a moderate decrease in the IR field during their propagation does not alter the escaping flux significantly. During the acceleration stage, the cosmic rays also exhibit some degree of motion. However, because of their lower energies, they are more confined and sample a smaller region compared to the subsequent escape stage. The strength of the IR field at acceleration (at $\rsh$) can therefore, in principle, differ from the average value by a factor of a few depending on the polar angle. Assuming an isotropic wind with an opening angle of $\SI{90}{\deg}$, approx.\ $13\%$ of cosmic rays are accelerated at polar angles less than $\SI{30}{\deg}$, and therefore encounter a weaker IR photon field.

    Here we investigate the maximum energy of cosmic rays injected at $\theta=\SI{0}{\deg}$ when the dust is distributed according to the ring model. The angular dependence of the dust ring photon field is treated accurately in \textsc{CRPropa} during the escape phase. The maximum acceleration energy is always at least as large as for the shell model due to the assumption of $\theta=0$. For nuclei, the maximum energy at acceleration is increased by, on average, approximately $(5\pm5)\%$ compared to the shell model (median: $2-3\%$); however, an increase of up to $\sim30\%$ is possible for individual UFOs. The largest increase is found for IRAS\,11119+3257 for all cosmic-ray species. For primary protons, the maximum energy remains largely unchanged (mean: $+1\%$, median: $+0\%$, max: $+10\%$). The maximum energy after escape is, on average, larger by a few percent for all nuclear species ($\sim20\%$ for protons and helium); however, it remains invariant within the uncertainty introduced by binning the escaping spectrum for between $30\%$ (helium) and $80\%$ (iron) of investigated UFOs. For a small subset of outflows ($\sim5\%$ for nuclei and $\sim1/3$ for protons) the maximum energy after escape is lower in the ring model. The distribution of maximum energies is shown in \cref{fig:results_sample_Emax_distribution_dustRing}. Our conclusions about the viability of ultra-fast outflows as the sources of ultra-high-energy protons and nuclei, derived for the fiducial dust-shell model, remain unchanged qualitatively.

\section{Analytical Escape Spectra}\label{apx:escape_analytical}
   \begin{figure}
        \centering
        \includegraphics[width=\linewidth]{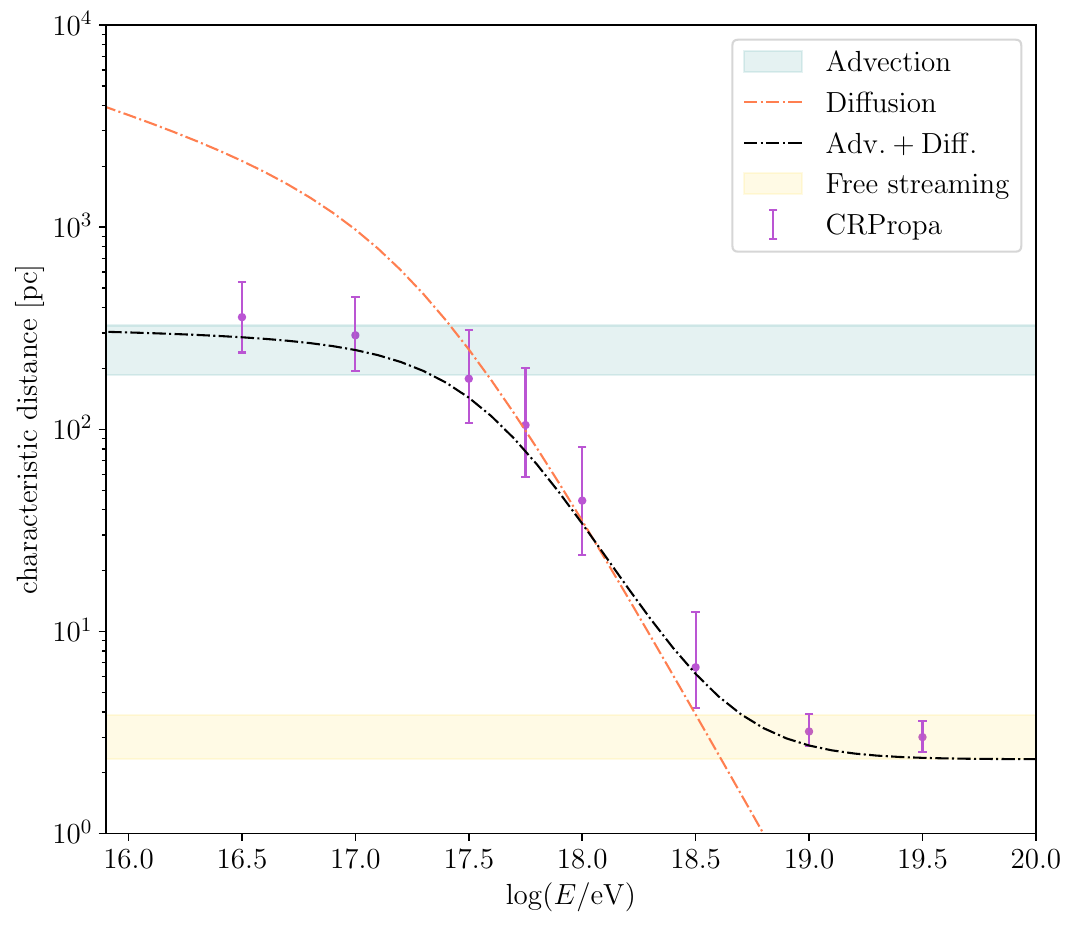}
        \caption{Comparison of the analytical predictions for advection, diffusion and free-streaming distance in the UFO environment and the typical distance particles propagate in the 3D simulation before escaping from the system (without interactions). The free-streaming escape distance is between $\rfs+\rsh$ and $\rfs-\rsh$ depending on the angle at which the particle was emitted at the wind termination shock. The analytical advection distance depends on how the radius-dependent velocity of the shocked wind is averaged.}\label{fig:characteristic_timescales}
    \end{figure}

    The suppression of the escaping cosmic-ray flux at $\rfs$ compared to the accelerated spectrum at $\rsh$ can be approximated with a simple semi-analytical approach by comparing the timescales relevant for escape and interactions, analogous to \cref{sec:interactions}. We assume, for simplicity, that the photon fields are constant everywhere, with the density fixed to the value calculated at the radius of the wind termination shock. As cosmic rays spend the majority of their time in the downstream, the predicted interaction rates represent a conservative upper limit. A more accurate result could be achieved by taking effective photon fields obtained from averaging the densities over the entire UFO. However, since we use the semi-analytical approach only to verify our numerical results, we adopt the constant photon fields.

    The (effective) photon fields can be used to derive the average energy-dependent energy loss length/time of the cosmic rays in the shocked wind due to photopion production, photodisintegration, and Bethe-Heitler pair production, and thus the total energy loss time from interactions as
    \begin{equation}
        \tau_\text{IA} = \left(\frac{1}{\tau_{A\gamma}} + \frac{1}{\tau_\text{dis}} + \frac{1}{\tau_\text{BH}}\right)^{-1}.
    \end{equation}
    The escape timescale is unchanged from \cref{eq:t_esc} since we assume a constant magnetic field in the downstream and the diffusion time is therefore independent of the distance from the centre, and $\tau_\text{adv}$ and $\tau_\text{free}$ are also constant. The characteristic escape timescale derived with this semi-analytical approach is in good agreement with the characteristic timescales that are obtained from the 3D simulations with \textsc{CRPropa} as shown in \cref{fig:characteristic_timescales}.

    Our model is different from the typical ``leaky-box'' approach, as e.g.\ used in the UFA model~\citep{Unger:2015laa}, where the cosmic rays are uniformly distributed and there is a nonzero probability of escape at all times. In our model, cosmic rays are injected at $\rsh$ and escape is only possible after they have propagated through the spatially dependent photon fields, advection flow, and magnetic field to the forward shock at $\rfs$. Furthermore, since we assume the same diffusion coefficient for all cosmic rays with a given energy, they all escape the system at the same time given by the diffusion timescale; see \cref{eq:t_diff}. The number of cosmic rays in the system changes as a function of time as
    \begin{equation}
        \frac{\dif N(E)}{\dif t} =
        \begin{cases}
            -\frac{N(E)}{t_\text{IA}(E)}\,,~t\leq t_\text{esc}(E) \\
            0\quad\quad\quad,\, t > t_\text{esc}(E)\,.
        \end{cases}
    \end{equation}
    The integration of this expression gives the number of cosmic rays at any point in time as
    \begin{equation}
        N(E,t) = N_0(E)\,\exp{\left(-\frac{t}{\tau_\text{IA}(E)}\right)}\,\mathcal{H}[\tau_\text{esc}(E)-t]\,,
    \end{equation}
    where $\mathcal{H}(\dots)$ is the Heaviside function. This does not include the potential production of secondary cosmic rays from the spallation of heavier primaries, i.e.\ inter-species migration. The number of cosmic rays that have successfully escaped the system after $t\to\infty$ is
    \begin{equation}
         N_\text{esc}(E) = N(E,t_\text{esc}) = N_0(E)\,\exp{\left(-\frac{\tau_\text{esc}(E)}{\tau_\text{IA}(E)}\right)}\,.
    \end{equation}
    The flux suppression in our model is generally stronger than in the leaky-box approach since the cosmic rays must always cross the entire downstream before escaping, and there is no probability of escape after only a fraction of that distance/time.

\section{Benchmark UFO results for all species}\label{apx:results_benchmark}
\begin{figure}
        \centering
        \includegraphics[width=0.87\linewidth]{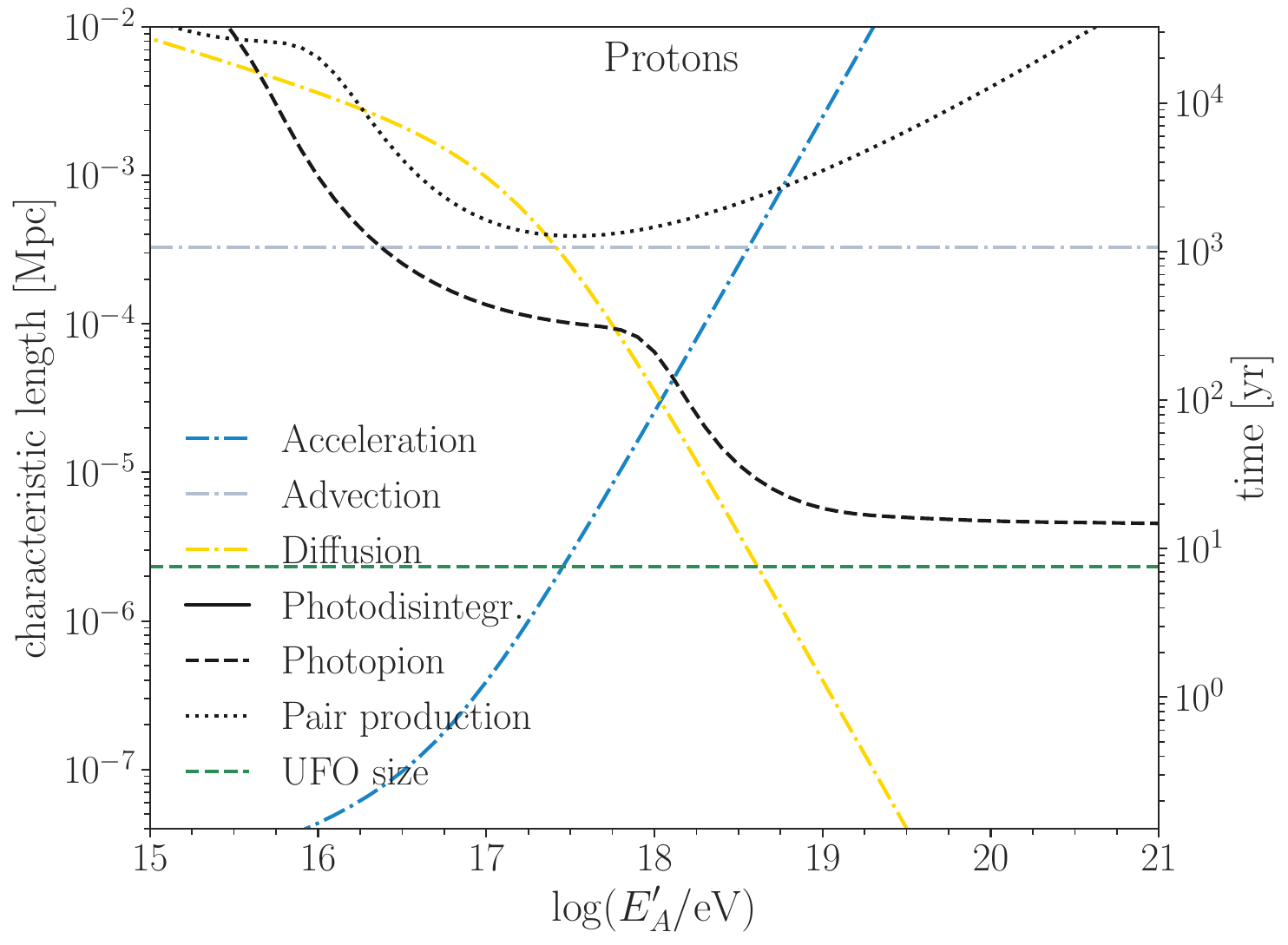}
        \caption{Same as \cref{fig:results_benchmark_nitrogen} but for protons.}\label{fig:CR_mfp_benchmark_protons}
    \end{figure}

\begin{figure}
        \centering
        \includegraphics[width=0.87\linewidth]{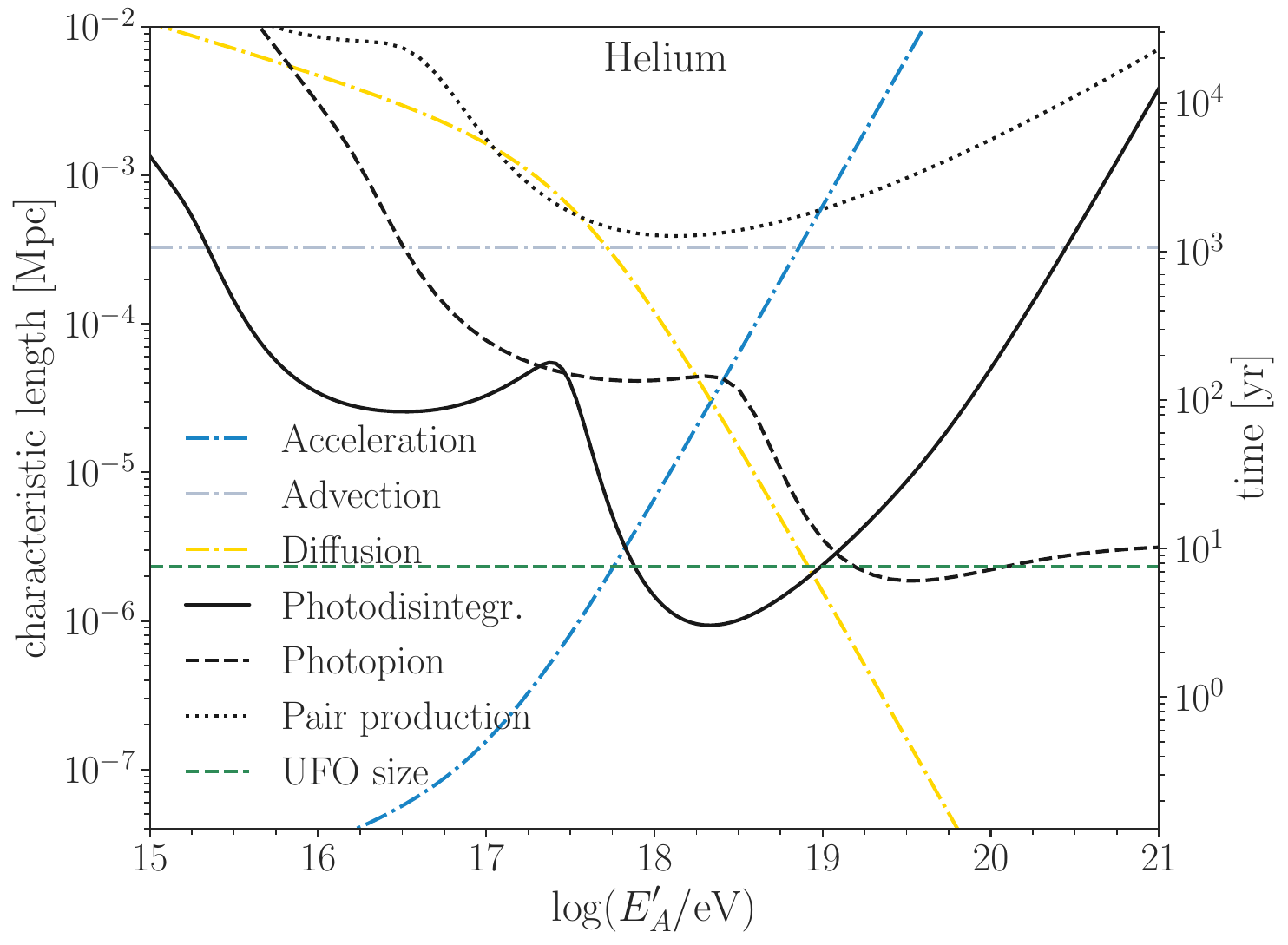}
        \caption{Same as \cref{fig:results_benchmark_nitrogen} but for helium.}\label{fig:CR_mfp_benchmark_helium}
    \end{figure}

\begin{figure}
        \centering
        \includegraphics[width=0.87\linewidth]{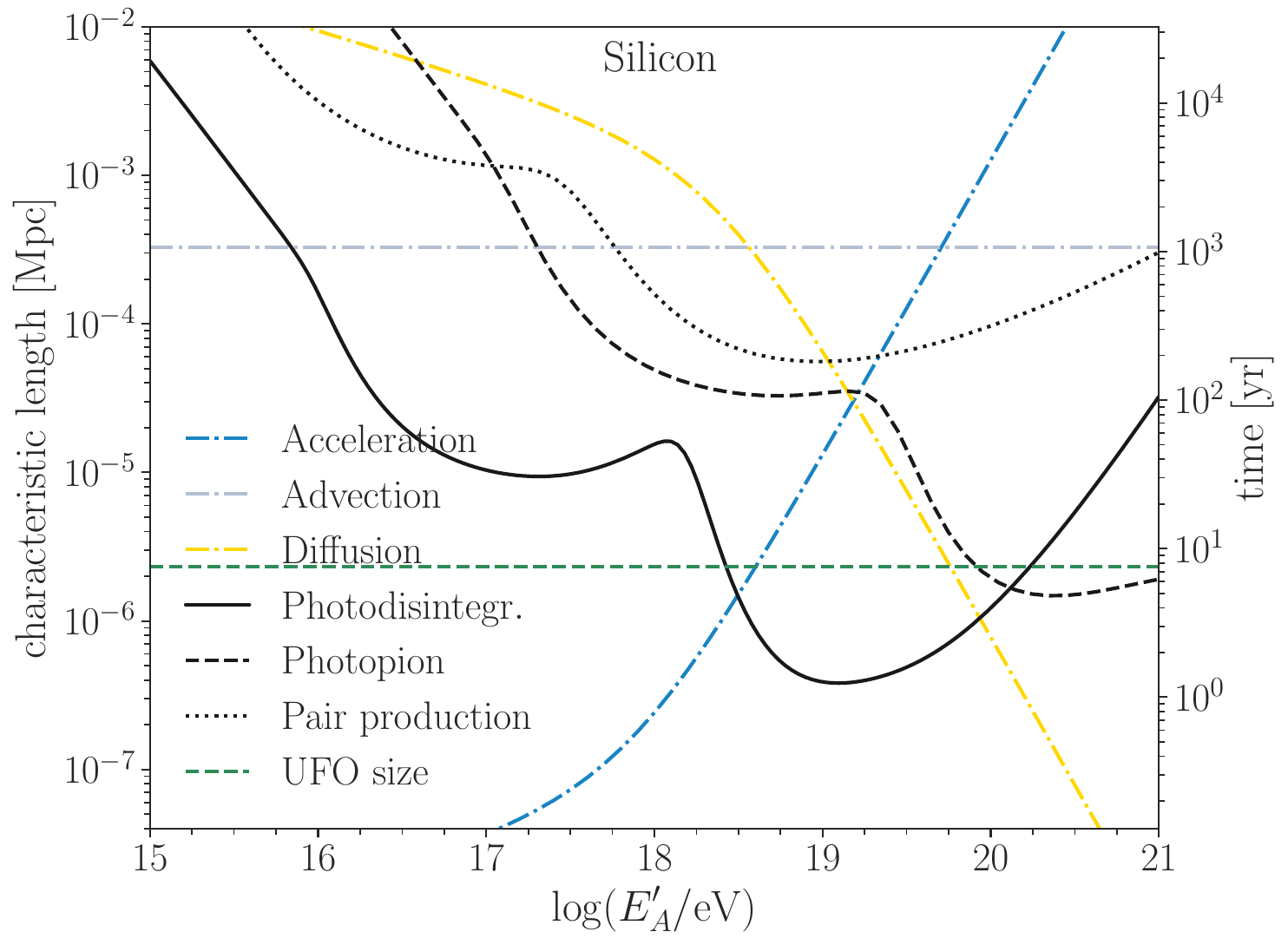}
        \caption{Same as \cref{fig:results_benchmark_nitrogen} but for silicon.}\label{fig:CR_mfp_benchmark_silicon}
    \end{figure}

\begin{figure}
        \centering
        \includegraphics[width=0.87\linewidth]{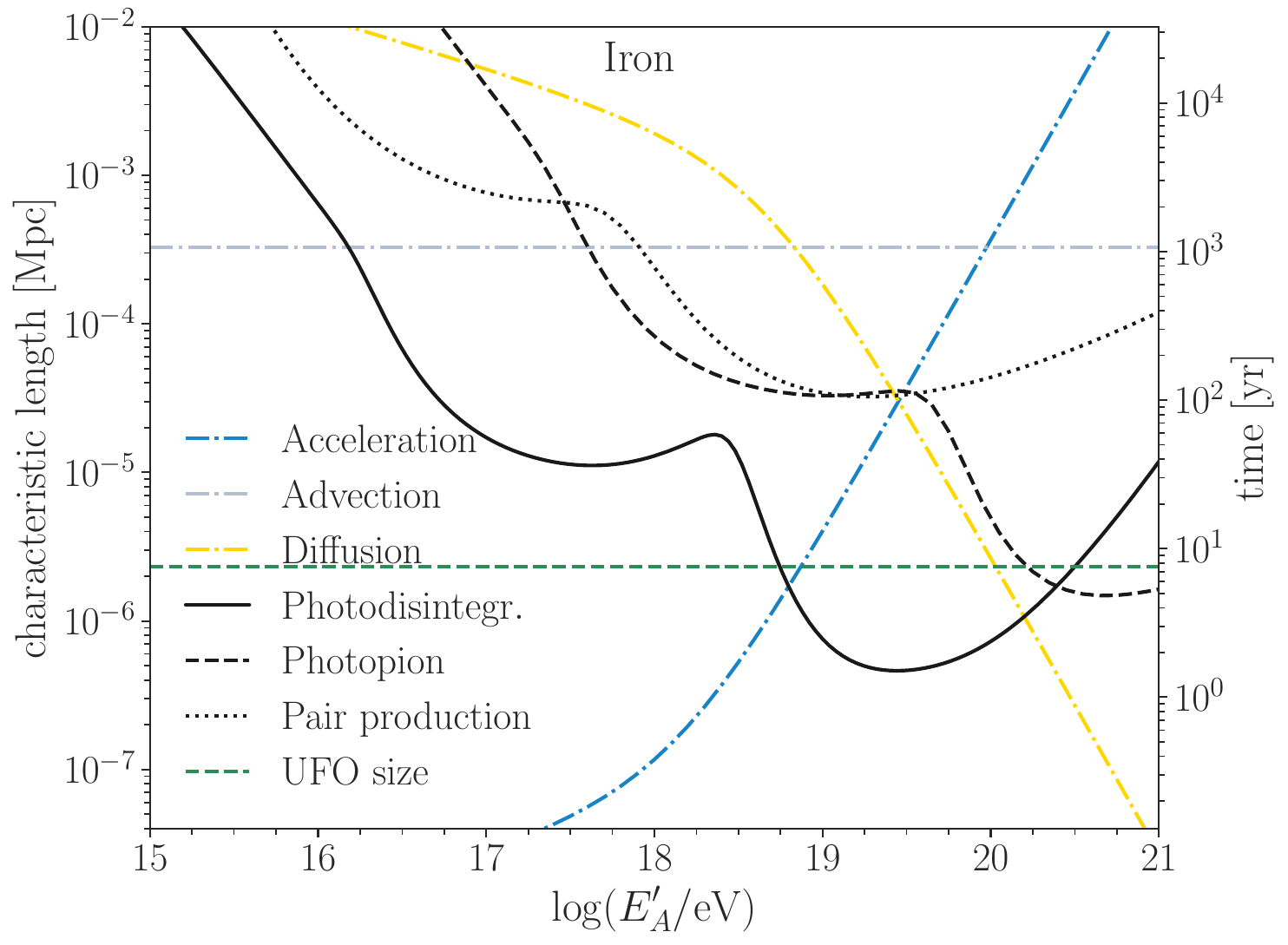}
        \caption{Same as \cref{fig:results_benchmark_nitrogen} but for iron.}\label{fig:CR_mfp_benchmark_iron}
    \end{figure}

    \cref{fig:CR_mfp_benchmark_protons,fig:CR_mfp_benchmark_helium,fig:CR_mfp_benchmark_silicon,fig:CR_mfp_benchmark_iron} show the characteristic distances in the environment of the benchmark UFO for protons, helium, silicon and iron, respectively. The maximum energy of nuclei is limited by photodisintegration due to the infrared field of the dust torus whereas protons are limited by confinement to the shock.
    
\section{Mass outflow rate}\label{apx:outflow_rate}
    For a wind that is launched from the disc at an angle $\phi$ with respect to the equatorial plane in an annular region with distance from the central object between $R$ and $R+\Delta R$, the mass outflow rate can be expressed as~\citep{Krongold:2007ApJ}
    \begin{equation}
        \dot{M}_\text{w} = \rho\,\left[\frac{\vw}{\cos{(\phi-\delta)}}\right]\,A(R,\Delta R)\,,
    \end{equation}
    where $\rho=n_Hm_p\mu$ is the mass density of the wind (with $\mu=n_H/n_e=1.2$~\citep{Nardini:2015Sci,Gianolli:2024jkq}), $\vw$ is the observed outflow velocity along the line of sight, $\delta$ is the viewing angle relative to the plane of the disc, and $A=\pi\left[(R+\Delta R)^2 - R^2\right]\cos^2\delta\sin\phi$ is the area of the disc annulus where the wind is launched. The thickness of the wind is related to the observed column density $N_H$ as $\Delta R = \mu N_H / n_e$. Assuming $\Delta R / R \ll1$, the expression can be simplified to~\citep{Krongold:2007ApJ}
    \begin{equation}
        \dot{M}_\text{w} \approx \pi\mu m_p \vw N_H R f(\delta,\phi)
    \end{equation}
    where $f(\delta,\phi)$ collects all angular dependencies; with $f=1.5$ for $\phi=\pi/2$ and $\delta=30^\circ$ -- as assumed throughout this paper. The final expression is similar to the one proposed by \citet{Nardini:2015Sci}, which replaces the angular factor with the solid angle covered by the wind. We apply the relativistic correction factor $\Psi=(1+\beta)/(1-\beta)$ to the column density~\citep{Luminari:2019rho}.

    Due to a lack of spatial resolution, the launching radius of the outflow is not observed directly. A lower limit is given by the radius of gravitational escape as
    \begin{equation}\label{eq:rLaunch_min}
        R_\text{min} = \left(\frac{c}{\vw}\right)^2 R_\text{S} = \frac{2GM_\text{BH}}{\vw^2}\,.
    \end{equation}
    A wind launched at a smaller radius has insufficient kinetic energy to overcome the gravitational potential of the central compact object, leading to an eventual fall-back and a failed wind.

    On the other hand, for a given hydrogen number density $n_H$, the width of the wind required to produce the observed column density cannot be greater than the distance from the source, i.e.\  $N_H = n_H\Delta R < n_HR$. Together with observations of the ionisation parameter, $\xi=L_\text{ion}/n_H\,R^2$, this allows to place an upper limit on the launching radius~\citep{Gofford:2015aka}
    \begin{equation}\label{eq:rLaunch_max}
        R_\text{max} = \frac{L_\text{ion}}{\xi N_H}\,.
    \end{equation}

    The difference between $R_\text{min}$ and $R_\text{max}$ can reach several orders of magnitude and, due to the approximately linear correlation of $\dot{M}_\text{w}$ with the launching radius, can lead to large uncertainties of the wind kinematics. Large outflow rates of $\dot{M}_\text{w}/\dot{M}_\text{Edd}\gg1$ are obtained for most UFOs in our sample when using $R_\text{max}$. We therefore use the minimal launching radius $R_\text{min}$ to derive conservative estimates of the mass outflow rate of all UFOs in our study.

\section{Limiting the mass outflow rate}\label{apx:limit_outflow_rate}
    \begin{figure*}
        \centering
        \includegraphics[width=0.48\linewidth]{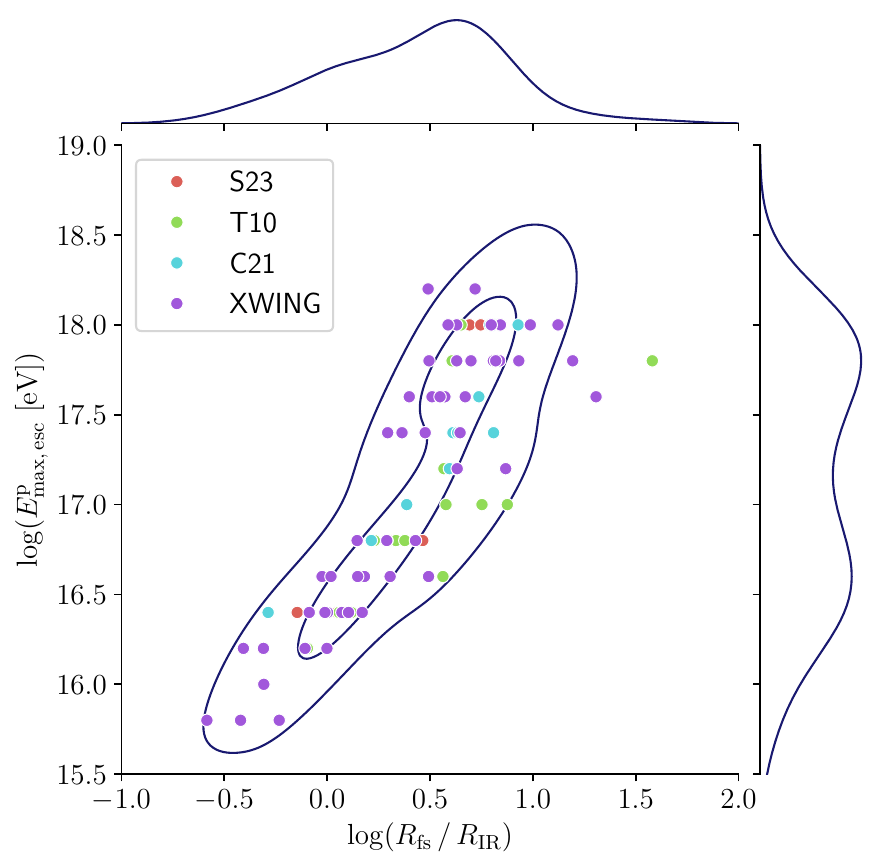}
        \includegraphics[width=0.48\linewidth]{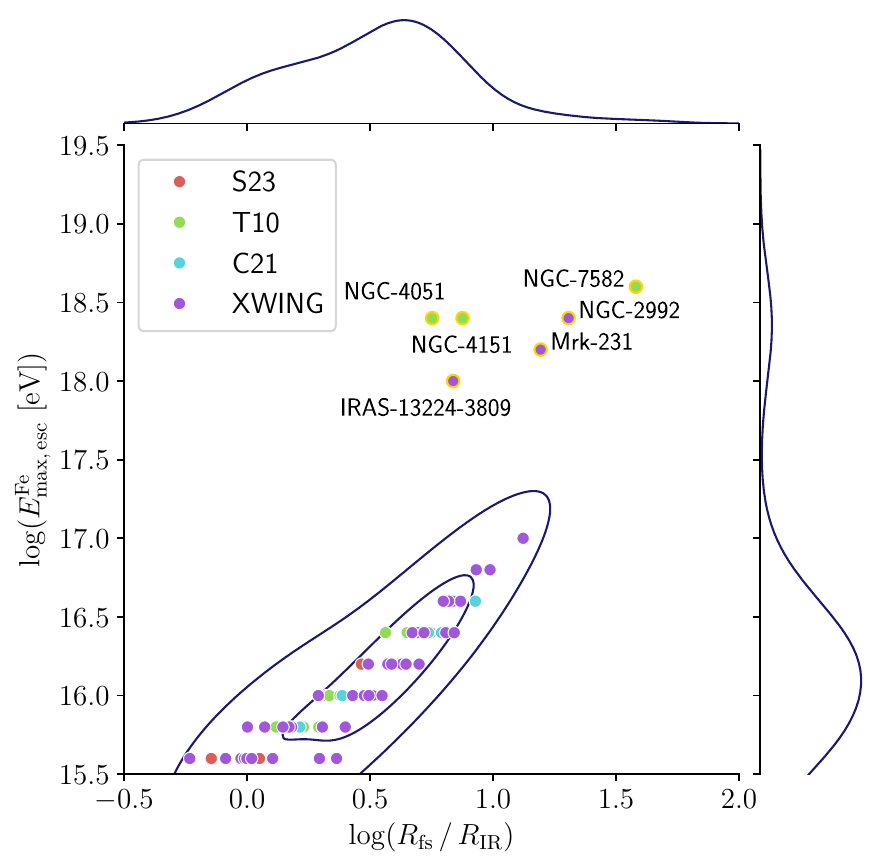}
        \caption{Same as \cref{fig:Emax_radius_correlation} but with the constraint that $L_\text{wind}/\lbol\leq1$. For primary protons (left) and iron nuclei (right).}\label{fig:Emax_radius_correlation_limLkin}
    \end{figure*}
    
    \begin{figure*}
        \centering
        \includegraphics[width=0.48\linewidth]{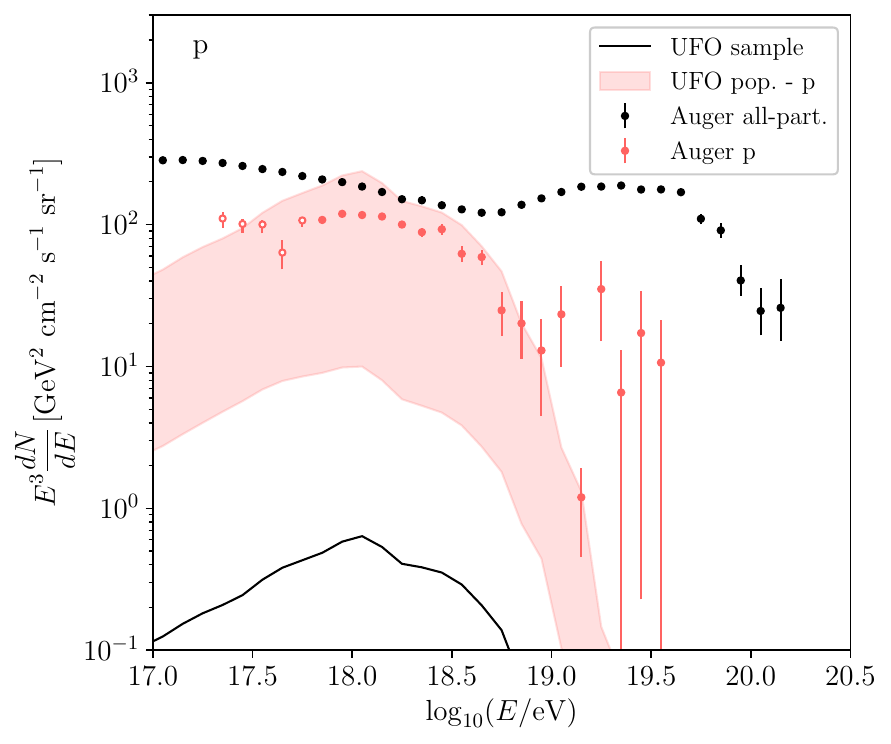}
        \includegraphics[width=0.48\linewidth]{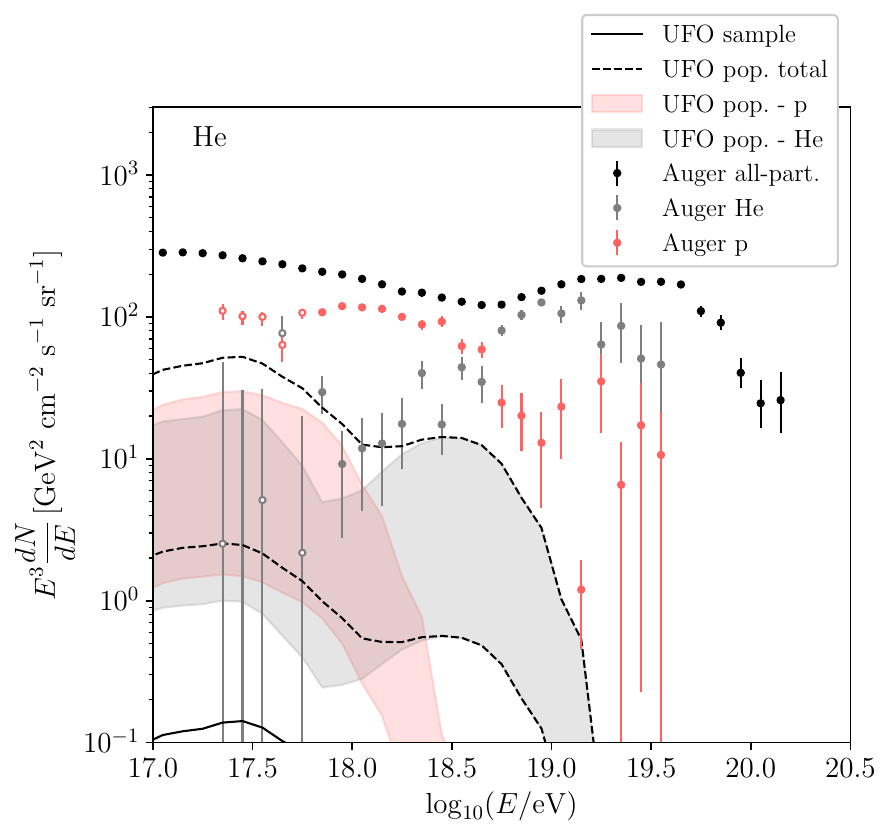}
        \includegraphics[width=0.48\linewidth]{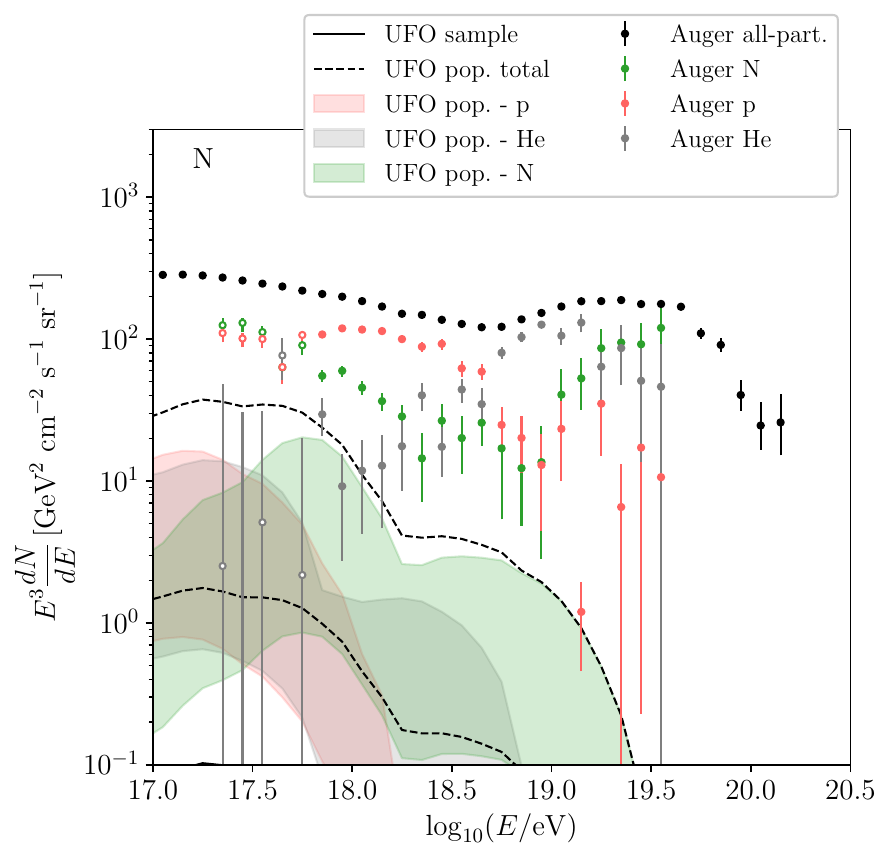}
        \includegraphics[width=0.48\linewidth]{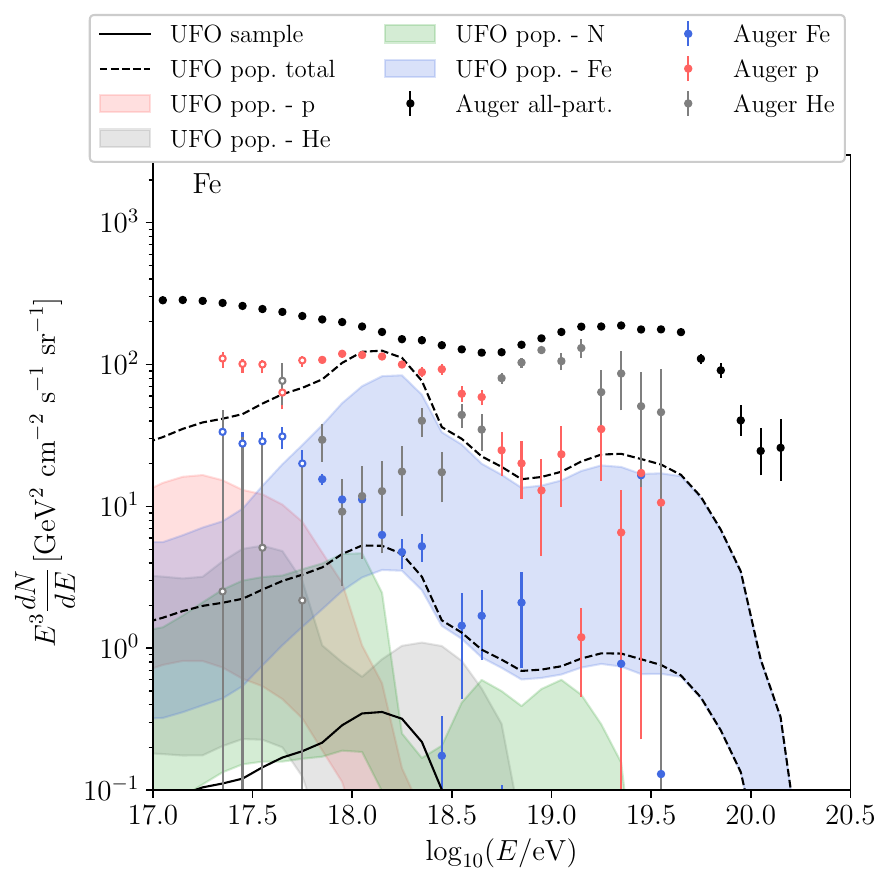}
        \caption{Same as \cref{fig:population_flux} but with the kinetic luminosity limited to at most the bolometric luminosity.}\label{fig:population_flux_limLkin}
    \end{figure*}
    
    To estimate the effect of the large mass outflow rates, we re-analyse all observed UFOs, imposing the constraint that the kinetic luminosity of the wind can not exceed the bolometric luminosity of the AGN. For UFOs where this is the case (approx.\ $25\%$ of the sample), we manually reduce the mass outflow rate until $L_\text{kin}/\lbol\leq1$. The number of UFOs with maximum energy after escape of $10^{18}\,\si{\electronvolt}$ or more is reduced from 23 to 16 for protons, but remains invariant for iron nuclei (six UFOs), see \cref{fig:Emax_radius_correlation_limLkin}. The maximum energy of the remaining candidate sources is, however, reduced.

    With the constrained mass outflow rate, the population of UFOs can still marginally supply the observed sub-ankle protons and comfortably the UHE iron; however, it is unable to explain the observed flux of UHE helium and nitrogen nuclei above the ankle (\cref{fig:population_flux_limLkin}). This is because, in our standard scenario, the expected nuclei flux at Earth was dominated by a limited number of CR-bright sources with large $L_\text{wind}/\lbol$ ratios in our fiducial model. The proton flux expected from our UFO sample is dominated by NGC\,1068, which shows only tentative evidence for an UFO~\citep[see][]{Yamada:2024lss}. Nevertheless, the predicted proton flux from the population of UFOs is reduced by no more than a factor of a few when this source is omitted.

\onecolumn
\section{The UFO sample}\label{apx:ufo_list}
\begin{longtable}{llrlrrrll}
\caption{List of investigated ultra-fast outflows and their nominal parameters: (1) AGN name, (2) reference publication, (3) redshift, (4) $2-10\,$keV X-ray and bolometric luminosity of the AGN [erg\,/\,s], (5) black hole mass [$M_\odot$], (6) mass outflow rate [g\,/\,s], (7) terminal wind velocity [c], (8 and 9) maximum energy at acceleration and after escape for primary protons and iron, respectively, and energy of 10-fold escaping flux suppression where relevant (only for iron), all in [eV]. The references are S23: \citet{Matzeu:2023A&A}, T10: \citet{Tombesi:2010a}, C21: \citet{Chartas:2021ApJ}, XW: \citet{Yamada:2024lss}. Parameters for the first three samples were taken from the meta-analysis by \citet{Gianolli:2024jkq}. The mass outflow rate was calculated for all UFOs based on the observables (see \cref{apx:outflow_rate}). An extended, electronic version of this table is available at \citet{ehlert_2025_zenodo}.} \label{tab:ufo_list} \\
\toprule
$\mathrm{Name}$ & $\mathrm{Ref.}$ & $z$ & $\log(L_\mathrm{X}|L_\mathrm{bol})$ & $\log(M_\mathrm{BH})$ & $\log(\dot{M}_\mathrm{w})$ & $v_\mathrm{out}$ & $\log\,E_\mathrm{max}^\mathrm{acc|esc}(\mathrm{p})$ & $\log\,E_\mathrm{max}^\mathrm{acc|esc}(\mathrm{Fe})$ \\
\midrule
\endfirsthead
\toprule
$\mathrm{Name}$ & $\mathrm{Ref.}$ & $z$ & $\log(L_\mathrm{X}|L_\mathrm{bol})$ & $\log(M_\mathrm{BH})$ & $\log(\dot{M}_\mathrm{w})$ & $v_\mathrm{out}$ & $\log\,E_\mathrm{max}^\mathrm{acc|esc}(\mathrm{p})$ & $\log\,E_\mathrm{max}^\mathrm{acc|esc}(\mathrm{Fe})$ \\
\midrule
\endhead
\midrule
\multicolumn{9}{r}{Continued on next page} \\
\midrule
\endfoot
\bottomrule
\endlastfoot
PG-1202+281 & S23 & 0.165 & 44.40 | 45.73 & 8.61 & 26.43 & 0.108 & 18.0 | 16.8 & 18.8 | 16.2 \\
PG-0947+396 & S23 & 0.205 & 44.21 | 45.50 & 8.68 & 26.07 & 0.305 & 18.4 | 17.4 & 19.2 | 16.4 \\
LBQS-1338-0038 & S23 & 0.237 & 44.52 | 45.87 & 7.77 & 24.80 & 0.152 & 17.6 | 16.4 & 18.6 | 15.6 \\
PG-1114+445 & S23 & 0.144 & 43.98 | 45.24 & 8.59 & 26.45 & 0.071 & 17.9 | 18.0 & 18.8 | 16.4 | 16.6 \\
PG-0804+761 & S23 & 0.100 & 44.45 | 45.78 & 8.31 & 26.16 & 0.130 & 18.0 | 16.8 & 18.8 | 16.0 \\
2MASX-J165315+2349 & S23 & 0.103 & 43.79 | 45.04 & 6.98 & 24.71 & 0.110 & 17.6 | 16.4 & 18.7 | 15.6 \\
2MASX-J105144+3539 & S23 & 0.159 & 43.69 | 44.93 & 8.40 & 24.94 & 0.237 & 18.1 | 18.0 & 18.9 | 16.4 \\
NGC-4151 & T10 & 0.033 & 42.34 | 43.53 & 7.36 & 23.22 & 0.106 & 17.2 | 17.0 & 18.5 | 16.8 | 18.6 \\
IC-4329A & T10 & 0.016 & 43.70 | 44.94 & 7.68 & 23.57 & 0.098 & 17.1 | 16.4 & 18.1 | 15.8 \\
NGC-4051 & T10 & 0.002 & 41.65 | 42.84 & 5.89 & 22.58 & 0.128 & 17.1 | 17.0 & 18.5 | 18.4 | 18.8 \\
PG-1211+143 & T10 & 0.081 & 43.70 | 44.94 & 7.61 & 24.38 & 0.151 & 17.7 | 16.8 & 18.7 | 16.0 \\
MCG-5-23-16 & T10 & 0.009 & 43.10 | 44.31 & 7.45 & 24.01 & 0.116 & 17.5 | 17.2 & 18.6 | 16.2 \\
NGC-4507 & T10 & 0.012 & 43.10 | 44.31 & 6.40 & 22.15 & 0.199 & 17.1 | 16.2 & 18.2 | $<15.5$ \\
NGC-7582 & T10 & 0.005 & 41.60 | 42.79 & 7.67 & 24.76 & 0.285 & 18.2 | 18.2 & 19.6 | 19.0 | 19.8 \\
Ark-120 & T10 & 0.033 & 44.00 | 45.27 & 8.07 & 23.63 & 0.306 & 17.8 | 16.8 & 18.7 | 15.8 | 16.0 \\
Mrk-509 & T10 & 0.034 & 43.97 | 45.23 & 8.04 & 24.33 & 0.182 & 17.7 | 16.6 & 18.7 | 16.0 \\
Mrk-79 & T10 & 0.022 & 43.40 | 44.62 & 7.61 & 24.95 & 0.091 & 17.6 | 17.0 & 18.7 | 16.2 \\
Mrk-766 & T10 & 0.013 & 42.73 | 43.93 & 6.14 & 22.89 & 0.085 & 16.9 | 16.4 & 18.1 | $<15.5$ | 15.6 \\
Mrk-841 & T10 & 0.036 & 43.50 | 44.72 & 8.52 & 24.94 & 0.034 & 17.0 | 16.8 & 18.1 | 16.4 | 16.6 \\
1H-0419-577 & T10 & 0.104 & 44.30 | 45.61 & 8.60 & 25.95 & 0.076 & 17.7 | 16.8 & 18.6 | 16.0 | 16.2 \\
Mrk-290 & T10 & 0.030 & 43.20 | 44.41 & 7.28 & 24.58 & 0.142 & 17.8 | 17.8 & 18.8 | 16.2 \\
Mrk-205 & T10 & 0.071 & 43.80 | 45.05 & 8.40 & 25.57 & 0.100 & 17.8 | 17.2 & 18.7 | 16.4 \\
APM-08279+5255 & C21 & 3.910 & 46.25 | 48.21 & 10.00 & 27.77 & 0.320 & 18.4 | 16.2 & 19.4 | $<15.5$ \\
HS-1700+6416 & C21 & 2.735 & 45.36 | 46.95 & 10.20 & 28.57 & 0.380 & 18.9 | 17.4 & 19.9 | 16.6 \\
MG-J0414+0534 & C21 & 2.640 & 44.50 | 45.85 & 9.00 & 26.78 & 0.280 & 18.5 | 18.2 & 19.3 | 16.4 \\
SDSS-J1442+4055 & C21 & 2.593 & 44.77 | 46.17 & 9.70 & 27.87 & 0.470 & 18.9 | 18.2 & 20.1 | 16.8 | 17.0 \\
SDSS-J1029+2623 & C21 & 2.197 & 44.05 | 45.33 & 8.80 & 26.70 & 0.580 & 18.9 | 18.4 & 20.0 | 17.0 | 17.2 \\
SDSS-J1529+1038 & C21 & 1.984 & 44.06 | 45.33 & 8.90 & 27.07 & 0.250 & 18.6 | 18.2 & 19.6 | 17.0 \\
PG-1115+080 & C21 & 1.720 & 44.17 | 45.46 & 8.80 & 26.96 & 0.230 & 18.6 | 18.0 & 19.4 | 16.6 | 16.8 \\
Q-2237+0305 & C21 & 1.695 & 44.14 | 45.42 & 9.10 & 27.26 & 0.180 & 18.5 | 17.8 & 19.4 | 16.8 | 17.0 \\
SDSS-J1353+1138 & C21 & 1.627 & 44.72 | 46.11 & 9.40 & 27.43 & 0.340 & 18.7 | 17.8 & 19.6 | 16.6 \\
SDSS-J1128+2402 & C21 & 1.608 & 44.34 | 45.65 & 8.70 & 26.74 & 0.560 & 18.9 | 18.2 & 19.8 | 16.6 \\
HS-0810+2554 & C21 & 1.510 & 43.65 | 44.88 & 8.60 & 26.45 & 0.430 & 18.8 | 18.4 & 19.9 | 17.4 | 18.2 \\
SDSS-J0921+2854 & C21 & 1.410 & 45.21 | 46.74 & 8.90 & 26.89 & 0.470 & 18.7 | 16.8 & 19.5 | 15.8 | 16.0 \\
1E-0754.6+3928 & XW & 0.096 & 43.70 | 44.94 & 8.02 & 24.89 & 0.231 & 18.1 | 17.4 & 18.8 | 16.2 | 16.4 \\
1ES-1927+654 & XW & 0.019 & 42.50 | 43.69 & 6.00 & 22.91 & 0.265 & 17.6 | 17.4 & 18.9 | 15.6 | 18.4 \\
1H-0707-495 & XW & 0.041 & 42.66 | 43.86 & 6.31 & 22.96 & 0.139 & 17.3 | 17.2 & 18.6 | 15.6 | 18.4 \\
1H-1934-063 & XW & 0.010 & 42.75 | 43.95 & 6.61 & 20.54 & 0.075 & 16.0 | 15.8 & 16.5 | 15.6 | 15.8 \\
2MASS-J1051+3539 & XW & 0.159 & 44.15 | 45.44 & 8.40 & 24.94 & 0.236 & 18.0 | 17.0 & 18.8 | 16.0 | 16.2 \\
2MASS-J1653+2349 & XW & 0.103 & 44.06 | 45.33 & 8.17 & 25.91 & 0.111 & 17.9 | 17.2 & 18.8 | 16.2 \\
3C-105 & XW & 0.089 & 44.30 | 45.61 & 8.59 & 24.66 & 0.226 & 17.9 | 16.8 & 18.7 | 15.8 | 16.0 \\
3C-111 & XW & 0.049 & 44.53 | 45.88 & 8.45 & 25.40 & 0.083 & 17.5 | 16.6 & 18.5 | 15.8 \\
3C-120 & XW & 0.033 & 43.98 | 45.24 & 7.74 & 23.73 & 0.101 & 17.1 | 16.2 & 18.0 | 15.6 | 15.8 \\
3C-390.3 & XW & 0.056 & 44.52 | 45.87 & 8.71 & 25.08 & 0.146 & 17.7 | 16.6 & 18.6 | 15.8 \\
3C-445 & XW & 0.056 & 44.25 | 45.55 & 7.89 & 25.58 & 0.034 & 17.1 | 16.2 & 18.0 | 15.8 \\
4C-+74.26 & XW & 0.104 & 44.87 | 46.30 & 9.83 & 26.09 & 0.114 & 17.8 | 16.8 & 18.6 | 16.0 \\
Ark-564 & XW & 0.024 & 43.34 | 44.56 & 6.27 & 21.89 & 0.186 & 16.7 | 15.8 & 17.6 | $<15.5$ \\
ESO-103-35 & XW & 0.013 & 43.36 | 44.58 & 7.37 & 23.49 & 0.057 & 16.9 | 16.4 & 17.8 | 15.8 | 16.0 \\
I-Zw-1 & XW & 0.061 & 43.65 | 44.88 & 6.97 & 23.70 & 0.264 & 17.8 | 16.4 & 18.9 | 15.6 \\
IC-5063 & XW & 0.011 & 43.02 | 44.22 & 7.74 & 24.53 & 0.311 & 18.2 | 18.0 & 19.1 | 16.8 | 18.0 \\
IRAS-F00183-7111 & XW & 0.327 & 44.32 | 45.63 & 8.66 & 26.49 & 0.179 & 18.3 | 17.2 & 19.0 | 16.4 \\
IRAS-00521-7054 & XW & 0.069 & 43.40 | 44.62 & 7.70 & 23.84 & 0.401 & 18.1 | 18.0 & 18.9 | 16.2 \\
IRAS-04416+1215 & XW & 0.089 & 43.41 | 44.63 & 6.78 & 22.99 & 0.100 & 16.9 | 16.2 & 17.9 | $<15.5$ | 15.6 \\
IRAS-05054+1718(W) & XW & 0.018 & 42.84 | 44.04 & 7.27 & 24.20 & 0.176 & 17.8 | 17.6 & 18.8 | 16.4 | 18.2 \\
IRAS-05189-2524 & XW & 0.043 & 43.40 | 44.62 & 7.40 & 24.81 & 0.111 & 17.7 | 17.2 & 18.7 | 16.0 | 16.2 \\
IRAS-11119+3257 & XW & 0.189 & 44.24 | 45.54 & 8.00 & 26.40 & 0.253 & 18.4 | 17.6 & 19.2 | 16.2 \\
IRAS-13224-3809 & XW & 0.066 & 42.75 | 43.95 & 6.82 & 24.45 & 0.252 & 18.0 | 17.8 & 19.1 | 18.2 | 18.4 \\
IRAS-13349+2438 & XW & 0.108 & 43.87 | 45.12 & 8.62 & 26.67 & 0.182 & 18.4 | 18.4 & 19.3 | 16.8 | 17.0 \\
IRAS-17020+4544 & XW & 0.060 & 43.70 | 44.94 & 6.77 & 21.88 & 0.086 & 16.2 | 15.8 & 16.4 | $<15.5$ \\
IRAS-18325-5926 & XW & 0.020 & 43.37 | 44.59 & 7.76 & 24.06 & 0.181 & 17.7 | 17.4 & 18.7 | 16.2 \\
MCG-01-24-12 & XW & 0.020 & 43.24 | 44.45 & 7.66 & 24.79 & 0.098 & 17.6 | 17.6 & 18.7 | 16.2 | 16.4 \\
MCG-03-58-007 & XW & 0.032 & 43.75 | 44.99 & 8.00 & 25.70 & 0.193 & 18.2 | 18.2 & 19.0 | 16.4 \\
MR-2251-178 & XW & 0.063 & 44.58 | 45.94 & 8.19 & 23.64 & 0.136 & 17.0 | 16.2 & 17.6 | $<15.5$ \\
Mrk-1044 & XW & 0.016 & 42.46 | 43.65 & 6.45 & 21.49 & 0.113 & 16.7 | 16.6 & 17.7 | 15.8 | 16.0 \\
Mrk-1048(=NGC-985) & XW & 0.043 & 43.78 | 45.02 & 7.33 & 24.59 & 0.038 & 16.9 | 16.4 & 17.8 | 15.6 | 15.8 \\
Mrk-231 & XW & 0.042 & 42.65 | 43.84 & 7.87 & 25.15 & 0.241 & 18.2 | 18.0 & 19.4 | 18.4 \\
Mrk-273 & XW & 0.038 & 43.07 | 44.27 & 8.35 & 26.35 & 0.265 & 18.6 | 18.6 & 19.8 | 18.4 \\
Mrk-279 & XW & 0.030 & 43.41 | 44.63 & 7.43 & 24.58 & 0.220 & 18.0 | 18.0 & 18.9 | 16.2 \\
Mrk-335 & XW & 0.025 & 43.21 | 44.42 & 7.23 & 24.18 & 0.121 & 17.5 | 17.4 & 18.7 | 16.0 \\
Mrk-590 & XW & 0.026 & 42.69 | 43.89 & 7.57 & 23.97 & 0.113 & 17.5 | 17.4 & 18.7 | 16.6 | 18.2 \\
NGC-1068 & XW & 0.004 & 43.04 | 44.24 & 7.23 & 24.74 & 0.277 & 18.2 | 18.0 & 19.1 | 16.4 | 18.2 \\
NGC-2992 & XW & 0.008 & 42.16 | 43.35 & 7.48 & 24.74 & 0.298 & 18.2 | 18.2 & 19.5 | 18.6 | 19.8 \\
NGC-5506 & XW & 0.006 & 43.08 | 44.29 & 7.24 & 24.21 & 0.247 & 18.0 | 18.0 & 18.9 | 16.2 | 16.4 \\
NGC-6240 & XW & 0.025 & 43.86 | 45.11 & 8.53 & 26.73 & 0.125 & 18.2 | 18.2 & 19.1 | 16.8 \\
PDS-456 & XW & 0.184 & 44.90 | 46.34 & 8.23 & 25.40 & 0.269 & 18.0 | 16.6 & 19.0 | 15.6 \\
PG-0844+349 & XW & 0.064 & 43.69 | 44.93 & 7.86 & 25.25 & 0.211 & 18.1 | 18.0 & 18.9 | 16.2 | 16.4 \\
PG-1126-041 & XW & 0.060 & 43.25 | 44.46 & 8.10 & 25.91 & 0.064 & 17.7 | 17.8 & 18.8 | 16.8 | 17.2 \\
PG-1402+261 & XW & 0.164 & 44.13 | 45.41 & 7.53 & 25.20 & 0.061 & 17.3 | 16.2 & 18.4 | 15.6 | 15.8 \\
PG-1448+273 & XW & 0.065 & 43.31 | 44.52 & 7.00 & 24.63 & 0.190 & 17.9 | 17.6 & 18.9 | 16.0 \\
PKS-1549-79 & XW & 0.152 & 44.72 | 46.11 & 8.43 & 25.29 & 0.354 & 18.2 | 16.8 & 19.1 | 15.8 \\
SWIFT-J2127.4+5654 & XW & 0.015 & 43.09 | 44.30 & 7.15 & 23.69 & 0.231 & 17.8 | 17.4 & 18.8 | 16.0 | 16.2 \\
Ton-28 & XW & 0.329 & 44.23 | 45.53 & 7.57 & 24.30 & 0.277 & 17.9 | 16.6 & 18.9 | 15.6 \\
Ton-S180 & XW & 0.062 & 43.65 | 44.88 & 7.06 & 21.71 & 0.196 & 16.7 | 16.0 & 17.4 | $<15.5$ \\
WKK-4438 & XW & 0.016 & 42.74 | 43.94 & 6.30 & 23.09 & 0.319 & 17.8 | 17.8 & 19.0 | 15.8 \\
PID352 & XW & 1.600 & 45.43 | 47.04 & 8.70 & 25.94 & 0.139 & 17.5 | 16.2 & 18.5 | $<15.5$ \\
SDSS-J0904+1512 & XW & 1.826 & 44.23 | 45.53 & 9.30 & 26.64 & 0.259 & 18.5 | 18.2 & 19.3 | 16.8 \\
\end{longtable}

\bsp	
\label{lastpage}
\end{document}